\documentclass[twocolumn,tighten,twocolappendix]{aastex63}
\pdfoutput=1 
\usepackage{amsmath,amstext}
\usepackage[T1]{fontenc}
\usepackage{ae,aecompl}
\usepackage[utf8]{inputenc}
\usepackage{apjfonts} 
\usepackage[figure,figure*]{hypcap}
\usepackage{enumitem}
\usepackage{bm}
\usepackage{graphicx}
\usepackage{lineno}
\usepackage{tikz}

\definecolor{wdm_3}{HTML}{FF9966}
\definecolor{wdm_4}{HTML}{FF7733}
\definecolor{wdm_5}{HTML}{FF5500}
\definecolor{wdm_6}{HTML}{CC4400}
\definecolor{wdm_6_5}{HTML}{993300}
\definecolor{wdm_10}{HTML}{662200}

\definecolor{fdm_25_9}{HTML}{E566FF}
\definecolor{fdm_69_4}{HTML}{DD33FF}
\definecolor{fdm_113}{HTML}{D400FF}
\definecolor{fdm_151}{HTML}{AA00CC}
\definecolor{fdm_185}{HTML}{7F0099}
\definecolor{fdm_490}{HTML}{550066}

\definecolor{idm_n2_1e-4_envelope}{HTML}{33BBFF}
\definecolor{idm_n2_1e-4_halfmode}{HTML}{33BBFF}
\definecolor{idm_n2_1e-2_envelope}{HTML}{0088CC}
\definecolor{idm_n2_1e-2_halfmode}{HTML}{0088CC}
\definecolor{idm_n2_1e0_envelope}{HTML}{004466}
\definecolor{idm_n2_1e0_halfmode}{HTML}{004466}

\definecolor{idm_n4_1e-4_envelope}{HTML}{33FFBB}
\definecolor{idm_n4_1e-4_halfmode}{HTML}{33FFBB}
\definecolor{idm_n4_1e-2_envelope}{HTML}{00CC88}
\definecolor{idm_n4_1e-2_halfmode}{HTML}{00CC88}
\definecolor{idm_n4_1e0_envelope}{HTML}{006644}
\definecolor{idm_n4_1e0_halfmode}{HTML}{006644}

\input{hyperlink-year-only-natbib-patch}

\newcommand*{\http}[1]{\href{http://#1}{#1}}
\newcommand*{\https}[1]{\href{https://#1}{#1}}

\shorttitle{COZMIC. I.}
\shortauthors{Nadler et al.}

\begin{document}

\title{COZMIC. I. Cosmological Zoom-in Simulations with Initial Conditions Beyond Cold Dark Matter}

\author[0000-0002-1182-3825]{Ethan O.~Nadler}
\affiliation{Department of Astronomy \& Astrophysics, University of California, San Diego, La Jolla, CA 92093, USA}
\affiliation{Carnegie Observatories, 813 Santa Barbara Street, Pasadena, CA 91101, USA}
\affiliation{Department of Physics $\&$ Astronomy, University of Southern California, Los Angeles, CA 90007, USA}

\author[0000-0001-9543-5012]{Rui An}
\affiliation{Department of Physics $\&$ Astronomy, University of Southern California, Los Angeles, CA 90007, USA}

\author[0000-0002-3589-8637]{Vera Gluscevic}
\affiliation{Department of Physics $\&$ Astronomy, University of Southern California, Los Angeles, CA 90007, USA}

\author[0000-0001-5501-6008]{Andrew Benson}
\affiliation{Carnegie Observatories, 813 Santa Barbara Street, Pasadena, CA 91101, USA}

\author[0000-0003-0728-2533]{Xiaolong Du}
\affiliation{Department of Physics and Astronomy, University of California, Los Angeles, 430 Portola Plaza, Los Angeles, CA 90095, USA}

\correspondingauthor{Ethan~O.~Nadler}
\email{enadler@ucsd.edu}

\label{firstpage}

\begin{abstract}
We present $72$ cosmological dark matter--only $N$-body zoom-in simulations with initial conditions beyond cold, collisionless dark matter (CDM), as the first installment of the COZMIC suite. We simulate Milky Way (MW) analogs with linear matter power spectra $P(k)$ for i) thermal-relic warm dark matter (WDM) with masses $m_{\mathrm{WDM}}\in [3,4,5,6,6.5,10]~\mathrm{keV}$, ii) fuzzy dark matter (FDM) with masses $m_{\mathrm{FDM}}\in [25.9,69.4,113,151,185,490]\times 10^{-22}~\mathrm{eV}$, and iii) interacting dark matter (IDM) with a velocity-dependent elastic proton scattering cross section $\sigma=\sigma_0 v^n$ relative particle velocity scaling $n\in [2,4]$, and dark matter mass $m_{\mathrm{IDM}}\in[10^{-4},~ 10^{-2},~ 1]~\mathrm{GeV}$. Subhalo mass function (SHMF) suppression is significantly steeper in FDM versus WDM, while dark acoustic oscillations in $P(k)$ can reduce SHMF suppression for IDM. We fit SHMF models to our simulation results and derive new bounds on WDM and FDM from the MW satellite population, obtaining $m_{\mathrm{WDM}}>5.9~\mathrm{keV}$ and $m_{\mathrm{FDM}}>1.4\times 10^{-20}~\mathrm{eV}$ at $95\%$ confidence; these limits are $\approx 10\%$ weaker and $5\times$ stronger than previous constraints owing to the updated transfer functions and SHMF models, respectively. We estimate IDM bounds for $n=2$ ($n=4$) and obtain $\sigma_0 < 1.0\times 10^{-27}$, $1.3\times 10^{-24}$, and $3.1\times 10^{-23}~\mathrm{cm}^2$ ($\sigma_0 < 9.9\times 10^{-27}$, $9.8\times 10^{-21}$, and $2.1\times 10^{-17}~\mathrm{cm}^2$) for $m_{\mathrm{IDM}}=10^{-4}$, $10^{-2}$, and $1$ GeV, respectively. Thus, future development of IDM SHMF models can improve IDM cross section bounds by up to a factor of $\sim 20$ with current data. COZMIC presents an important step toward accurate small-scale structure modeling in beyond-CDM cosmologies, critical to upcoming observational searches for dark matter physics.
\end{abstract}

\keywords{\href{http://astrothesaurus.org/uat/353}{Dark matter (353)}; 
\href{http://astrothesaurus.org/uat/574}{Galaxy abundances (574)};
\href{http://astrothesaurus.org/uat/1049}{Milky Way dark matter halo (1049)}; 
\href{http://astrothesaurus.org/uat/1083}{$N$-body simulations (1083)};
\href{http://astrothesaurus.org/uat/1787}{Warm dark matter (1787)};
\href{http://astrothesaurus.org/uat/1880}{Galaxy dark matter halos (1880)}}

\section{Introduction}\label{sec:intro}

Low-mass dark matter (DM) halos are a key cosmological probe. In the standard cold, collisionless (CDM) paradigm, DM is at most weakly coupled to the thermal plasma; in canonical weakly interacting massive particle (WIMP) models, this allows halos to form down to Earth-mass scales  \citep{Green0309621,Diemand0501589}. Cosmologies that feature DM physics beyond gravity generically alter small-scale linear density perturbations and (sub)halo populations, often on significantly larger scales than in WIMP models. We collectively refer to these as ``beyond-CDM'' scenarios.

Beyond-CDM model building has often been driven by potential tensions between CDM predictions and small-scale structure data. A popular scenario is thermal-relic warm dark matter (WDM), which free streams on relevant scales if particle masses are $\mathcal{O}(\mathrm{keV})$. WDM was proposed as a solution to the ``missing satellites'' problem \citep{Klypin9901240,Moore9907411} because it can reduce low-mass halo abundances (e.g., \citealt{Goetz0210599,Lovell13081399}).

Other forms of DM microphysics can alter matter clustering on small scales. For example, wave interference in fuzzy dark matter (FDM) models, featuring ultralight scalar fields with particle masses of $\mathcal{O}(10^{-22}~\mathrm{eV})$, suppresses small-scale power \citep{Hu008506,Marsh151007633,Hui179504}. Collisional damping in interacting dark matter (IDM) models---which feature nongravitational elastic scattering between DM particles and baryons or radiation---also suppress density perturbations \citep{Boehm0012504,Boehm0410591,McDermott10112907, Dvorkin13112937,Boddy189808,Gluscevic1812108}. Some IDM scenarios lead to a suppression of the linear matter power spectrum, $P(k)$, that resembles the smooth cutoff seen in thermal-relic WDM \citep{Boehm0112522,Nadler190410000}, while others produce prominent dark acoustic oscillations (DAOs; e.g., \citealt{Boddy189808, Maamari201002936}); thus, structure formation is a leading probe of such DM interactions (see \citealt{Gluscevic190305140} for a review). In WDM, FDM, and IDM, the effects of DM microphysics are most prominent on small scales. 

Our understanding of small-scale structure has dramatically advanced in recent years. For example, CDM predictions for the observable population of Milky Way (MW) satellite galaxies were shown to depend sensitively on baryonic physics, including photoionization feedback and tidal stripping due to the Galactic disk, as well as observational incompleteness (see, e.g., \citealt{Bullock10094505} and \citealt{Bullock170704256} for reviews). Studies that account for these effects and their uncertainties find that CDM predictions are consistent with current MW satellite observations \citep{Kim1812121,Nadler191203303}. As a result, MW satellite abundances now place stringent constraints on WDM, FDM, and IDM  \citep{Nadler190410000,Nadler200800022,Maamari201002936,Newton201108865,Nguyen210712380,Dekker211113137,Newton240816042}. Complementary probes of small-scale structure including the Ly$\alpha$ forest \citep{Viel1308804,Irsic179602,Irsic1711903,Irsic230904533,Rogers200712705,Rogers211110386,Villasenor220914220}, strong gravitational lensing \citep{Gilman190806983,Hsueh190504182,Powell230210941,Keeley240501620}, stellar streams \citep{Banik191102663}, and combinations thereof \citep{Enzi201013802,Nadler210107810} have also been used to constrain beyond-CDM scenarios.

Predictions for (sub)halo populations in beyond-CDM cosmologies are a key input to many of these studies. However, only a handful of cosmological simulations with sufficient resolution to accurately model low-mass (sub)halos have been performed in the beyond-CDM cosmologies described above. To illustrate this point, we consider zoom-in simulations of MW-mass systems, which are generally needed to achieve sufficient resolution for studies of the MW satellite galaxy population.\footnote{Although we focus on zoom-ins, we note that several large-volume cosmological simulations have been run in the beyond-CDM scenarios we consider (e.g., \citealt{Angulo13042406,Bose150701998,Murgia170407838,Stucker210909760,May220914886,Meshveliani231200150,Rose240500766,Zhang240218880}).} For WDM, \cite{Lovell13081399} ran $N$-body simulations of one MW-mass system in four thermal-relic models, and \cite{Lovell161100010,Lovell181005168} and \cite{Lovell191111785} presented hydrodynamic simulations of six Local Group--like pairs in two sterile neutrino models each. For FDM, \cite{Elgamal231103591} ran 16 simulations of MW-mass systems including both $P(k)$ suppression and FDM dynamics. For IDM, \cite{Schewtschenko151206774} ran $N$-body simulations of four Local Group--like pairs in one DM--radiation scattering IDM model, with additional IDM models for one pair, while \cite{Vogelsberger151205349} and subsequent work in the Effective Theory of Structure Formation (ETHOS) framework \citep{Cyr-Racine151505344} simulated a handful of MW-mass systems with DM--dark radiation interaction initial conditions (ICs). To our knowledge, no zoom-in simulations with ICs appropriate for DM--baryon elastic scattering IDM models have been published.

Thus, beyond-CDM parameter space is sparsely sampled by current zoom-in simulations. As a result, beyond-CDM models are thus far mostly constrained by leveraging simulations of well-studied scenarios like thermal-relic WDM. For example, constraints have been derived by matching the wavenumber where $P(k)$ is suppressed by a characteristic amount (e.g., \citealt{Nadler190410000,Nguyen210712380}), or by matching integrals of $P(k)$ (e.g., \citealt{Schneider160107553,Dienes211209105}) to thermal-relic WDM. However, the uncertainties associated with such techniques are difficult to quantify (see \citealt{Schutz200105503} for an example in the context of FDM). As a result, studies often conservatively map WDM constraints to beyond-CDM scenarios (e.g., \citealt{Maamari201002936,Nadler200800022}), leaving orders of magnitude in parameter space untested owing to a lack of precise theoretical predictions for (sub)halo populations. Furthermore, due to the limited number of existing simulations even for benchmark models like thermal-relic WDM, the uncertainties associated with commonly used fitting functions (e.g., for the suppression of (sub)halo abundances relative to CDM) have not been systematically quantified. Dedicated beyond-CDM simulation suites are therefore timely as the community prepares to robustly analyze next-generation small-scale structure data \citep{Banerjee220307049,Nadler240110318}.

As a step toward this goal, we present the first installment of COZMIC: \textbf{CO}smological \textbf{Z}oo\textbf{M}-in simulations with \textbf{I}nitial \textbf{C}onditions beyond CDM. In particular, we run $72$ DM-only zoom-in simulations of three MW-mass systems in the WDM, FDM, and IDM scenarios described above. Following the approach of the recent Symphony \citep{Nadler220902675} and Milky Way-est \citep{Buch240408043} compilations of CDM zoom-in simulations, two of our zoom-in hosts resemble the MW in detail and include LMC-analog subhalos and realistic merger histories. We simulate 6 WDM models, 6 FDM models, and 12 DM--proton scattering IDM models, for each of these three systems, yielding $72$ new simulations; we also present $24$ higher-resolution resimulations of one host across all models to assess convergence. Our simulations span models that bracket current observational constraints for all three beyond-CDM scenarios, allowing us to derive accurate subhalo population predictions that will facilitate robust constraints using upcoming data.

We choose to simulate WDM, FDM, and IDM to study the effects of the $P(k)$ cutoff shape (by comparing FDM to WDM) and DAO features (by comparing IDM to WDM). We show that both effects impact subhalo populations, which suggests that $P(k)$ can be reconstructed using future small-scale structure data (see, e.g., \citealt{Nadler240110318}). To isolate the effects of ICs on small-scale structure, we only modify $P(k)$ when generating ICs and perform standard $N$-body simulations for all scenarios. We do not include effects such as thermal velocities of WDM particles (which are expected to be small for the models we simulate; \citealt{Leo170607837}) or Schr{\"o}dinger--Poisson dynamics of FDM particles (which can affect halo abundances at a level similar to modified ICs; \citealt{May220914886}). Thus, in each scenario, we assume that $100\%$ of the DM is a non-CDM species that only affects $P(k)$ and subsequently evolves like CDM; we relax each of these assumptions in upcoming COZMIC papers.

To demonstrate the utility of COZMIC, we derive new subhalo mass function (SHMF) suppression models from our simulation results and use these to update constraints on WDM, FDM, and IDM. Specifically, we incorporate our SHMF suppression models into a forward model of the MW satellite galaxy population observed by the Dark Energy Survey (DES) and Pan-STARRS1 (PS1), as compiled by \cite{Drlica-Wagner191203302}, using the framework from \cite{Nadler191203303,Nadler200800022}. For FDM, this yields a factor of $\approx 5$ improvement over the constraint from \cite{Nadler200800022}. Furthermore, we conservatively estimate new upper bounds on the velocity-dependent DM--proton scattering IDM cross section by matching subhalo abundances in these scenarios to WDM models. These limits improve upon those from \cite{Maamari201002936} by one order of magnitude, on average.

Two studies accompany this work. In \defcitealias{An241103431}{Paper~II} An et al.\ (\citeyear{An241103431}, hereafter \citetalias{An241103431}), we present $24$ simulations with a fractional non-CDM component that features a suppression and plateau in the ratio of $P(k)$ relative to CDM. \citetalias{An241103431} derives a fitting function for the SHMF suppression as a function of the $P(k)$ suppression scale and plateau height, and presents new bounds from the MW satellite population on these parameters and on fractional thermal-relic WDM models. In \defcitealias{Nadler241213065}{Paper~III} Nadler et al.\ (\citeyear{Nadler241213065}, hereafter \citetalias{Nadler241213065}), we present eight high-resolution simulations of beyond-CDM models that feature both strong, velocity-dependent self-interacting dark matter (SIDM) and $P(k)$ suppression; we study the interplay between these two effects and gravothermal core collapse for the first time. \citetalias{Nadler241213065} builds on recent SIDM simulations (see, e.g., \citealt{Tulin170502358,Adhikari220710638} for reviews) by simultaneously modeling its impact on linear matter perturbations and (sub)halo abundances and density profiles.

This paper is organized as follows: In Section~\ref{sec:model}, we describe the beyond-CDM scenarios and models we consider; in Section~\ref{sec:simulations}, we describe our IC and simulation pipeline; in Section~\ref{sec:basic}, we present SHMF measurements; in Section~\ref{sec:shmf}, we model SHMF suppression in our beyond-CDM scenarios. We derive updated bounds on WDM, FDM, and IDM from the MW satellite population in Section~\ref{sec:limits}, discuss caveats and areas for future work in Section~\ref{sec:caveats}, and conclude in Section \ref{sec:conclusions}. Appendices present transfer function calculations (Appendix~\ref{sec:tk_details}), additional simulation results (Appendix~\ref{sec:simulation_table}), convergence tests (Appendix~\ref{sec:convergence}), a study of artificial fragmentation (Appendix~\ref{sec:fragmentation}), a study of SHMF suppression universality across our three hosts (Appendix~\ref{sec:shmf_details}), and WDM and FDM MW satellite inference posteriors (Appendix~\ref{sec:full_posteriors}).

We adopt cosmological parameters used for the Symphony Milky Way and Milky Way-est CDM simulations: $h = 0.7$, $\Omega_{m} = 0.286$, $\Omega_{b} = 0.049$, $\Omega_{\Lambda} = 0.714$, $\sigma_8 = 0.82$, and $n_s=0.96$ \citep{Hinshaw_2013}. Halo masses are defined via the \cite{Bryan_1998} virial overdensity, which corresponds to $\Delta_{\mathrm{vir}}\approx 99\times \rho_{\mathrm{crit}}$ in our cosmology, where $\rho_{\mathrm{crit}}$ is the critical density of the universe at $z=0$. We refer to DM halos within the virial radius of our MW hosts as ``subhalos,'' and we refer to halos that are not within the virial radius of any larger halo as ``isolated halos.'' Throughout, ``log'' always refers to the base-$10$ logarithm and we work in natural units with $\hbar=c=1$.


\section{Beyond-CDM Scenarios}
\label{sec:model}

We begin with a general overview of WDM (Section~\ref{sec:wdm_model}), FDM (Section~\ref{sec:fdm_model}), and IDM (Section~\ref{sec:idm_model}) scenarios, including relevant scaling relations. We then describe the specific models we simulate and our procedure for generating ICs in Section~\ref{sec:simulations}.

Many beyond-CDM models only affect matter clustering at early times, when perturbations are small, while late-time, nonlinear evolution is indistinguishable from CDM. In such scenarios, structure formation can be modeled using simulation techniques developed for CDM, with ICs quantified by
\begin{equation}
    \mathcal{T}^2_{\mathrm{beyond-CDM}}(k) \equiv \frac{P_{\mathrm{beyond-CDM}}(k)}{P_{\mathrm{CDM}}(k)},
\end{equation}
where $k$ is the comoving wavenumber, $\mathcal{T}_{\mathrm{beyond-CDM}}(k)$ is the transfer function, and $P_{\mathrm{beyond-CDM}}(k)$ and $P_{\mathrm{CDM}}(k)$ are linear matter power spectra in a beyond-CDM model and in CDM, respectively. Note that $\mathcal{T}_{\mathrm{beyond-CDM}}(k)$ depends on parameters specific to each beyond-CDM scenario we consider. 

The transfer functions we consider feature suppression of power small scales, so we define the half-mode wavenumber where the power drops to a quarter of that in CDM
\begin{equation}
    \mathcal{T}^2_{\mathrm{beyond-CDM}}(k_{\mathrm{hm}}) \equiv 0.25.
\end{equation}
Halo mass and wavenumber are related in linear theory via \citep{Nadler190410000}
\begin{equation}
    M(k) \equiv \frac{4\pi}{3}\Omega_m\bar{\rho}\left(\frac{\pi}{k}\right)^3=5.1\times 10^9~M_{\mathrm{\odot}}\times\left(\frac{k}{10~\mathrm{Mpc}^{-1}}\right)^{-3},\label{eq:m_k}
\end{equation}
where $\Omega_{m}$ is the DM density fraction, $\bar{\rho}$ is the average density of the universe at $z=0$, and we have evaluated the expression numerically in our fiducial cosmology. The half-mode mass associated with $k_{\mathrm{hm}}$ is $M_{\mathrm{hm}}\equiv M(k_{\mathrm{hm}})$.

\subsection{Warm Dark Matter}
\label{sec:wdm_model}

WDM refers to particles with masses of $\mathcal{O}(\mathrm{keV})$ that decouple while still (semi)relativistic, leading to a nonnegligible free streaming length and suppression of small-scale density perturbations \citep{Bond1983,Bode0010389}. We consider the simplest case of thermal-relic WDM, for which $P(k)$ is determined solely by the WDM particle mass,  $m_{\mathrm{WDM}}$. Note that nonthermal production mechanisms (e.g., in the case of sterile neutrino models; \citealt{Boyarsky180707938}) introduce additional model dependence; we leave simulations of specific WDM particle models beyond the thermal-relic paradigm to future work.

We use the WDM transfer function \citep{Viel0501562}
\begin{equation}
    \mathcal{T}^2_{\mathrm{WDM}}(k,m_{\mathrm{WDM}}) = \left[1+(\alpha(m_{\mathrm{WDM}})\times k)^{2\nu}\right]^{-10/\nu}.\label{eq:transfer_wdm}
\end{equation}
We use $\nu=1.049$ and \citep{Vogel221010753}
\begin{equation}
\alpha(m_{\mathrm{WDM}})=a \left(\frac{m_{\mathrm{WDM}}}{1~\mathrm{keV}}\right)^b \left(\frac{\omega_{\mathrm{WDM}}}{0.12}\right)^{\eta}\left(\frac{h}{0.6736}\right)^{\theta}h^{-1}~\mathrm{Mpc},\label{eq:alpha_wdm}
\end{equation}
where $a=0.0437$, $b=-1.188$, $\theta=2.012$, and $\eta=0.2463$ for spin-$1/2$ particles.\footnote{\cite{Vogel221010753} find that the commonly used \cite{Viel0501562} $\alpha(m_{\mathrm{WDM}})$ fit yields transfer functions that are $\approx 10\%$ too for~$m_{\mathrm{WDM}}\gtrsim 3\ \mathrm{keV}$ (see also \citealt{Decant211109321}).} Here $\omega_{\mathrm{WDM}}\equiv \Omega_{\mathrm{WDM}}h^2$ is the full matter density. Combining Equations~\ref{eq:transfer_wdm} and \ref{eq:alpha_wdm} yields
\begin{equation}
    M_{\mathrm{hm}}(m_{\mathrm{WDM}}) = 4.3\times 10^8~M_{\mathrm{\odot}}\times\left(\frac{m_{\mathrm{WDM}}}{3~\mathrm{keV}}\right)^{-3.564}\label{eq:Mhm_mwdm}
\end{equation}
in our fiducial cosmology.

We also quote WDM free streaming wavenumbers defined by \citep{Schneider11120330}
\begin{equation}
    k_{\mathrm{fs}} \equiv 13.93\times k_{\mathrm{hm}}.\label{eq:fs}
\end{equation}
The free streaming scale, $\lambda_{\mathrm{fs}}\equiv 2\pi/k_{\mathrm{fs}}$, is $\approx 20~\mathrm{kpc}$, for the warmest model we simulate, $m_{\mathrm{WDM}}=3~\mathrm{keV}$ (see Table~\ref{tab:summary}). Because $\lambda_{\mathrm{fs}}\ll \lambda_{\mathrm{hm}}$, the direct effects of WDM free streaming (e.g., thermal velocities of WDM particles) are negligible on the scales we simulate, consistent with previous studies (e.g., \citealt{Maccio12021282,Angulo13042406,Leo170607837}). Quantitatively, the mean thermal velocity at $z=99$, when our simulations are initialized, is $\approx 1~\mathrm{km\ s}^{-1}$ for $m_{\mathrm{WDM}}=3~\mathrm{keV}$. This is much smaller than the internal velocities induced of the smallest halos our simulations resolve, justifying our choice to neglect thermal velocities.

\subsection{Fuzzy Dark Matter}
\label{sec:fdm_model}

FDM refers to scalar field DM with an $\mathcal{O}(10^{-22}~\mathrm{eV})$ particle mass and $\mathcal{O}(\mathrm{kpc})$ de Broglie wavelength \citep{Hu008506,Hui179504}. FDM is produced cold by a nonthermal mechanism such as axion misalignment (see, e.g., \citealt{Marsh151007633}). Small-scale power is suppressed in FDM cosmologies owing to wave interference, resulting in a $P(k)$ cutoff steeper than in the case of WDM. We assume that the FDM particle mass, $m_{\mathrm{FDM}}$, is the only parameter that determines the transfer function and the shape of $P(k)$. Some ultralight DM models can violate this assumption and lead to $P(k)$ enhancement on certain scales, along with a small-scale cutoff (e.g., due to self-interactions; \citealt{Arvanitaki190911665}); we leave simulations of these scenarios to future studies.

We use the FDM transfer function \citep{Passaglia220110238}
\begin{equation}
    \mathcal{T}^2_{\mathrm{FDM}}(k,m_{\mathrm{FDM}}) = \left[\frac{\sin(x^m)}{x^m(1+Bx^{6-m})}\right]^2,\label{eq:transfer_fdm}
\end{equation}
where $m=5/2$, $x\equiv A(k/k_J)$, and the Jeans wavenumber is
\begin{equation}
  k_J \equiv 9~\mathrm{Mpc}^{-1}\times m_{\mathrm{FDM,22}}^{1/2}.\label{eq:kj}
\end{equation} 
Here we have defined
\begin{equation}
    m_{\mathrm{FDM,22}} \equiv \frac{m_{\mathrm{FDM}}}{10^{-22}~\mathrm{eV}},
\end{equation}
and we use $A = 2.22\left(m_{\mathrm{FDM,22}}\right)^{1/25-\ln(m_{\mathrm{FDM,22}})/1000}$ and $B =0.16 \left(m_{\mathrm{FDM,22}}\right)^{-1/20}$, following \cite{Passaglia220110238}.\footnote{Several small-scale structure analyses adopt the \cite{Hu008506} FDM transfer function fit, which underestimates suppression by $\approx 5\%$ compared to the fit in \cite{Passaglia220110238}.} As for WDM, it is again useful to define the half-mode mass, 
\begin{equation}
    M_{\mathrm{hm}}(m_{\mathrm{FDM,22}}) = 4.5\times 10^{10}~M_{\mathrm{\odot}}\times m_{\mathrm{FDM,22}}^{-1.41}\label{eq:mhm_mfdm}.
\end{equation}

Finally, the de Broglie wavelength sets another relevant scale, given by \citep{Hui179504}
\begin{equation}
    \lambda_{\mathrm{dB}} = 1.92~\mathrm{kpc}\times m_{\mathrm{FDM,22}}^{-1}\times\left(\frac{v}{10~\mathrm{km\ s}^{-1}}\right)^{-1}.\label{eq:dB}
\end{equation}
The de Broglie wavelength is $\approx 70~\mathrm{pc}$ for the lightest FDM model we simulate, $m_{\mathrm{FDM,22}}=25.9$. This is smaller than the scales that our simulations resolve; for example, our fiducial gravitational softening length is $\approx 243~\mathrm{pc}$ (Section~\ref{sec:zoom-in}).  Furthermore, Table~\ref{tab:summary} shows that $k_{\mathrm{hm}}\approx k_J$ for the FDM models we simulate, as noted in previous studies (e.g., \citealt{Passaglia220110238}). Thus, the scales we aim to model are mainly affected by Jeans suppression of linear density perturbations rather than late-time wave interference. This hierarchy of scales justifies our use of $N$-body simulations to capture the impact of FDM on subhalos with masses down to $\approx 10^8~M_{\mathrm{\odot}}$.

\subsection{Interacting Dark Matter}
\label{sec:idm_model}

IDM generally refers to a family of models that feature nongravitational interactions between DM and baryons or radiation. Here we focus on interactions between a single DM species, with particle mass $m_{\mathrm{IDM}}$, and protons. Such interactions can arise from a variety of dark sector models; we follow \cite{Boddy180800001} and \cite{Gluscevic1812108} by parameterizing effective velocity-dependent interactions in terms of the momentum transfer cross section
\begin{equation}
    \sigma = \sigma_0 v^n,\label{eq:sigma_mt}
\end{equation}
where $n$ is a power-law exponent, $\sigma_0$ sets the scattering amplitude, and $v$ is the relative particle velocity. Interactions with $n=0$ lead to a small-scale $P(k)$ cutoff that is similar to thermal-relic WDM \citep{Nadler190410000}, so we only run simulations in velocity-dependent scattering models with $n=2$ or $n=4$. However, we return to the $n=0$ case when deriving updated bounds in Section~\ref{sec:limits}. Models with $n=2$ or $n=4$ can yield prominent DAOs on small scales, with an amplitude that depends on $n$ and $m_{\mathrm{IDM}}$ \citep{Maamari201002936}.

Unlike WDM and FDM, fitting functions for $\mathcal{T}^2_{\mathrm{IDM}}(k)$ as a function of $m_{\mathrm{IDM}}$ and $n$ have not been developed. Instead, we summarize our IDM models by their half-mode scales, and---similar to the parameterization in \cite{Bohr200601842} for ETHOS models---we define $h_{\mathrm{peak}}$ and $k_{\mathrm{peak}}$ as the amplitude and wavenumber, respectively, of the first DAO peak in the squared transfer function relative to CDM.

To estimate a characteristic scale below which density perturbations are suppressed in our IDM models, we follow \cite{Nadler190410000} and \cite{Maamari201002936} by calculating the size of the particle horizon when DM and protons kinetically decouple. We write the DM--proton momentum transfer rate \citep{Dvorkin13112937,Gluscevic1812108}
\begin{equation}
    R_{\mathrm{IDM}} = \mathcal{N}_n a \rho_b Y_p\frac{\sigma_0}{m_{\mathrm{IDM}}+m_p}\left(\frac{T_b}{m_p}+\frac{T_{\mathrm{IDM}}}{m_{\mathrm{IDM}}}\right)^{(n+1)/2},
\end{equation}
where $\mathcal{N}_n\equiv2^{(n+5)/2}\Gamma(n/2 + 3)/3\sqrt{\pi}$, $a$ is the scale factor, $\rho_b=\Omega_b\bar{\rho}$ is the baryon energy density, $Y_p=0.75$ is the proton mass fraction, $m_p=0.938~\mathrm{GeV}$ is the proton mass, and $T_b=T_0 (1+z)$ is the baryon temperature, where $T_0=2.73~\mathrm{K}$ is the cosmic microwave background temperature at $z=0$. The IDM temperature, $T_{\mathrm{IDM}}$, is strongly coupled to $T_b$ until the redshift of thermal decoupling, $z_{\mathrm{th}}$, which occurs when the heat transfer rate, $R'_{\mathrm{IDM}}=[m_{\mathrm{IDM}}/(m_{\mathrm{IDM}}+m_p)]R_{\mathrm{IDM}}$, drops below the Hubble rate. After thermal decoupling, DM cools adiabatically via $T_{\mathrm{IDM}}=[(1+z)^2/(1+z_{\mathrm{th}})]T_0$. 

We then calculate the redshift, $z_{\mathrm{dec}}$, at which $R_{\mathrm{IDM}}$ falls below the Hubble rate by solving
\begin{equation}
    aH = R_{\mathrm{IDM}}\rvert_{z=z_{\mathrm{dec}}}.
\end{equation}
The horizon size at $z_{\mathrm{dec}}$ sets the scale below which $\mathcal{T}^2_{\mathrm{IDM}}(k)$ is suppressed. We define a critical wavenumber that undergoes one full oscillation within the horizon at decoupling,
\begin{equation}
    k_{\mathrm{crit}} \equiv 2\left(\frac{1}{aH}\right)^{-1}\bigg\rvert_{z=z_{\mathrm{dec}}}.\label{eq:lambda_dec}
\end{equation}
The decoupling scale for each IDM model we simulate is listed in Table~\ref{tab:summary}. Our analytic calculation predicts suppression scales that match the half-mode scales from our \textsc{CLASS} calculations in Section~\ref{sec:idm_ics} reasonably well, although we systematically find $k_{\mathrm{crit}}\approx 0.5 k_{\mathrm{hm}}$, consistent with the results of \cite{Maamari201002936} for $n=2$ and $n=4$ models. This difference may be related to the number of subhorizon oscillations modes undergo before density perturbations are significantly suppressed on the corresponding scales \citep{Nadler190410000}. We leave a more detailed comparison with these analytic estimates to future work.

Note that $R_{\mathrm{IDM}}$ drops steeply with decreasing redshift and freezes out at $z\approx 10^6$ for the $n\geq0$ models and cross sections we consider. At late times, the momentum transfer cross section per unit DM mass evaluated at a characteristic velocity dispersion in the lowest-mass subhalos our simulations resolve ($v=10~\mathrm{km\ s}^{-1}$) is $\mathcal{O}(10^{-10}~\mathrm{cm}^2~\mathrm{g}^{-1})$. Thus, compared to SIDM models studied in simulations, with typical cross sections of $\mathcal{O}(1~\mathrm{cm}^2~\mathrm{g}^{-1})$ (e.g., \citealt{Tulin170502358}), momentum transfer in the IDM models we consider is small. This justifies our use of DM-only simulations with modified ICs to capture the leading-order effects of IDM on subhalos with masses down to $\approx 10^8~M_{\mathrm{\odot}}$.

We do not simulate $n<0$ IDM models because they suppress power over a wide range of scales, with nonnegligible late-time interactions (e.g., \citealt{Dvorkin13112937,Driskell220904499}); these effects may not be accurately captured by our zoom-in simulations. In particular, each high-resolution region we simulate only encompasses the $\approx (3~\mathrm{Mpc})^3$ Lagrangian volume of the corresponding MW host. Furthermore, DM-only simulations may not capture the full effects of late-time scattering present in $n<0$ IDM models. We leave studies that address these effects to future work.

\vspace{10mm}


\section{Simulation Pipeline}
\label{sec:simulations}

Next, we describe the COZMIC simulation pipeline, including our method for generating ICs (Section~\ref{sec:ics}), running zoom-in simulations (Section~\ref{sec:zoom-in}), and post-processing and analysis (Section~\ref{sec:post-process}). Appendix~\ref{sec:tk_details} provides technical details of our transfer function calculations, and Table~\ref{tab:summary} summarizes all models we simulate in this work.

\subsection{Initial Conditions}
\label{sec:ics}

\subsubsection{Warm Dark Matter}
\label{sec:wdm_ics}

To compute WDM transfer functions, we use the linear Boltzmann solver \textsc{CLASS}~\citep{class}.\footnote{\url{https://github.com/lesgourg/class_public/tree/master}} We generate transfer functions for $m_{\mathrm{WDM}}\in [3,4,5,6,6.5,10]~\mathrm{keV}$. These WDM masses are chosen to bracket current small-scale structure constraints, with $m_{\mathrm{WDM}}=3~\mathrm{keV}$ being excluded by several probes (e.g., see \citealt{Drlica-Wagner190201055} for a review), $m_{\mathrm{WDM}}=6.5~\mathrm{keV}$ corresponding to the \cite{Nadler200800022} bound from the MW satellite population (which we update in this work), and $m_{\mathrm{WDM}}=10~\mathrm{keV}$ roughly corresponding to the most stringent reported WDM limit to date, derived from a combination of strong lensing and MW satellite galaxies \citep{Nadler210107810}.

\begin{figure*}[t!]
\centering
\includegraphics[width=\textwidth]{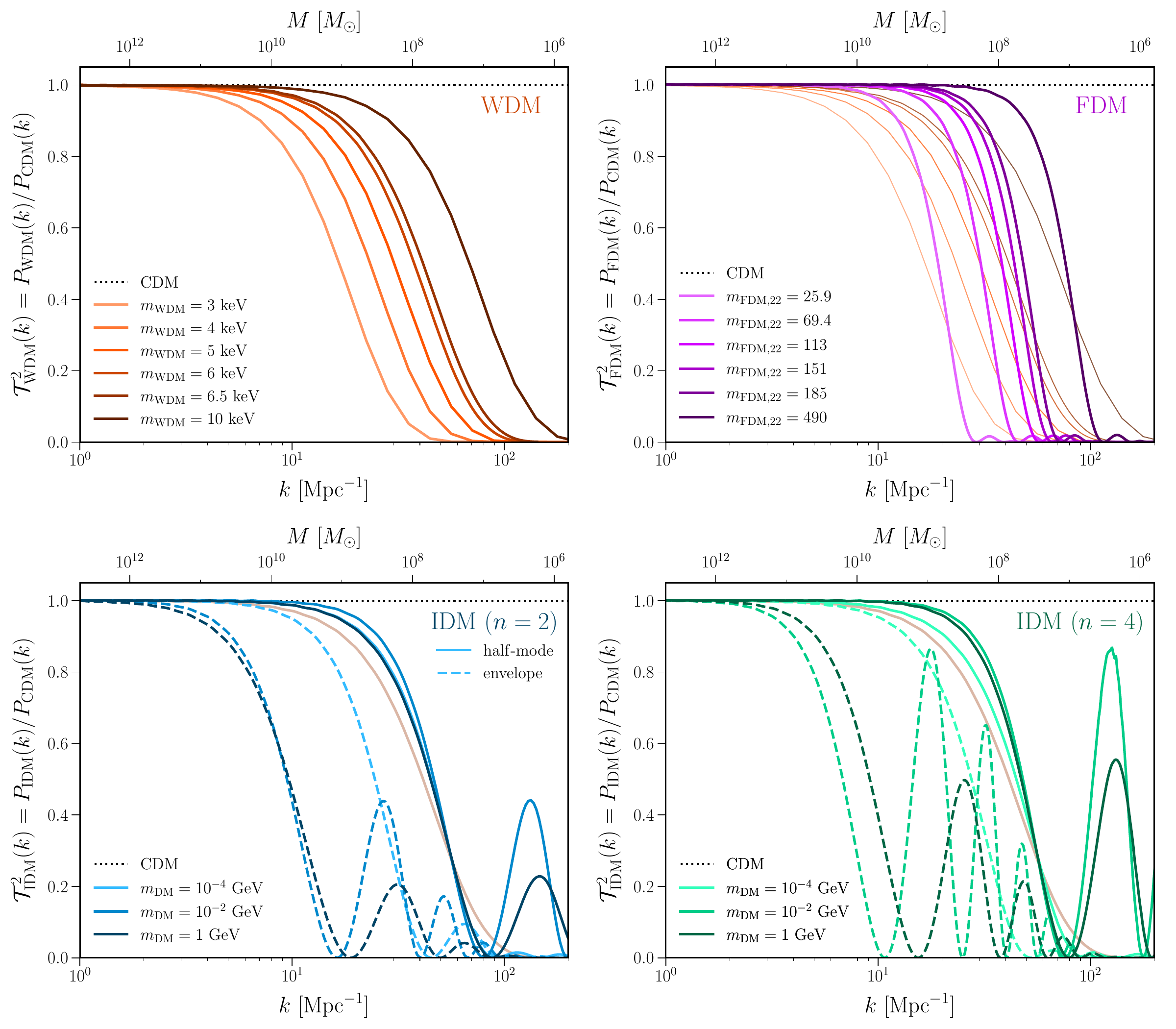}
    \caption{Ratio of the linear matter power spectrum for all beyond-CDM models represented in the COZMIC I suite. Top ticks show halo masses (in units of solar mass) associated with given wavenumbers in linear theory (Equation~\ref{eq:m_k}). FDM models are chosen to match the half-mode wavenumber for each WDM model we simulate, which are shown as faint lines in the top right panel (see Section~\ref{sec:fdm_ics}). In the IDM panels, solid (dashed) lines show models matched to the half-mode scale of (models that are strictly more suppressed than) the $m_{\mathrm{WDM}}=6.5~\mathrm{keV}$ transfer function, which is shown as a faint orange line (see Section~\ref{sec:idm_ics}).}
    \label{fig:transfers}
\end{figure*}

Half-mode and free streaming scales for these WDM models are listed in Table~\ref{tab:summary}; the corresponding transfer functions are shown in the top left panel of Figure~\ref{fig:transfers} and cut off smoothly near the half-mode scale. This suppression is broadly characteristic of many WDM particle models, including sterile neutrinos.

\begin{deluxetable*}{{ccccc}}[t!]
\centering
\tablecolumns{5}
\tablecaption{Summary of COZMIC I Simulations.}
\tablehead{\colhead{Scenario} & \colhead{Input Parameter(s)} & \colhead{Transfer Function Feature(s)} & \colhead{Derived Parameter(s)} & \colhead{Color and Linestyle}}
\startdata 
\hline \hline
CDM &
-- &
-- &
-- &
\begin{tikzpicture}[yscale=0.5] \draw [line width=0.45mm,dotted,black] (0,-1) -- (1,-1) node[right]{};; \end{tikzpicture}
\\
\hline \hline
\phantom{.} & 
$m_{\mathrm{WDM}}~[\mathrm{keV}]$ &
$k_{\mathrm{hm}}~[\mathrm{Mpc}^{-1}],\ M_{\mathrm{hm}}~[M_{\mathrm{\odot}}]$ &
$k_{\mathrm{fs}}~[\mathrm{Mpc}^{-1}]$ &
\phantom{.} \\ 
\hline
\phantom{.} & 
3 & 
22.8, $4.3\times 10^8$ & 
316.7 &
\begin{tikzpicture}[yscale=0.5] \draw [line width=0.25mm,wdm_3] (0,-1) -- (1,-1) node[right]{};; \end{tikzpicture}
\\
\phantom{.} & 
4 & 
32.1, $1.5\times 10^8$ & 
445.5 & 
\begin{tikzpicture}[yscale=0.5] \draw [line width=0.25mm,wdm_4] (0,-1) -- (1,-1) node[right]{};; \end{tikzpicture}
\\
Thermal-relic & 
5 & 
41.8, $7.0\times 10^7$ & 
580.9 &
\begin{tikzpicture}[yscale=0.5] \draw [line width=0.25mm,wdm_5] (0,-1) -- (1,-1) node[right]{};; \end{tikzpicture}
\\
WDM & 
6 & 
52.0, $3.6\times 10^7$ & 
721.1 & 
\begin{tikzpicture}[yscale=0.5] \draw [line width=0.25mm,wdm_6] (0,-1) -- (1,-1) node[right]{};; \end{tikzpicture}
\\
\phantom{.} & 
6.5 & 
57.1, $2.7\times 10^7$ & 
793.1 & 
\begin{tikzpicture}[yscale=0.5] \draw [line width=0.25mm,wdm_6_5] (0,-1) -- (1,-1) node[right]{};; \end{tikzpicture}
\\
\phantom{.} & 
10 & 
95.3, $5.9\times 10^6$ &
1323.5 &
\begin{tikzpicture}[yscale=0.5] \draw [line width=0.25mm,wdm_10] (0,-1) -- (1,-1) node[right]{};; \end{tikzpicture}
\\
\hline \hline
\phantom{.} & 
$m_{\mathrm{FDM}}~[10^{-22}~\mathrm{eV}]$ &
$k_{\mathrm{hm}}~[\mathrm{Mpc}^{-1}],\ M_{\mathrm{hm}}~[M_{\mathrm{\odot}}]$ &
$k_{J}~[\mathrm{Mpc}^{-1}]$ &
\phantom{.} \\ 
\hline
\phantom{.} & 
25.9 & 
22.4, $1.5\times 10^8$ & 
45.8 &
\begin{tikzpicture}[yscale=0.5] \draw [line width=0.25mm,fdm_25_9] (0,-1) -- (1,-1) node[right]{};; \end{tikzpicture}
\\
\phantom{.} & 
69.4 & 
35.5, $1.1\times 10^8$ & 
75.0 & 
\begin{tikzpicture}[yscale=0.5] \draw [line width=0.25mm,fdm_69_4] (0,-1) -- (1,-1) node[right]{};; \end{tikzpicture}
\\
Ultralight & 
113 & 
44.6, $5.7\times 10^7$ & 
95.7 &
\begin{tikzpicture}[yscale=0.5] \draw [line width=0.25mm,fdm_113] (0,-1) -- (1,-1) node[right]{};; \end{tikzpicture}
\\
FDM & 
151 & 
51.2, $3.8\times 10^7$ & 
110.6 & 
\begin{tikzpicture}[yscale=0.5] \draw [line width=0.25mm,fdm_151] (0,-1) -- (1,-1) node[right]{};; \end{tikzpicture}
\\
\phantom{.} & 
185 & 
56.3, $2.9\times 10^7$ & 
122.4 & 
\begin{tikzpicture}[yscale=0.5] \draw [line width=0.25mm,fdm_185] (0,-1) -- (1,-1) node[right]{};; \end{tikzpicture}
\\
\phantom{.} & 
490 & 
89.4, $7.1\times 10^6$ &
199.2 &
\begin{tikzpicture}[yscale=0.5] \draw [line width=0.25mm,fdm_490] (0,-1) -- (1,-1) node[right]{};; \end{tikzpicture}
\\
\hline \hline
\phantom{.} & 
$m_{\mathrm{IDM}}~[\mathrm{GeV}]$, $\sigma_0~[\mathrm{cm}^2]$ &
$k_{\mathrm{hm}}~[\mathrm{Mpc}^{-1}],\ M_{\mathrm{hm}}~[M_{\mathrm{\odot}}]$, $h_{\mathrm{peak}}$, $k_{\mathrm{peak}}$ &
$k_{\mathrm{crit}}~[\mathrm{Mpc}^{-1}],\ m_{\mathrm{WDM,eff}}~[\mathrm{keV}]$ &
\phantom{.} \\ 
\hline
\phantom{.} & 
$10^{-4}$, $4.2\times 10^{-28}$ & 
58.0, $2.6\times 10^7$, 115.8, 0.01 & 
35.5, 7.5 &
\begin{tikzpicture}[yscale=0.5] \draw [line width=0.25mm,idm_n2_1e-4_halfmode] (0,-1) -- (1,-1) node[right]{};; \end{tikzpicture}
\\
\phantom{.} & 
$10^{-4}$, $2.8\times 10^{-27}$ & 
30.8, $1.7\times 10^8$ 65.1, 0.1 & 
16.6, 4.1 & 
\begin{tikzpicture}[yscale=0.5] \draw [line width=0.35mm,dashed,idm_n2_1e-4_envelope] (0,-1) -- (1,-1) node[right]{};; \end{tikzpicture}
\\
DM--proton scattering & 
$10^{-2}$, $1.3\times 10^{-25}$ & 
58.0, $2.6\times 10^7$, 133.7, 0.44 & 
28.7, 10.4 &
\begin{tikzpicture}[yscale=0.5] \draw [line width=0.25mm,idm_n2_1e-2_halfmode] (0,-1) -- (1,-1) node[right]{};; \end{tikzpicture}
\\
IDM ($n=2$) & 
$10^{-2}$, $7.1\times 10^{-24}$ & 
21.8, $4.9\times 10^8$, 26.7, 0.44 & 
5.7, 2.7 & 
\begin{tikzpicture}[yscale=0.5] \draw [line width=0.35mm,dashed,idm_n2_1e-2_envelope] (0,-1) -- (1,-1) node[right]{};; \end{tikzpicture}
\\
\phantom{.} & 
$1$, $1.6\times 10^{-23}$ & 
58.0, $2.6\times 10^7$, 145.8, 0.23 & 
30.6, 6.6 & 
\begin{tikzpicture}[yscale=0.5] \draw [line width=0.25mm,idm_n2_1e0_halfmode] (0,-1) -- (1,-1) node[right]{};; \end{tikzpicture}
\\
\phantom{.} & 
$1$, $8.0\times 10^{-22}$ & 
12.6, $2.5\times 10^9$, 31.7, 0.21 &
6.4, 2.2 &
\begin{tikzpicture}[yscale=0.5] \draw [line width=0.35mm,dashed,idm_n2_1e0_envelope] (0,-1) -- (1,-1) node[right]{};; \end{tikzpicture}
\\
\hline \hline
\phantom{.} & 
$m_{\mathrm{IDM}}~[\mathrm{GeV}]$, $\sigma_0~[\mathrm{cm}^2]$ &
$k_{\mathrm{hm}}~[\mathrm{Mpc}^{-1}],\ M_{\mathrm{hm}}~[M_{\mathrm{\odot}}]$, $h_{\mathrm{peak}}$, $k_{\mathrm{peak}}$ &
$k_{\mathrm{crit}}~[\mathrm{Mpc}^{-1}],\ m_{\mathrm{WDM,eff}}~[\mathrm{keV}]$ &
\phantom{.} \\ 
\hline
\phantom{.} & 
$10^{-4}$, $2.2\times 10^{-27}$ & 
58.2, $2.6\times 10^7$, 112.9, 0.003 & 
44.9, 7.6 &
\begin{tikzpicture}[yscale=0.5] \draw [line width=0.25mm,idm_n4_1e-4_halfmode] (0,-1) -- (1,-1) node[right]{};; \end{tikzpicture}
\\
\phantom{.} & 
$10^{-4}$, $3.4\times 10^{-26}$ & 
34.7, $1.2\times 10^8$, 103.6, 0.01 & 
20.5, 4.5 & 
\begin{tikzpicture}[yscale=0.5] \draw [line width=0.35mm,dashed,idm_n4_1e-4_envelope] (0,-1) -- (1,-1) node[right]{};; \end{tikzpicture}
\\
DM--proton scattering & 
$10^{-2}$, $1.7\times 10^{-22}$ & 
58.2, $2.6\times 10^7$, 126.7, 0.87 & 
27.7, 9.4 &
\begin{tikzpicture}[yscale=0.5] \draw [line width=0.25mm,idm_n4_1e-2_halfmode] (0,-1) -- (1,-1) node[right]{};; \end{tikzpicture}
\\
IDM ($n=4$) & 
$10^{-2}$, $1.7\times 10^{-19}$ & 
8.2, $9.2\times 10^9$, 17.6, 0.87 & 
3.9, 3.5 & 
\begin{tikzpicture}[yscale=0.5] \draw [line width=0.35mm,dashed,idm_n4_1e-2_envelope] (0,-1) -- (1,-1) node[right]{};; \end{tikzpicture}
\\
\phantom{.} & 
$1$, $8.6\times 10^{-19}$ & 
58.2, $2.6\times 10^7$, 132.3, 0.56 & 
28.4, 6.6 & 
\begin{tikzpicture}[yscale=0.5] \draw [line width=0.25mm,idm_n4_1e0_halfmode] (0,-1) -- (1,-1) node[right]{};; \end{tikzpicture}
\\
\phantom{.} & 
$1$, $2.8\times 10^{-16}$ & 
11.3, $3.5\times 10^9$, 25.6, 0.5 &
5.5, 2.8 &
\begin{tikzpicture}[yscale=0.5] \draw [line width=0.35mm,dashed,idm_n4_1e0_envelope] (0,-1) -- (1,-1) node[right]{};; \end{tikzpicture}
\\
\hline \hline
\enddata
{\footnotesize \tablecomments{The first column lists the names of DM scenarios and the second column lists input parameter(s) used to generate ICs. The third column lists the half-mode wavenumber and mass of the transfer function (and, for IDM, the height and wavenumber of the first DAO peak). The fourth column lists the free streaming wavenumber for WDM (Equation~\ref{eq:fs}), the Jeans wavenumber for FDM (Equation~\ref{eq:kj}), and the decoupling wavenumber (Equation~\ref{eq:lambda_dec}) and effective WDM mass based on subhalo abundance matching (Section~\ref{sec:idm_shmf}) for IDM. The fifth column shows the color and linestyle used for each model in figures throughout this work. For IDM, solid (dashed) lines indicate half-mode (envelope) cross sections; see Section~\ref{sec:idm_ics} for details.}}
\label{tab:summary}
\end{deluxetable*}

\subsubsection{Fuzzy Dark Matter}
\label{sec:fdm_ics}

To compute FDM transfer functions, we use a lightly modified version of \textsc{axionCAMB} \citep{Hlozek:2014lca,Grin2203026}, described in Appendix~\ref{sec:tk_details}.\footnote{\url{https://github.com/Xiaolong-Du/axionCAMB_patch}} We generate transfer functions for $m_{\mathrm{FDM,22}}\in [25.9,69.4,113,151,185,490]$. These values are chosen such that the corresponding half-mode scale $k_{\mathrm{hm}}$ matches that for each WDM scenario we simulate at the percent level. 

Half-mode scales and de Broglie scales for these FDM models are listed in Table~\ref{tab:summary}. The corresponding transfer functions in the top right panel of Figure~\ref{fig:transfers} cut off more sharply than the thermal-relic WDM models with matched half-mode scales. The impact of this difference on nonlinear modeling was noted in previous studies (e.g., \citealt{Armengaud170309126,Nadler190410000,Schutz200105503}), and we study the effects of this difference directly in this work.

\subsubsection{Interacting Dark Matter}
\label{sec:idm_ics}

To compute IDM transfer functions, we use a modified version of \textsc{CLASS}~\citep{Boddy180800001,Gluscevic1812108}.\footnote{\url{https://github.com/kboddy/class_public/tree/dmeff}} For each $m_{\mathrm{IDM}}\in[10^{-4},~ 10^{-2},~ 1]~\mathrm{GeV}$, we choose $\sigma_0$ so that the resulting transfer function either ($i$) matches the half-mode scale of the $m_{\mathrm{WDM}}=6.5~\mathrm{keV}$ transfer function, or ($ii$) is strictly more suppressed than the $m_{\mathrm{WDM}}=6.5~\mathrm{keV}$ transfer function. We refer to these as ``half-mode'' and ``envelope'' models, respectively.\footnote{``Envelope'' refers to the fact that the transfer function for the reference WDM model presents a tight upper envelope to the IDM transfer function.} We choose $m_{\mathrm{WDM}}=6.5~\mathrm{keV}$ as a reference model because it corresponds to the current WDM bound from the MW satellite galaxy population \citep{Nadler200800022}. By comparing to this bound, \cite{Maamari201002936} have shown that the envelope models are robustly disfavored by MW satellite abundances, while the half-mode models must be simulated directly to derive more stringent bounds; the latter is thus a goal of this work.

The parameters of the IDM models we simulate are summarized in Table~\ref{tab:summary}, including their half-mode scales and first DAO peak characteristics. The corresponding transfer functions are shown in the bottom panels of Figure~\ref{fig:transfers}, featuring a variety of shapes that depend on model parameters, as discussed in previous studies (e.g., \citealt{Nadler190410000,Maamari201002936}). In particular, we note that the DAO peak height is a nonmonotonic function of $m_{\mathrm{IDM}}$, which (among the IDM parameter values we consider) reaches a maximum amplitude for $m_{\mathrm{IDM}}=10^{-2}~\mathrm{GeV}$. DAO features are generally more prominent for $n=4$ models.

\subsubsection{Generating Zoom-in Initial Conditions}

We generate ICs by passing the density and velocity transfer functions generated for each beyond-CDM scenario into \textsc{MUSIC} \citep{Hahn11036031}. We use identical random seeds in all cases, such that phases of density perturbations are fixed in our beyond-CDM simulations; thus, the only difference relative to CDM is that the amplitude of each mode is multiplied by $\mathcal{T}^2_{\mathrm{beyond-CDM}}(k)$. This procedure ensures that the same high-mass subhalos form in each run and that suppression of low-mass subhalo abundances is due to $P(k)$ suppression rather than stochasticity from resampling of small-scale density perturbations.

We resimulate two Milky Way-est hosts (Halo004 and Halo113; \citealt{Buch240408043}) and one Symphony Milky Way host (Halo023; \citealt{Nadler220902675}). We refer to the Milky Way-est hosts as ``MW-like'' because they contain LMC-analog subhalos and merge with Gaia--Sausage--Enceladus (GSE) analogs at early times ($z\sim 2$); we refer to the Symphony Milky Way host as ``MW-mass'' because it is only constrained to have a host halo mass similar to the MW. We choose these hosts because of their small Lagrangian volumes, which make them relatively inexpensive to resimulate. For each simulation, we initialize a region at $z=99$ that corresponds to the Lagrangian volume of particles within $10$ times the virial radius of the host halo in the parent box at $z=0$, following \cite{Nadler220902675}. Thus, the high-resolution region has a radius of $\sim 3~\mathrm{Mpc}$ for each host. The models we simulate do not suppress $P(k)$ on scales larger than the zoom-in region, which corresponds to a wavenumber $k_L = 2\pi/(3~\mathrm{Mpc})\approx 2~\mathrm{Mpc}^{-1}$.

Our $72$ fiducial-resolution beyond-CDM simulations (i.e., $24$ simulations per each of our three hosts) use four refinement regions relative to the parent box, yielding an equivalent of $8192$ particles per side in the highest-resolution region and a mean interparticle spacing of $22~\mathrm{kpc}$, corresponding to a Nyquist frequency $k_{\mathrm{Ny}}=\pi/(22~\mathrm{kpc})\approx 143~\mathrm{Mpc}^{-1}$. For these fiducial-resolution simulations, the DM particle mass in the highest-resolution regions is $m_{\mathrm{part}}=4.0\times 10^5~M_{\mathrm{\odot}}$. We also perform $24$ high-resolution beyond-CDM simulations of one host (Halo004) using an additional refinement region; see Appendix~\ref{sec:convergence} for details.

Figure~\ref{fig:delta_ics} compares the distribution of density contrasts, $\delta$, for high-resolution particles in the zoom-in region for Halo004 at $z=99$ in CDM and our WDM models. Local densities are computed in \textsc{Pynbody} at the position of each particle with a smoothed particle hydrodynamics (SPH) kernel, using $64$ nearest neighbors. The WDM density contrast distributions are suppressed relative to CDM at both large positive and negative values of $\delta$, indicating that small-scale overdensities and underdensities are smoothed out by $P(k)$ suppression. Interestingly, even the $m_{\mathrm{WDM}}=10~\mathrm{keV}$ model's overdensity distribution noticeably differs from CDM, although we will show that subhalo abundances in these models are statistically consistent above our fiducial-resolution limit. The results in Figure~\ref{fig:delta_ics} are qualitatively similar for our FDM and IDM models.

\begin{figure}[t!]
\centering
\hspace{-5mm}
\includegraphics[trim={0 0.35cm 0 0cm},width=0.49\textwidth]{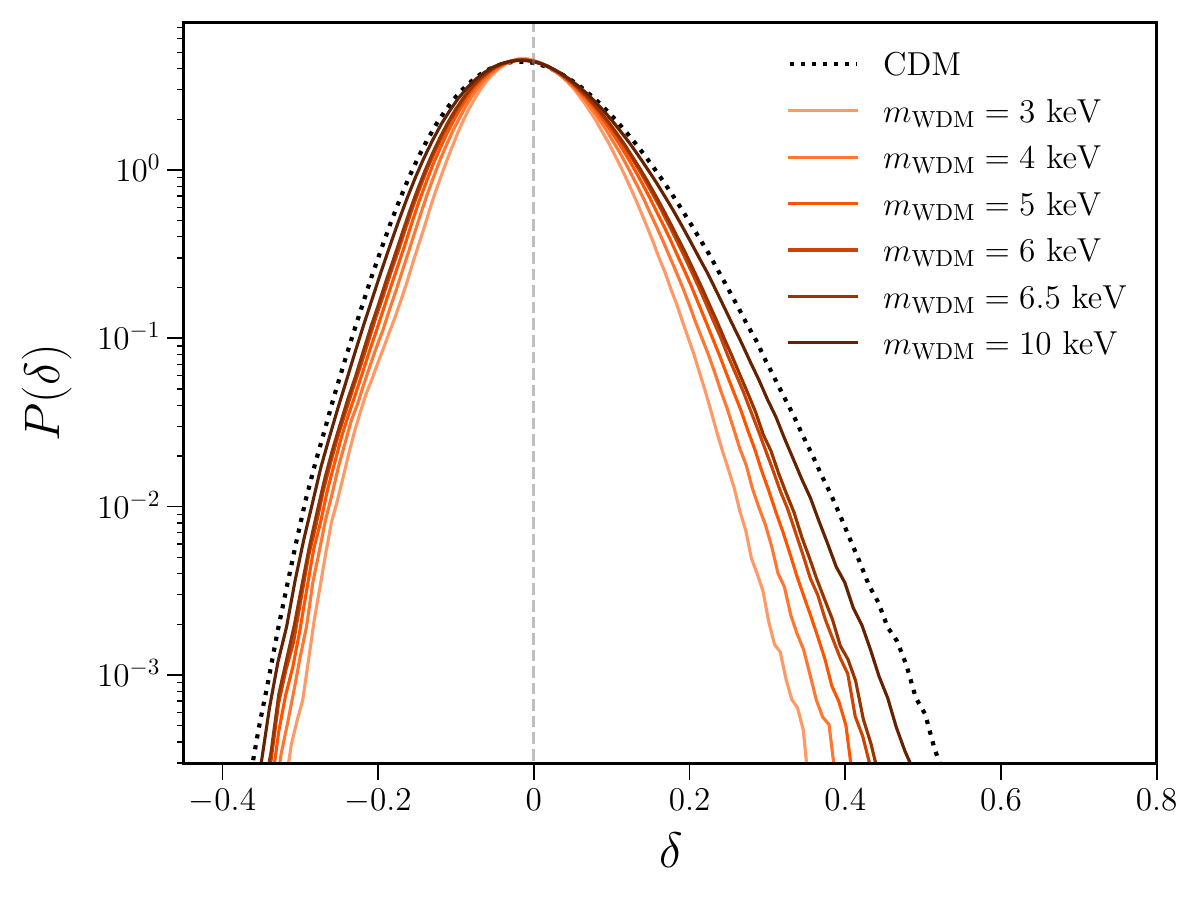}
    \caption{Distribution of local density contrast evaluated at the position of each high-resolution particle at $z=99$ in CDM (dotted black line) and each of our WDM models (solid orange lines), generated by \textsc{MUSIC} for one of our MW-like zoom-ins (Halo004). The dashed gray vertical line marks $\delta=0$, differentiating overdensities from underdensities. Density contrasts are computed using \textsc{pynbody} \citep{pynbody}.}
    \label{fig:delta_ics}
\end{figure}

\subsection{Zoom-in Simulations}
\label{sec:zoom-in}

\begin{figure*}[t!]
\centering
\hspace{-1.25cm}
\includegraphics[trim={0 0.75cm 0 -0.15cm},width=0.885\textwidth]{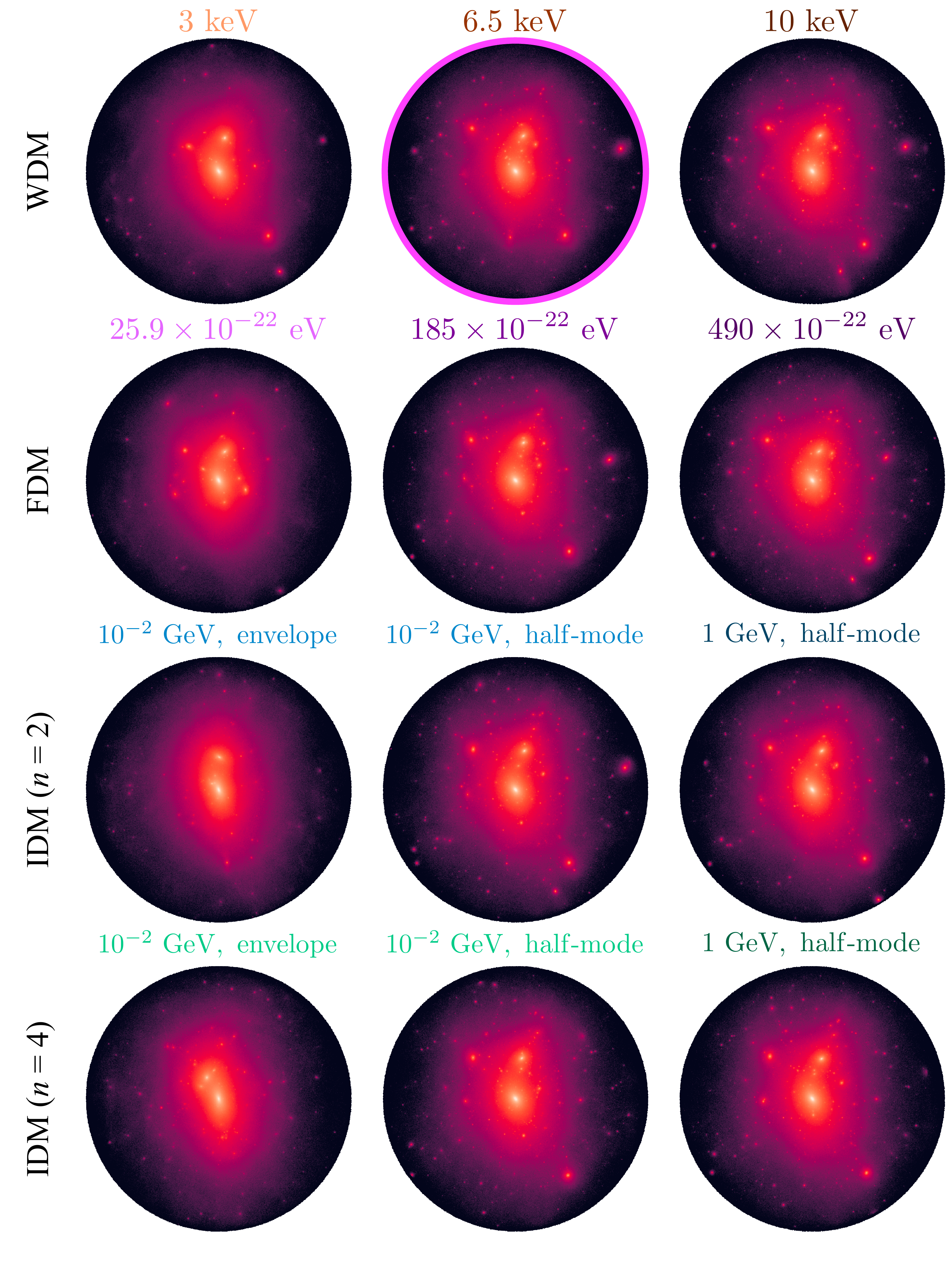}
    \caption{Projected DM density maps at $z=0$ for a subset of the high-resolution beyond-CDM simulations of an MW-like system (Halo004). For each simulation, the visualization is centered on the host halo and spans $1.5$ times its virial radius. The half-mode scale is the same for every model within each column, except for the IDM models in the left column, which correspond to envelope cross sections for $m_{\mathrm{IDM}}=10^{-2}~\mathrm{GeV}$ with $n=2$ (third row) and $n=4$ (fourth row). The $m_{\mathrm{WDM}} = 6.5~\mathrm{keV}$ visualization is highlighted as a reference model used throughout the paper. Note that the $m_{\mathrm{WDM}}=10~\mathrm{keV}$ and $m_{\mathrm{FDM,22}}=490$ density maps are visually similar to CDM. Visualizations were created using \textsc{meshoid} (\url{https://github.com/mikegrudic/meshoid}).}
    \label{fig:vis_main}
\end{figure*}

We run the simulations using \textsc{Gadget-2} \citep{Springel0505010}. We save $236$ output snapshots per run, starting at $z\approx 10$, with a typical output cadence of $\approx 25~\mathrm{Myr}$ near $z=0$. We set the time-stepping criterion to $\eta=0.01$ and the comoving Plummer-equivalent gravitational softening to $\epsilon=170~\mathrm{pc}~h^{-1}$, following Symphony and Milky Way-est settings \citep{Nadler220902675,Buch240408043}.

Figure~\ref{fig:vis_main} shows projected DM density maps from the high-resolution runs of an MW-like host (Halo004), for various beyond-CDM models. Even though the high-resolution region in each zoom-in extends to $\approx 10$ times the virial radius of the zoom-in host $R_{\mathrm{vir}}$, the visualizations shown in the figure show a region spanning only $1.5 R_{\mathrm{vir}}$, in order to highlight substructure. Small-scale structure is clearly suppressed in the beyond-CDM scenarios, and the amount of suppression depends on the ICs. For example, very little substructure is visible in the $m_{\mathrm{WDM}}=3~\mathrm{keV}$ simulation, while the $m_{\mathrm{WDM}}=6.5~\mathrm{keV}$ simulation only subtly differs from $m_{\mathrm{WDM}}=10~\mathrm{keV}$ based on visual inspection. In turn, the $m_{\mathrm{WDM}}=10~\mathrm{keV}$ case is nearly visually indistinguishable from CDM. At a fixed half-mode scale, substructure is visibly affected by both the slope of the transfer function (e.g., compare the WDM and FDM rows in Figure~\ref{fig:vis_main}) and the presence of DAOs (e.g., compare the WDM and IDM rows in Figure~\ref{fig:vis_main}). Substructure in our envelope IDM scenarios is more suppressed than in the corresponding half-mode scenarios, which visually confirms that the IDM constraints derived in \cite{Maamari201002936} are conservative. We quantify these findings below.

\subsection{Post-processing and Analysis}
\label{sec:post-process}

We generate halo catalogs and merger trees by running {\sc Rockstar} and {\sc consistent-trees} \citep{Behroozi11104372,Behroozi11104370} on the highest-resolution particles from each simulation's output snapshots. We analyze all simulations at $z=0$ to ensure that SHMF suppression relative to CDM is measured at the same cosmic time in all cases. For one of our two MW-like halos (Halo004), this matches the analysis snapshot in \cite{Buch240408043}.

Following the Symphony convergence tests in \cite{Nadler220902675}, we only analyze subhalos with at least $300$ particles at $z=0$. In our fiducial-resolution simulations, this corresponds to a virial mass threshold of $M_{\mathrm{res}}=1.2\times 10^8~M_{\mathrm{\odot}}$. Throughout, we calculate SHMFs using peak virial mass
\begin{equation}
    M_{\mathrm{sub,peak}}\equiv \max(M_{\mathrm{sub}}(z)),\label{eq:mpeak}
\end{equation}
because $M_{\mathrm{sub,peak}}$ most directly connects to the scale of the linear density perturbation that formed a given subhalo, and thus to the mass associated with a given wavenumber in linear theory (Equation~\ref{eq:m_k}).\footnote{Note that $M_{\mathrm{sub,peak}}$ measured before infall is equal to (or within $10\%$ of) $M_{\mathrm{sub,peak}}$ from Equation~\ref{eq:mpeak} for nearly all subhalos we study.}

In the absence of additional cuts, SHMFs are less well converged as a function of peak (vs.\ present-day) mass because subhalos of a given peak mass can be heavily stripped (e.g., \citealt{Nadler220902675,Mansfield230810926}). We therefore restrict our peak SHMF measurements to subhalos above our fiducial present-day mass threshold of $M_{\mathrm{sub}}>1.2\times 10^8~M_{\mathrm{\odot}}$. Given this cut, convergence of the $M_{\mathrm{peak}}$ SHMF requires that (1) the SHMF measured using present-day mass is converged and (2) $M_{\mathrm{peak}}$ values are consistent across resolution levels. The former criterion is demonstrated in \cite{Nadler220902675}, and we have explicitly checked that the latter holds in all of our simulations. Furthermore, in Appendix~\ref{sec:convergence}, we show that the \emph{suppression} of the peak SHMF relative to CDM is converged in our beyond-CDM simulations. Together, these tests thoroughly demonstrate that our measurements are robust to simulation resolution.

We do not impose additional cuts (beyond the resolution cut described above) to remove spurious halos formed through artificial fragmentation, which have hampered previous WDM simulations (e.g., \citealt{Wang0702575}). Even though we use a standard $N$-body code, rather than a simulation technique that mitigates artificial fragmentation (e.g., based on evolving the phase-space sheet; \citealt{Angulo13042406,Stucker210909760}), spurious objects contribute negligibly to the population of well-resolved subhalos in our simulations. Specifically, the lower limit on halo mass derived in previous studies to remove spurious halos is defined as $\kappa M_{\mathrm{lim}}$ \citep{Lovell13081399}, where $\kappa\approx 0.5$ and $M_{\mathrm{lim}}$ is a characteristic mass scale below which artificial fragmentation is a significant effect. For \emph{all} of our simulations, $\kappa M_{\mathrm{lim}}<M_{\mathrm{res}}=1.2\times 10^8~M_{\mathrm{\odot}}$. Indeed, for most of the models we simulate, $M_{\mathrm{lim}}\ll M_{\mathrm{res}}$. In Appendix~\ref{sec:fragmentation}, we show that spurious halos contribute negligibly to subhalo populations above $M_{\mathrm{res}}$ in all beyond-CDM simulations we present based on the shapes of their protohalo particle distributions in the ICs.

\section{Subhalo Mass Functions}
\label{sec:basic}

\begin{figure*}[t!]
\centering
\includegraphics[width=\textwidth]{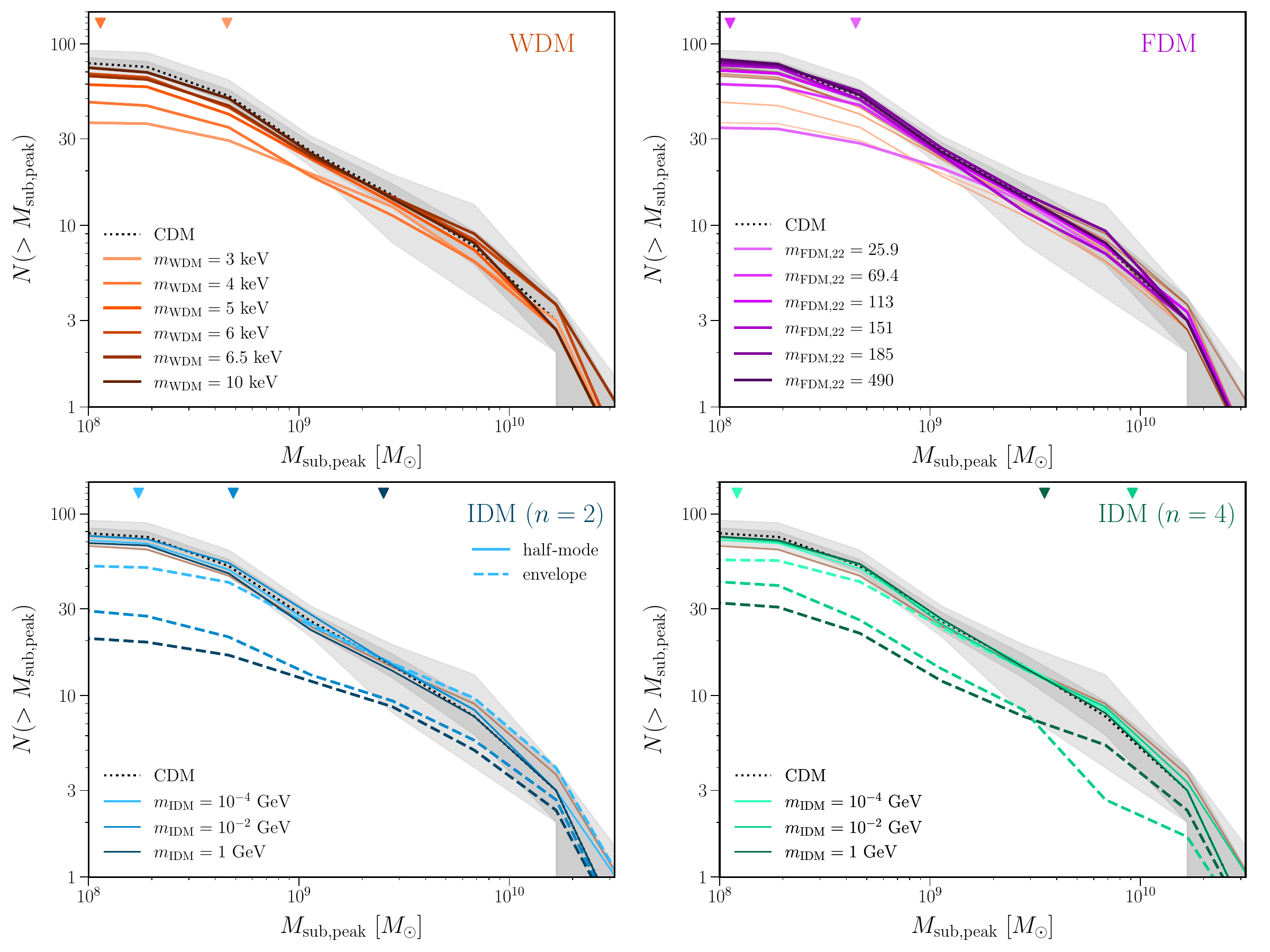}
    \caption{Cumulative SHMFs, averaged over our three MW hosts, shown in four different beyond-CDM cosmologies (solid colored lines) and in CDM (dotted black lines). We only consider subhalos with present-day virial masses $M_{\mathrm{sub}}>1.2\times 10^8~M_{\mathrm{\odot}}$, corresponding to $>300$ particles in the fiducial-resolution simulations. Each pair of WDM and FDM models has a matched half-mode scale. In the IDM panels, solid lines show models that have the same half-mode scale as the $m_{\mathrm{WDM}}=6.5~\mathrm{keV}$ model, while the dashed lines show IDM models where the transfer function is strictly more suppressed than the $m_{\mathrm{WDM}}=6.5~\mathrm{keV}$ model at all $k$. Markers indicate the half-mode masses of the most extreme models in each panel (in the IDM panels, markers correspond to envelope models); the half-mode masses for less extreme models are below $10^8~M_{\mathrm{\odot}}$. Dark-gray bands show the $1\sigma$ Poisson uncertainty on the cumulative CDM SHMF, and light-gray bands show the range of host-to-host variation.}
    \label{fig:shmf}
\end{figure*}

Figure~\ref{fig:shmf} shows SHMFs averaged over the three MW hosts, along with the associated Poisson uncertainty on the CDM SHMF (dark bands) and the range of host-to-host scatter (light bands). Subhalo abundances are clearly suppressed at the low-mass end in all of our beyond-CDM simulations. For a number of models, the suppression is statistically significant, exceeding both the Poisson uncertainty on the CDM SHMF and the host-to-host variance. The total amount of suppression is primarily determined by the $P(k)$ cutoff scale, such that models with lower WDM masses, lower FDM masses, and higher IDM cross sections yield fewer subhalos. SHMFs are most suppressed below each model's half-mode mass, indicated by the triangles in Figure~\ref{fig:shmf}. However, even for cases where $M_{\mathrm{hm}}<10^8~M_{\mathrm{\odot}}$, SHMF suppression can be significant at the lowest $M_{\mathrm{sub,peak}}$ we resolve, indicating that even the onset of $P(k)$ suppression can affect subhalo abundances.

In addition to the cutoff scale, the shape of $P(k)$ affects the SHMF. For example, all FDM models (except the most suppressed case) yield more subhalos than WDM models with matched half-mode scales. Meanwhile, IDM scenarios with the same half-mode scale as $m_{\mathrm{WDM}}=6.5~\mathrm{keV}$ yield slightly higher subhalo abundances, suggesting that either the slope of the suppression or the DAOs in $P(k)$, can reduce SHMF suppression. For example, the $m_{\mathrm{IDM}}=10^{-2}~\mathrm{GeV}$ envelope run has a smaller $k_{\mathrm{hm}}$ than the $m_{\mathrm{IDM}}=1~\mathrm{GeV}$ run, yet forms more substructure. Among the beyond-CDM models we consider, the shape of the SHMF differs most from CDM in the envelope IDM scenarios. We interpret all of these findings in Section~\ref{sec:shmf} by modeling the suppression of the SHMF relative to CDM in WDM and FDM cosmologies and by comparing IDM results to the other beyond-CDM scenarios.

At $M_{\mathrm{sub,peak}}\gtrsim 10^{10}~M_{\mathrm{\odot}}$, beyond-CDM SHMFs are consistent with CDM within the Poisson uncertainty, except for certain envelope IDM models. In some cases (e.g., the $m_{\mathrm{WDM}}=6~\mathrm{keV}$ model in the top left panel of Figure~\ref{fig:shmf}), beyond-CDM simulations have a few more high-mass subhalos than CDM, which may be related to orbital phase shifts discussed below. For $M_{\mathrm{sub,peak}}\gtrsim 3\times 10^{10}~M_{\mathrm{\odot}}$, beyond-CDM SHMFs converge to the CDM result. Furthermore, the masses of LMC analogs are nearly identical in all simulations.

All of these conclusions hold when the present-day halo masses are used to measure SHMFs, rather than peak virial masses. Furthermore, SHMFs are converged as a function of simulation resolution (see Appendix~\ref{sec:convergence}). Note that Figure~\ref{fig:shmf} supports our claim from Section~\ref{sec:post-process} that spurious halos do not significantly impact our results, since the cumulative SHMF stops increasing significantly for $M_{\mathrm{sub,peak}}\lesssim 2\times 10^8~M_{\mathrm{\odot}}$ and flattens for both CDM and all beyond-CDM models shown here. This flattening is largely due to the cut on present-day subhalo mass, since surviving subhalos are typically tidally stripped in mass by a factor of $\gtrsim 2$ relative to their peak mass (e.g., \citealt{Nadler220902675}).


\section{Modeling Subhalo Mass Functions beyond CDM}
\label{sec:shmf}

In this section, we treat the COZMIC subhalo populations as ``data'' that provide measurements of the SHMF in beyond-CDM scenarios. We present a parametric model for the SHMF and perform probabilistic inference to reconstruct model parameters that fit the simulated data. We describe the SHMF model and likelihood in Section~\ref{sec:fit_method}; we then present results for WDM (Section~\ref{sec:wdm_shmf}), FDM (Section~\ref{sec:fdm_shmf}), and IDM (Section~\ref{sec:idm_shmf}) in subsequent subsections. We summarize key takeaways in Section~\ref{sec:shmf_summary}.

Before we begin, we emphasize the following key advances of our SHMF modeling approach:
\begin{enumerate}
    \item We measure SHMF suppression using peak (rather than present-day) subhalo mass, since $M_{\mathrm{sub,peak}}$ more directly traces the relevant scales in $P(k)$. Furthermore, $M_{\mathrm{peak}}$ is more directly related to the luminosities of MW satellites galaxies than $M_{\mathrm{vir}}$, since these systems form stars at high redshifts and generally do not lose a significant fraction of their stars during tidal evolution~\citep{Nadler191203303}.
    \item We explicitly model the effect of stripping below our simulations' present-day mass resolution limit. As a result, our SHMF models are only fit down to present-day virial masses of $1.2\times 10^8~M_{\mathrm{\odot}}$. Note that this is slightly below the minimum peak halo mass of MW satellite galaxies detected in current data sets \citep{Nadler191203303}.
    \item We fit SHMF models probabilistically to our simulation data and report the resulting uncertainties in and degeneracies between the model parameters, along with the list of best-fit values.
\end{enumerate}

\subsection{Subhalo Mass Function Model}
\label{sec:fit_method}

To proceed, we write the differential SHMF in host $i$ and beyond-CDM scenario $j$ as
\begin{align}
    \frac{\mathrm{d}N_{ij}(\Theta)}{\mathrm{d}\log(M_{\mathrm{sub,peak}})} &= \left(\frac{a_i}{100}\right)\left(\frac{M_{\mathrm{sub,peak}}}{M_{\mathrm{host,}ij}}\right)^{-b_i}e^{-50\left(\frac{M_{\mathrm{sub,peak}}}{M_{\mathrm{host,}ij}}\right)^4}\nonumber& \\ &\times p_{ij}(M_{\mathrm{sub}}>M_{\mathrm{min}}\lvert M_{\mathrm{sub,peak}})&\nonumber \\ &\times f_{\mathrm{beyond-CDM}}(M_{\mathrm{sub,peak}},\theta_j,\alpha,\beta,\gamma),&\label{eq:shmf_model}
\end{align}
where $\vec{\Theta}=\{a_i,~b_i,~\alpha,~\beta,~\gamma\,~ \theta_j\}$. The first three terms model the CDM SHMF following \cite{vandenBosch14036835}, where the normalization $a_i$ and power-law slope $b_i$ are free parameters. The fourth term models the probability that a subhalo with a given $M_{\mathrm{sub,peak}}$ has a present-day virial mass $M_{\mathrm{sub}}$ above the resolution limit, $M_{\mathrm{min}}=M_{\mathrm{res}}=1.2\times 10^8~M_{\mathrm{\odot}}$, and is therefore accounted for in our measurements of the SHMF from simulations; thus, this term models the effects of the simulation resolution limit on derived SHMFs measured using $M_{\mathrm{peak,sub}}$. For each host and DM scenario, we use the simulated subhalo population to reconstruct the joint probability distribution of $M_{\mathrm{sub}}$ and $M_{\mathrm{sub,peak}}$. We use kernel density estimation (KDE) for this purpose, which allows us to integrate the normalized KDE above $M_{\mathrm{min}}$ and recover the probability that a subhalo is stripped below the resolution limit as a function of $M_{\mathrm{sub,peak}}$.

The final term in Equation~\ref{eq:shmf_model} models the SHMF suppression relative to CDM, which depends on a parameter $\theta$ specific to each beyond-CDM scenario (i.e., $m_{\mathrm{WDM}}$ or $m_{\mathrm{FDM}}$). We adopt a model that was previously shown to describe halo and SHMF suppression accurately in WDM and FDM (e.g., \citealt{Benito200111013,Lovell200301125}),
\begin{equation}
    f_{\mathrm{beyond-CDM}}(M_{\mathrm{sub,peak}},\theta,\alpha,\beta,\gamma) = \left[1+\left(\frac{\alpha M_{\mathrm{hm}}(\theta)}{M_{\mathrm{sub,peak}}}\right)^{\beta}\right]^{-\gamma},\label{eq:f_beyond-CDM}
\end{equation}
where $M_{\mathrm{hm}}(\theta)$ is the half-mode mass and $\alpha$, $\beta$, and $\gamma$ are free parameters. In the following, we show that a single distribution of $\{\alpha,\beta,\gamma\}$ accurately describes all WDM models, while a different distribution describes the FDM scenarios.

For each beyond-CDM scenario, we fit the model defined by Equation~\ref{eq:shmf_model} to our simulations, using a Poisson likelihood that describes the probability of measuring a set of subhalos with peak subhalo masses $M_{\mathrm{sub,peak}}$ and present-day masses $M_{\mathrm{sub}}>M_{\mathrm{min}}$. Specifically, the likelihood function reads
\begin{align}
&\mathcal{L}(\vec{\Theta}) =  \prod_{i}^{N_{\mathrm{hosts}}}\prod_{j}^{N_{\mathrm{models}}}\mathrm{Pois}(N_{\mathrm{sim,}ij}\lvert N_{ij}(\vec{\Theta}))\prod_{k}^{N_{\mathrm{sim,}ij}} P_{ij}(M_{\mathrm{sub,peak,}k}\lvert \vec{\Theta})& \nonumber \\ 
&= \prod_{i}^{N_{\mathrm{hosts}}}\prod_{j}^{N_{\mathrm{models}}}\frac{N_{ij}(\vec{\Theta})^{N_{\mathrm{sim,}ij}}e^{-N_{ij}(\vec{\Theta})}}{N_{\mathrm{sim,}ij}!} \prod_{k}^{N_{\mathrm{sim,}ij}} \frac{\left(\frac{\mathrm{d}N_{ij}(\vec{\Theta})}{\mathrm{d}\log M_{\mathrm{sub,peak}}}\right)\big\rvert_{M_{\mathrm{sub,peak,}k}}}{\int\frac{\mathrm{d}N_{ij}(\vec{\Theta})}{\mathrm{d}\log M_{\mathrm{sub,peak}}}~ \mathrm{d}\log M_{\mathrm{sub,peak}}}& \nonumber \\ 
& = \prod_{i}^{N_{\mathrm{hosts}}}\prod_{j}^{N_{\mathrm{models}}} \frac{e^{-N_{ij}(\vec{\Theta})}}{N_{\mathrm{sim,}ij}!} \prod_{k}^{N_{\mathrm{sim,}ij}} \left(\frac{\mathrm{d}N_{ij}(\vec{\Theta})}{\mathrm{d}\log M_{\mathrm{sub,peak}}}\right)\bigg\rvert_{M_{\mathrm{sub,peak,}k}}.& \label{eq:like}
\end{align}
Here $N_{\mathrm{sim},ij}$ is the total number of subhalos in a given simulation, appearing above the present-day mass resolution limit, in the host $i$, for DM scenario $j$. $P_{ij}(M_{\mathrm{sub,peak,}k}\lvert \vec{\Theta})$ is the probability of finding a subhalo with a specific mass of $M_{\mathrm{sub,peak,}k}$, for a given parameter vector $\vec\Theta$ (Equation~\ref{eq:shmf_model}). Finally,  
\begin{equation}
N_{ij}(\vec{\Theta}) \equiv \int(\mathrm{d}N_{ij}(\vec{\Theta})/\mathrm{d}\log M_{\mathrm{sub,peak}})~ \mathrm{d}\log M_{\mathrm{sub,peak}}
\end{equation} 
is the expected total number of subhalos for a given $\vec\Theta$. 

Taking the logarithm of Equation~\ref{eq:like} yields
\begin{align}
\ln \mathcal{L}(\vec{\Theta}) &= -N_{\mathrm{tot}}(\vec{\Theta}) - \sum_{i}^{N_{\mathrm{hosts}}}\sum_{j}^{N_{\mathrm{models}}}\ln(N_{\mathrm{sim,}ij}!)& \nonumber \\ &+ \sum_{i}^{N_{\mathrm{hosts}}}\sum_{j}^{N_{\mathrm{models}}}\sum_{k}^{N_{\mathrm{sim,}ij}} \ln \left(\frac{\mathrm{d}N_{ij}(\vec{\Theta})}{\mathrm{d}\log M_{\mathrm{sub,peak}}}\right)\bigg\rvert_{M_{\mathrm{sub,peak,}k}},&
\end{align}
where $N_{\mathrm{sim,tot}} = \sum_{i}^{N_{\mathrm{hosts}}} \sum_{j}^{N_{\mathrm{models}}} N_{\mathrm{sim,}ij}$ is the total number of simulated subhalos above $M_{\mathrm{res}}$ across all hosts and DM models.
The posterior probability distribution for $\vec{\Theta}$ is given by Bayes's theorem,
\begin{align}
    P(\vec{\Theta}\lvert \vec{M}_{\mathrm{sub,peak}}) = \frac{P(\vec{M}_{\mathrm{sub,peak}}\lvert \vec{\Theta})P(\vec{\Theta})}{P(\vec{M}_{\mathrm{sub,peak}})},\label{eq:wdm_posterior}
\end{align}
where $\vec{M}_{\mathrm{sub,peak}}$ represents the set of $M_{\mathrm{sub,peak}}$ values across all hosts and beyond-CDM models and $P(\vec{M}_{\mathrm{sub,peak}}\lvert \vec{\Theta})\equiv\mathcal{L}(\vec{\Theta})$, while $P(\vec{\Theta})$ represents the prior probability for each model parameter. For both our FDM and WDM fits, we implement linear uniform prior probability distributions for the normalizations $0.1<a_i<10$ and $0<\alpha<50$, and Jeffreys priors for the slopes $0.1<b_i<1.5$, $0<\beta<10$, and $0<\gamma<5$. We sample the nine-dimensional posterior in Equation~\ref{eq:wdm_posterior} by running the Markov Chain Monte Carlo sampler \textsc{emcee} \citep{emcee} for $10^5$ steps with $100$ walkers, discarding $10^4$ burn-in steps. This yields well-converged posteriors with hundreds of independent samples as presented for WDM in Section~\ref{sec:wdm_shmf} and FDM in Section~\ref{sec:fdm_shmf}, respectively.

We ultimately marginalize over $a_i$ and $b_i$ to infer posterior distributions for $\{\alpha,~\beta,~\gamma\}$, thereby capturing covariances between these parameters. Because of the significant difference in SHMF shapes, we perform separate inference of the SHMF model parameters for WDM and for FDM. We include the CDM simulations in the likelihood for each scenario because CDM is captured in the $f_{\mathrm{beyond-CDM}}\rightarrow 1$ limit. For IDM, we compare SHMF suppression to WDM models that produce matching total subhalo abundances and assess the impact of DAOs and the transfer function cutoff shape on the resulting suppression. In Appendix~\ref{sec:shmf_details}, we demonstrate that our three hosts yield consistent posteriors for the SHMF suppression parameters and that the inferred CDM SHMF normalization $a$ and slope $b$ is consistent across our beyond-CDM models. Furthermore, we quantify host-to-host variations in the SHMF normalization and slope.

\subsection{Warm Dark Matter}
\label{sec:wdm_shmf}

\begin{figure}[t!]
\centering
\hspace{-5mm}
\includegraphics[trim={0 0.35cm 0 0cm},width=0.49\textwidth]{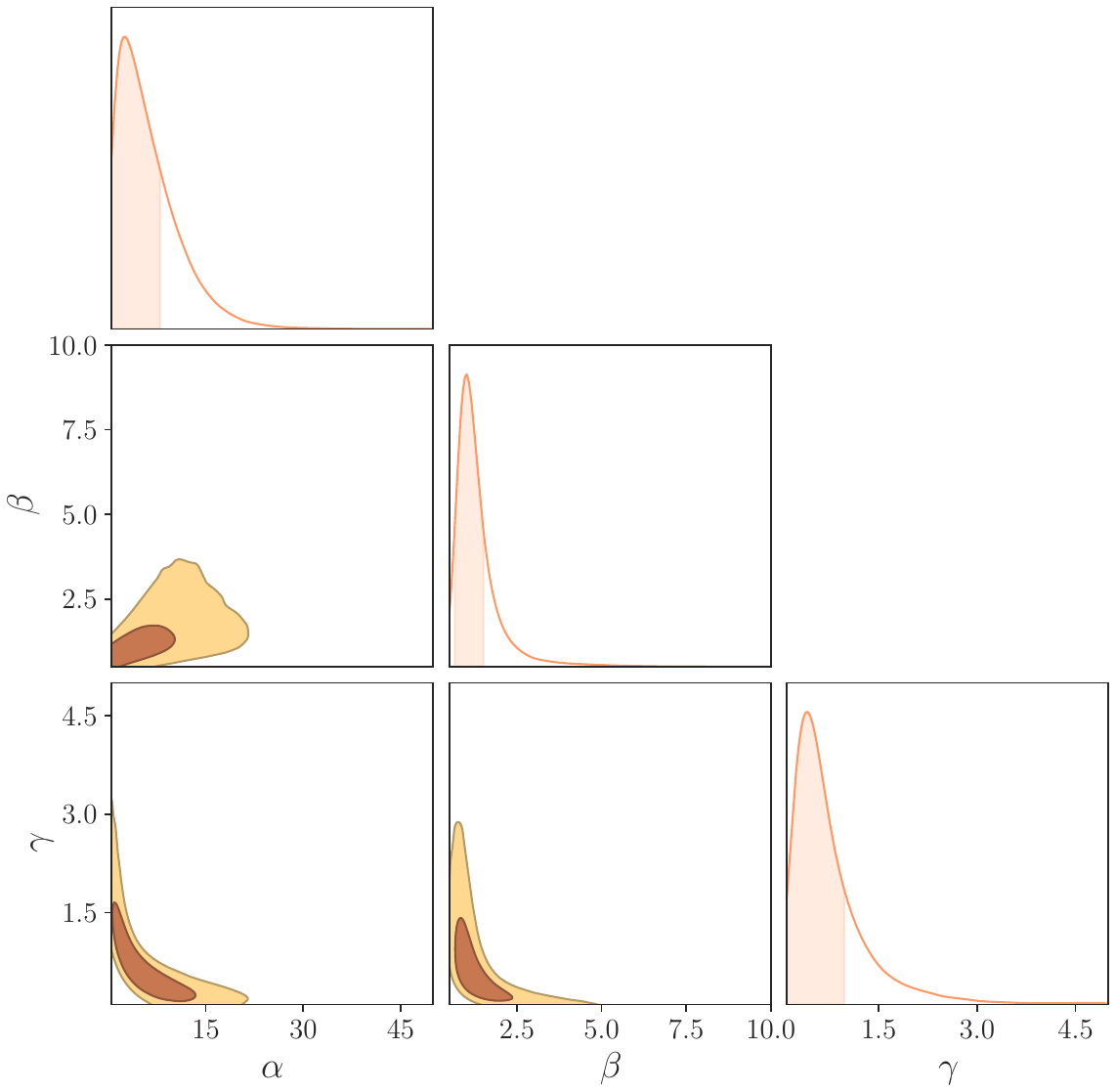}
    \caption{Marginalized posterior for our WDM SHMF suppression model. Dark (light) two-dimensional contours show $68\%$ ($95\%$) confidence intervals; top and side panels show marginal posteriors with shaded $68\%$ confidence intervals.}
    \label{fig:wdm_shmf_posterior}
\end{figure}

We define WDM SHMF suppression by inserting $M_{\mathrm{hm}}(m_{\mathrm{WDM}})$ (Equation~\ref{eq:Mhm_mwdm}) into Equation~\ref{eq:f_beyond-CDM}. The resulting posterior, marginalized over the normalizations $a_i$ and slopes $b_i$, is shown in Figure~\ref{fig:wdm_shmf_posterior}. We infer $\alpha_{\mathrm{WDM}}=2.1^{+5.9}_{-1.9}$, $\beta_{\mathrm{WDM}}=1.0^{+0.5}_{-0.3}$, and $\gamma_{\mathrm{WDM}}=0.4^{+0.6}_{-0.3}$ at $68\%$ confidence. All three parameters have upper bounds, but $\alpha_{\mathrm{WDM}}$ and $\beta_{\mathrm{WDM}}$ are only constrained at the $\approx 1\sigma$ level from below. The best-fit parameters that maximize our posterior probability distribution are as follows:
\begin{align}
    \alpha_{\mathrm{WDM}}=2.5,\nonumber\\
    \beta_{\mathrm{WDM}}=0.9,\nonumber\\
 \gamma_{\mathrm{WDM}}=1.0.\label{eq:wdm_best_fit}
\end{align}
The inferred values of $\alpha_{\mathrm{WDM}}$, $\beta_{\mathrm{WDM}}$, and $\gamma_{\mathrm{WDM}}$ are consistent with previous measurements of the WDM SHMF suppression from \cite{Lovell13081399}, which reported $\alpha_{\mathrm{WDM}}=2.7$, $\beta_{\mathrm{WDM}}=1$, $\gamma_{\mathrm{WDM}}=0.99$. We discuss other results from the WDM literature below and in Section~\ref{sec:caveats}.

Studies that model the WDM SHMF suppression routinely report a single set of best-fit values for $\alpha_{\mathrm{WDM}}$, $\beta_{\mathrm{WDM}}$, and $\gamma_{\mathrm{WDM}}$ (e.g., \citealt{Lovell13081399,Lovell200301125}). However, as shown by Figure~\ref{fig:wdm_shmf_posterior}, we find significant uncertainties on parameter values, as well as correlations between parameters; this holds even though we use a relatively large collection of WDM zoom-in simulations to constrain the SHMF model. The uncertainty is particularly large for $\alpha_{\mathrm{WDM}}$, which controls the mass scale of the SHMF suppression onset. This is partially due to a degeneracy with $\gamma_{\mathrm{WDM}}$, which controls the slope of the suppression for $M_{\mathrm{sub,peak}}\lesssim \alpha_{\mathrm{WDM}} M_{\mathrm{hm}}(m_{\mathrm{WDM}})$. In particular, a less steep suppression (smaller $\gamma_{\mathrm{WDM}}$) is degenerate with an earlier onset of suppression (at higher halo masses and larger $\alpha_{\mathrm{WDM}}$). The asymptotic slope is given by $\beta_{\mathrm{WDM}}\times\gamma_{\mathrm{WDM}}$, and thus $\gamma_{\mathrm{WDM}}$ is degenerate with $\beta_{\mathrm{WDM}}$.

\begin{figure*}[t!]
\includegraphics[width=0.5\textwidth]{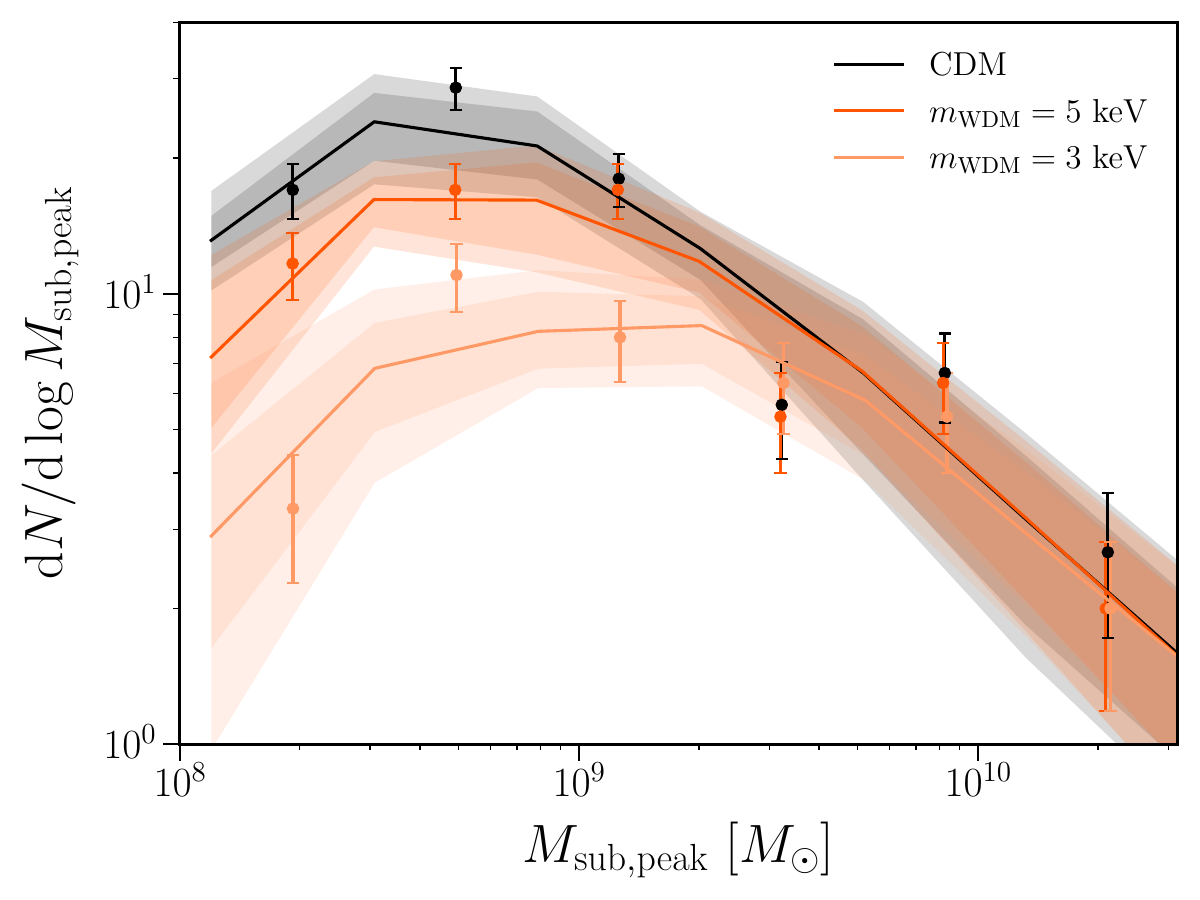}
\includegraphics[width=0.5\textwidth]{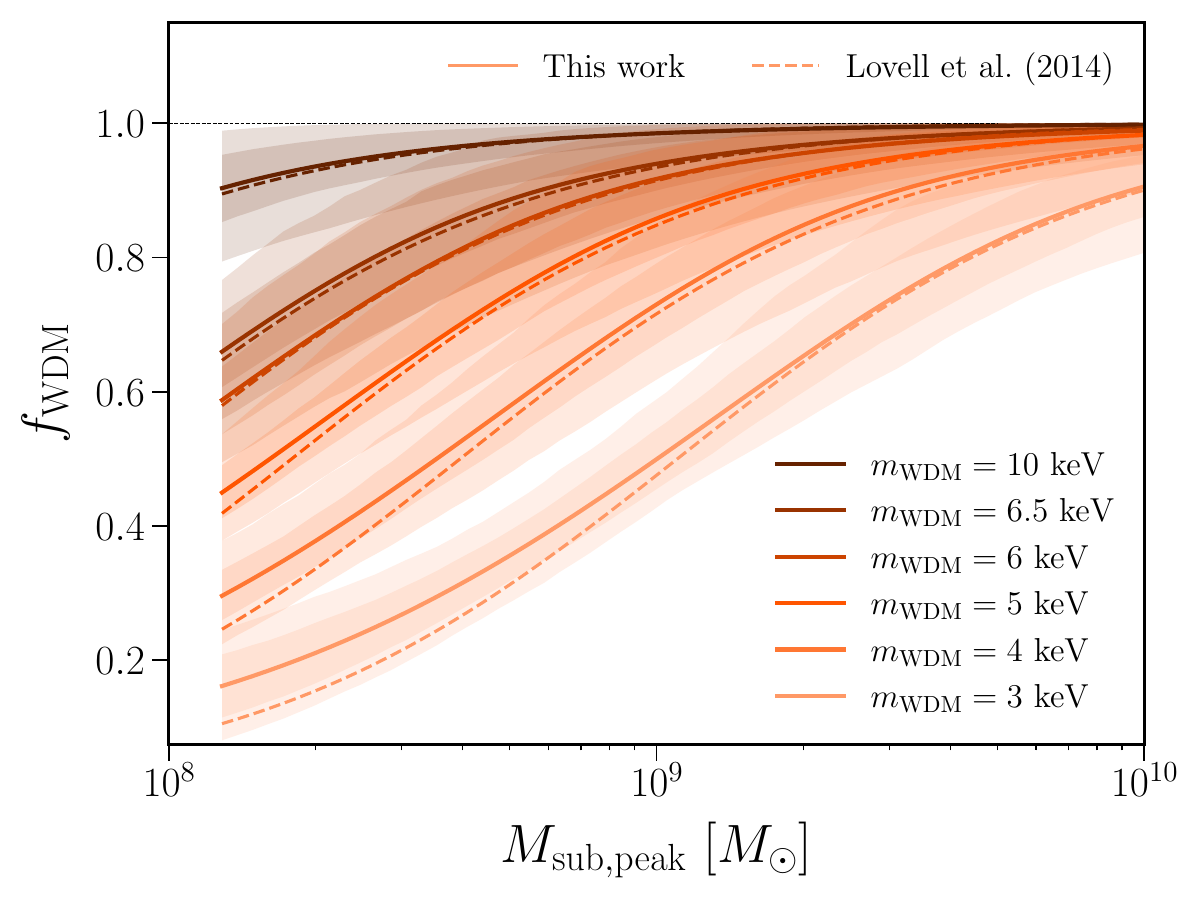}
    \caption{Left: differential SHMF as a function of peak virial mass, measured as the mean over the three host halos simulated in this work, for WDM models with $m_{\mathrm{WDM}}=3~\mathrm{keV}$ (light orange), $5~\mathrm{keV}$ (orange), and CDM (black). Error bars show the $1\sigma$ Poisson error on the mean. Right: corresponding SHMF suppression functions derived from our likelihood analysis (solid lines and bands) compared to the \cite{Lovell13081399} model (dashed lines). In both panels, dark (light) bands show $68\%$ ($95\%$) confidence intervals around the best-fit model.}
    \label{fig:wdm_shmf_pred}
\end{figure*}

The left panel of Figure~\ref{fig:wdm_shmf_pred} illustrates the best-fit SHMF model and its suppression function, as well as the associated uncertainties. We find that the model describes simulation data well: the reduced $\chi^2$ statistic for each model (assuming Poisson errors on the simulation measurements) is $\chi^2=[0.76,~ 0.72,~ 0.73,~ 1.1,~ 1.0,~ 0.72]$ for $m_{\mathrm{WDM}}= [3,~ 4,~ 5,~ 6,~ 6.5,~ 10]~\mathrm{keV}$, respectively, indicating a good fit in all cases. Note that the differential SHMF turns over at low $M_{\mathrm{sub,peak}}$, even in CDM, due to our $M_{\mathrm{sub}}$ resolution cut. As $m_{\mathrm{WDM}}$ increases, the SHMF suppression becomes less severe (at fixed $M_{\mathrm{sub,peak}}$) and flatter (as a function of $M_{\mathrm{sub,peak}}$).

We directly compare our results with \cite{Lovell13081399}, derived from a suite of four thermal-relic WDM zoom-in simulations with comparable resolution to ours. Specifically, this previous study fixed $\beta_{\mathrm{WDM}}=1$ and showed that $\alpha_{\mathrm{WDM}}=2.7$, $\gamma_{\mathrm{WDM}}=0.99$ accurately describes the SHMF suppression in their simulations, measured using present-day subhalo mass. As shown by the dashed lines in the right panel of Figure~\ref{fig:wdm_shmf_pred}, our $f_{\mathrm{WDM}}$ posteriors agree well with the \cite{Lovell13081399} fit. Our predictions are slightly less suppressed at $10^8~M_{\mathrm{\odot}}\lesssim M_{\mathrm{sub,peak}}\lesssim 10^9~M_{\mathrm{\odot}}$ for $m_{\mathrm{WDM}}\leq 5~\mathrm{keV}$, although the difference is within the $1\sigma$ uncertainties of our model.

We have also compared our results with the (sub)halo mass function suppression fits from \cite{Lovell200301125} and \cite{Stucker210909760}. Interestingly, our $f_{\mathrm{WDM}}$ fit agrees well with the isolated halo mass function suppression from Lovell et al.\ (\citeyear{Lovell200301125}; i.e., $\alpha_{\mathrm{WDM}}=2.3,\ \beta_{\mathrm{WDM}}=0.8,\ \gamma_{\mathrm{WDM}}=1.0$), while their SHMFs are significantly less suppressed than ours (i.e., $\alpha_{\mathrm{WDM}}=4.2,\ \beta_{\mathrm{WDM}}=2.5,\ \gamma_{\mathrm{WDM}}=0.2$), even after accounting for the uncertainty on our $f_{\mathrm{WDM}}$ reconstruction (see Section 4.3 of \citealt{Stucker210909760} for a related discussion). Meanwhile, \cite{Stucker210909760} simulate WDM-like transfer function cutoffs using a hybrid phase-space sheet plus $N$-body simulation technique. Instead of fitting for $\alpha_{\mathrm{WDM}}$, $\beta_{\mathrm{WDM}}$, and $\gamma_{\mathrm{WDM}}$, these authors infer the mass scales at which the halo mass function is suppressed by $20\%$, $50\%$, and $80\%$ relative to CDM. Both the halo and SHMF suppression models from \cite{Stucker210909760} are significantly less suppressed than our result.

These comparisons are complicated by several factors, and a detailed assessment is beyond the scope of our work. For example, \cite{Lovell200301125} simulated sterile neutrino WDM transfer functions, which differ from $P(k)$ in our thermal-relic WDM models (see, e.g., Figure 1 of \citealt{Lovell191111785}), while \cite{Stucker210909760} simulated generalized $\alpha$--$\beta$--$\gamma$ transfer functions, which span (but do not exactly overlap with) the transfer functions we simulate. Furthermore, \cite{Lovell200301125} ran cosmological WDM simulations in addition to zoom-ins, while \cite{Stucker210909760} exclusively ran cosmological simulations. Our comparison to \cite{Lovell13081399} is the most straightforward in terms of both ICs and simulation technique and yields good agreement, despite our use of peak (rather than present-day) subhalo mass. Indeed, we infer a similar $f_{\mathrm{WDM}}$ when the SHMF is measured using present-day mass, suggesting that the mass-loss rates of surviving, well-resolved subhalos do not significantly differ between our CDM and WDM simulations. This should be studied directly in future work (e.g., building on \citealt{Du240309597}).

\subsection{Fuzzy Dark Matter}
\label{sec:fdm_shmf}

\begin{figure}[t!]
\centering
\hspace{-5mm}
\includegraphics[trim={0 0.35cm 0 0cm},width=0.49\textwidth]{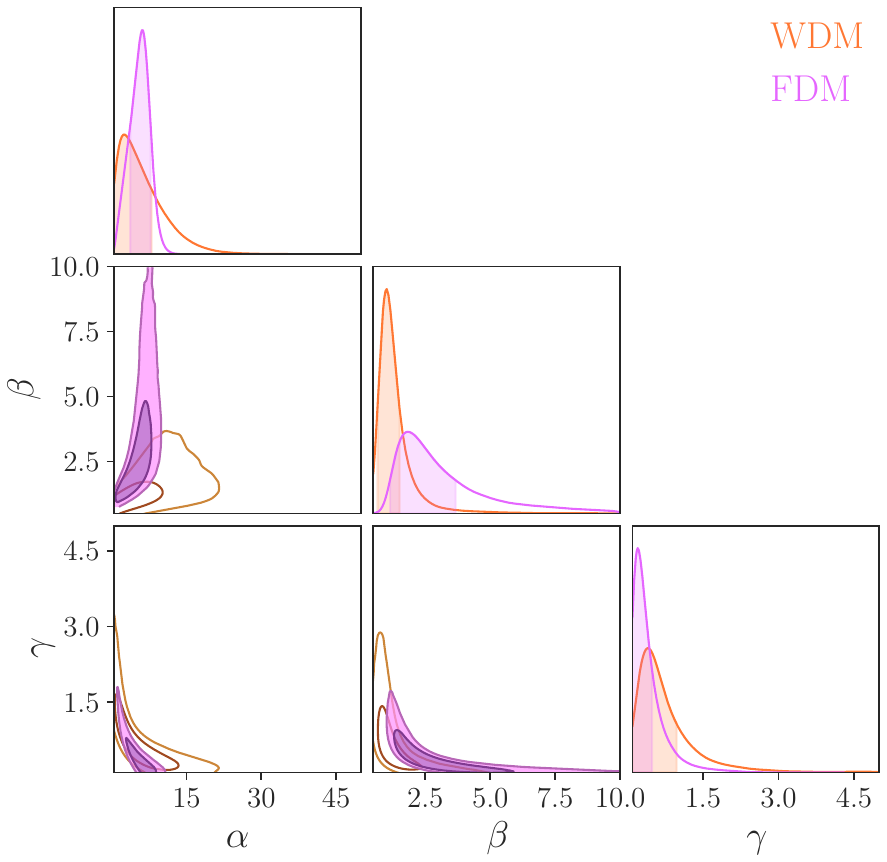}
    \caption{Marginalized posterior for our FDM SHMF suppression model (filled purple). Dark (light) two-dimensional contours show $68\%$ ($95\%$) confidence intervals; top and side panels show marginal posteriors with shaded $68\%$ confidence intervals. The WDM posterior from Figure~\ref{fig:wdm_shmf_posterior} is shown for reference (unfilled orange).}\label{fig:fdm_shmf_posterior}
\end{figure}

We define FDM SHMF suppression by inserting $M_{\mathrm{hm}}(m_{\mathrm{FDM,22}})$ (Equation~\ref{eq:mhm_mfdm}) into Equation~\ref{eq:f_beyond-CDM}. The marginalized posterior for our FDM SHMF suppression parameters is shown in Figure~\ref{fig:fdm_shmf_posterior}. We infer $\alpha_{\mathrm{FDM}}=6.2^{+1.6}_{-2.5}$, $\beta_{\mathrm{FDM}}=1.8^{+1.9}_{-0.7}$, and $\gamma_{\mathrm{FDM}}=0.2^{+0.3}_{-0.1}$ at $68\%$ confidence. These parameters are bounded from above but are only marginally constrained from below, and they display the same qualitative degeneracies as in our WDM fit. We again provide the best-fit parameters that maximize our FDM SHMF posterior:
\begin{align}
    \alpha_{\mathrm{FDM}}=5.5\nonumber\\
    \beta_{\mathrm{FDM}}=2.5,\nonumber\\
 \gamma_{\mathrm{FDM}}=0.3.\label{eq:fdm_best_fit}
\end{align}
Note that authors often define FDM SHMF suppression using Equation~\ref{eq:f_beyond-CDM} with $M_0\approx M_{\mathrm{hm}}/3$ in place of $M_{\mathrm{hm}}$ (see Section~\ref{sec:fdm_model}). This choice would correspondingly decrease our inferred values of $\alpha_{\mathrm{FDM}}$ by a factor of $\approx 3$.

\begin{figure*}[t!]
\includegraphics[width=0.5\textwidth]{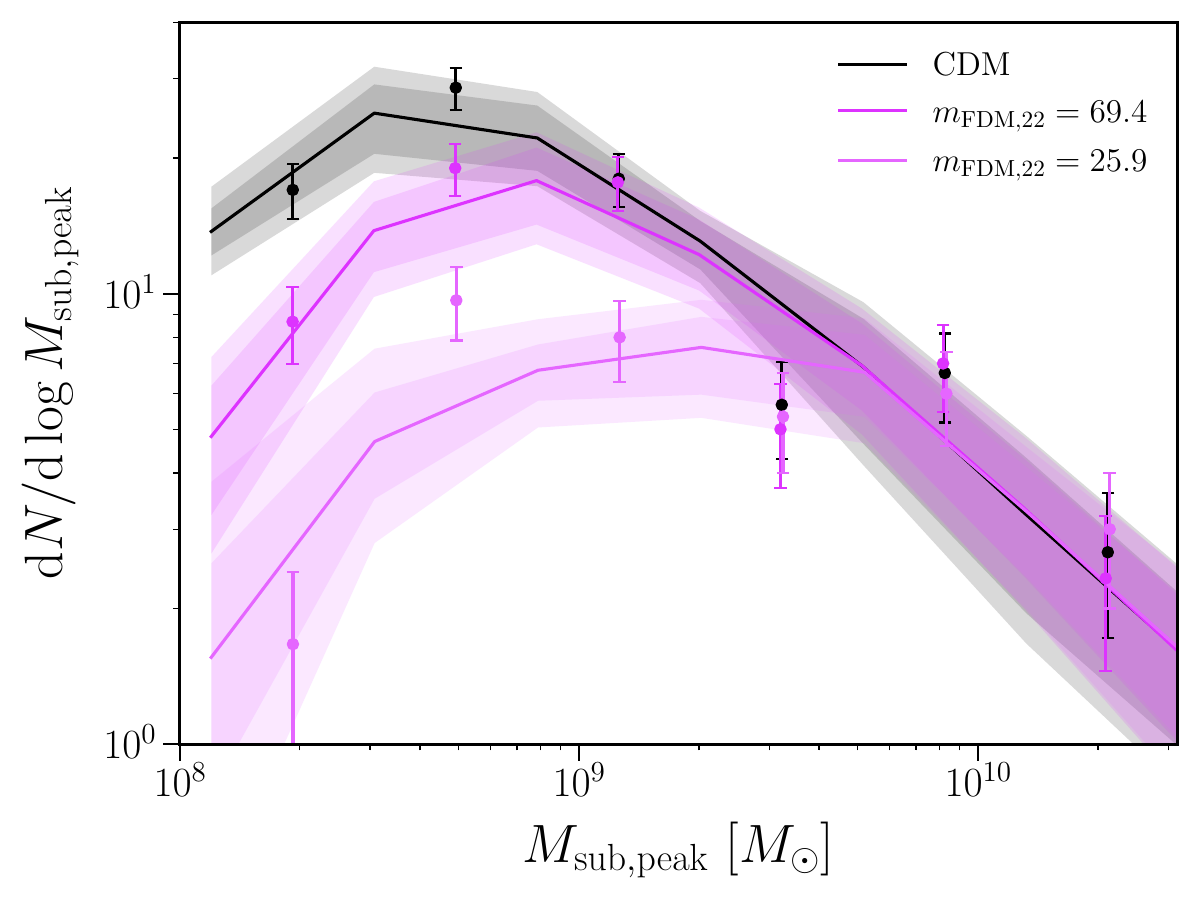}
\includegraphics[width=0.5\textwidth]{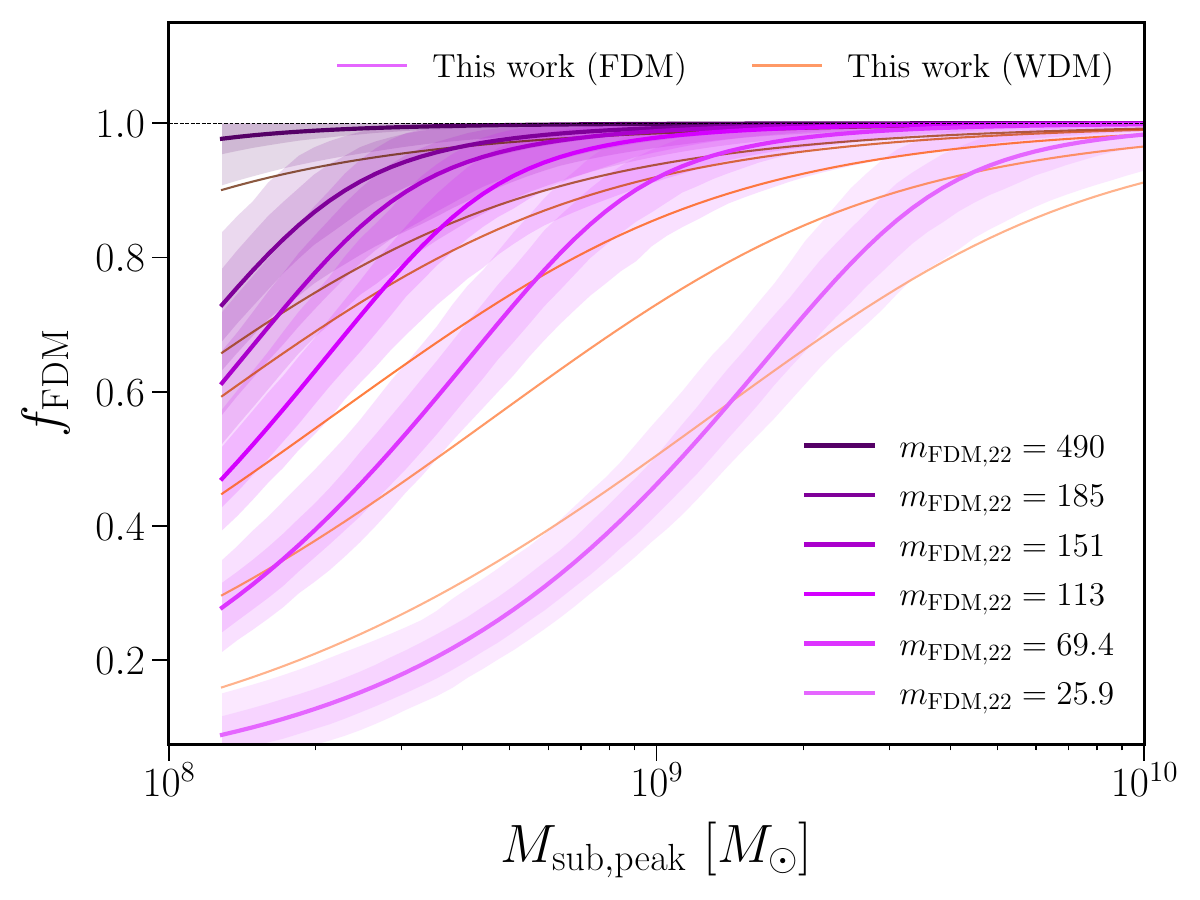}
    \caption{Same as Figure~\ref{fig:wdm_shmf_pred}, but for our FDM SHMF model. In the right panel, FDM SHMF suppression functions are compared to WDM models with matched half-mode wavenumbers (faint orange lines).}
    \label{fig:fdm_shmf_pred}
\end{figure*}

Figure~\ref{fig:fdm_shmf_pred} shows our FDM SHMF fits, which again are in good agreement with our simulation results, with $\chi^2=[0.43,~ 0.74,~ 0.93,~ 0.69,~ 1.2,~ 0.61]$ for $m_{\mathrm{FDM,22}}= [25.9,~ 69.4,~ 113,~ 151,~ 185,~ 490]$. The inferred SHMF becomes less suppressed (at fixed $M_{\mathrm{sub,peak}}$) and flatter (as a function of $M_{\mathrm{sub,peak}}$) as $m_{\mathrm{FDM,22}}$ increases. In comparison to our WDM results, we find the following:
\begin{enumerate}
    \item The best-fit value of $\alpha_{\mathrm{FDM}}$ is larger than $\alpha_{\mathrm{WDM}}$ by a factor of $\approx 2$; however, the marginalized posteriors are consistent at the $1\sigma$ level.
    \item The best-fit value of $\beta_{\mathrm{FDM}}$ is larger than $\beta_{\mathrm{WDM}}$ by a factor of $\approx 3$, corresponding to a $1\sigma$ shift in the marginalized posteriors, indicating that FDM SHMF suppression is steeper than in WDM.
    \item The best-fit values and marginalized posteriors for $\gamma_{\mathrm{FDM}}$ and $\gamma_{\mathrm{WDM}}$ are statistically consistent.
    \item Half-mode-matched FDM and WDM models yield consistent suppression at $M_{\mathrm{sub,peak}}=10^8~M_{\mathrm{{\odot}}}$. For each half-mode-matched pair, the probability that we draw an $f_{\mathrm{FDM}}(M_{\mathrm{sub,peak}}=10^8~M_{\mathrm{\odot}})$ consistent with our inferred $f_{\mathrm{WDM}}(M_{\mathrm{sub,peak}}=10^8~M_{\mathrm{\odot}})$ at the $1\sigma$ level is $p=0.09,~ 0.4,~ 0.08,~ 0.14,~ 0.11$, and $0.18$, for $m_{\mathrm{FDM,22}}=25.9,~ 69.4,~ 113,~ 151,~ 185$, and $ 490$, respectively.

    \item At $M_{\mathrm{sub,peak}}=10^9~M_{\mathrm{\odot}}$, the same procedure yields $p=0.16,~ 0.03,~ 0.04,~ 0.1,~ 0.13$, and $0.21$ for $m_{\mathrm{FDM,22}}=25.9,~ 69.4,~ 113,~ 151,~ 185$, and $ 490$, respectively. Thus, the suppression significantly differs at $M_{\mathrm{sub,peak}}=10^9~M_{\mathrm{\odot}}$ between $m_{\mathrm{FDM,22}}=69.4$ and $m_{\mathrm{WDM}}=4~\mathrm{keV}$, and between $m_{\mathrm{FDM,22}}=113$ and $m_{\mathrm{WDM}}=5~\mathrm{keV}$.
    \item All half-mode-matched models yield suppression consistent at the $2\sigma$ level at both of these mass scales; however, two-sample KS tests using our $f_{\mathrm{FDM}}$ and $f_{\mathrm{WDM}}$ posteriors indicate significant differences between the distributions at each $M_{\mathrm{sub,peak}}$.
    \end{enumerate}
These differences in SHMF suppression amplitude and shape reflect the sharper $P(k)$ cutoff for FDM compared to WDM. This follows because $M_{\mathrm{hm}}$ is the only parameter that enters our SHMF suppression model (Equation~\ref{eq:f_beyond-CDM}) and we compare WDM and FDM models with matched $M_{\mathrm{hm}}$.

We also compare our $f_{\mathrm{FDM}}$ predictions to various literature results. First, the generalized $\alpha$--$\beta$--$\gamma$ models simulated by \cite{Stucker210909760} contain FDM-like transfer functions. These authors' FDM halo and SHMFs are less suppressed than ours; the magnitude of the difference is similar to that between our $f_{\mathrm{WDM}}$ result and their WDM fit. As discussed above, interpreting this comparison is beyond the scope of our work owing to differences in the underlying transfer functions and simulation techniques. Meanwhile, \cite{Elgamal231103591} ran zoom-in simulations that incorporate both FDM transfer functions and wave dynamics using an SPH solver. Their SHMF fit predicts nearly zero suppression relative to CDM for $m_{22,\mathrm{FDM}}>16$ (i.e., the coldest FDM model they simulate). These authors measured the mass function of subhalos with $M_{\mathrm{sub}}\gtrsim 3\times 10^{9}~M_{\mathrm{\odot}}$, where our fit also predicts very little suppression; at lower masses, our relatively high resolution allows us to resolve the prominent cutoff in $f_{\mathrm{FDM}}$.

\begin{figure*}[t!]
\centering
\includegraphics[width=\textwidth]{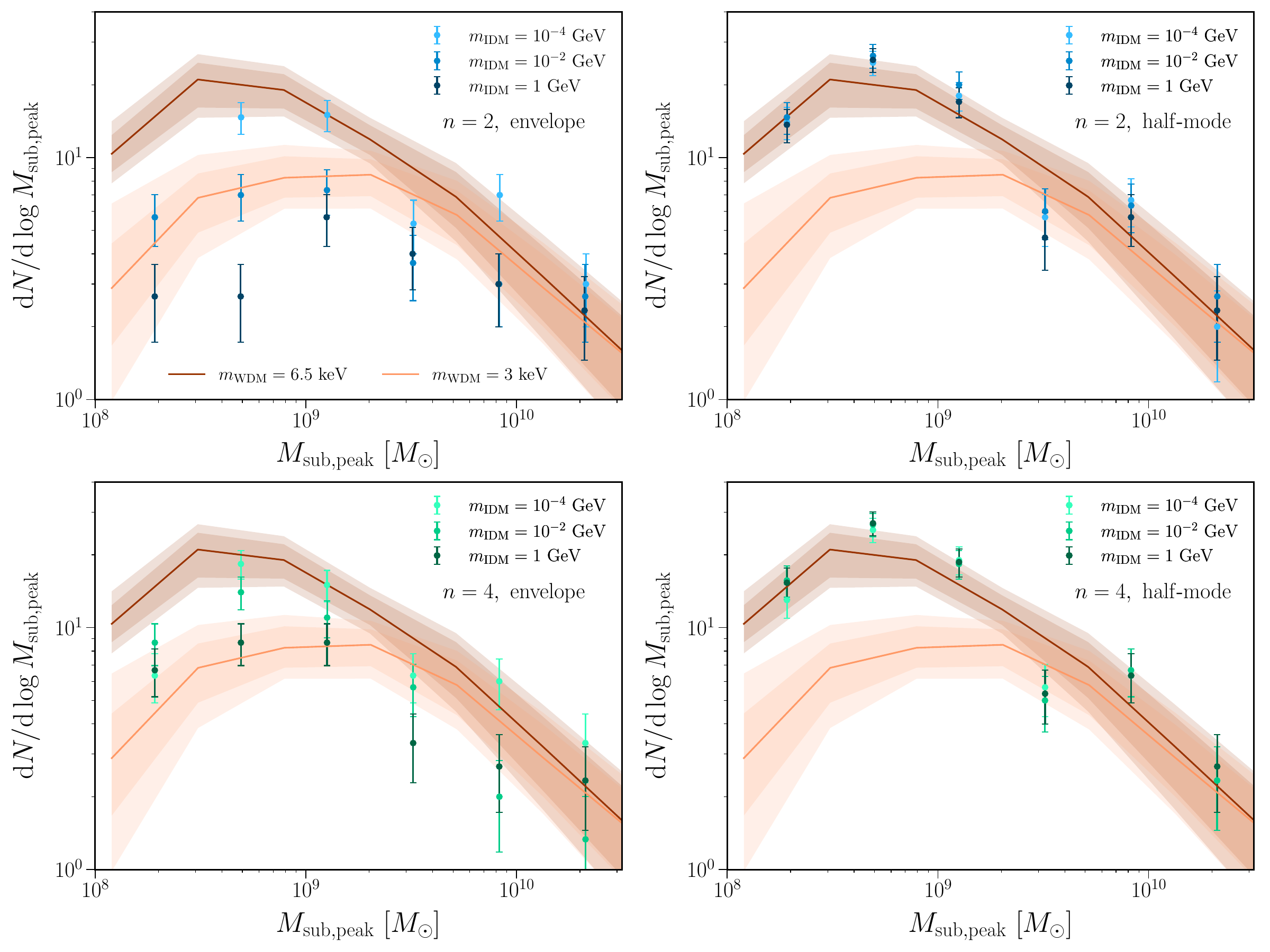}
    \caption{Mean SHMFs, measured using peak virial mass, for our IDM models. ICs for our half-mode (envelope) IDM in the left column (right column) are chosen to match the half-mode wavenumber of (be strictly more suppressed than) the transfer function for $m_{\mathrm{WDM}}=6.5~\mathrm{keV}$. The mean SHMF for this WDM model, predicted by our WDM SHMF fit, is shown as a dark-orange line in each panel; light-orange lines and bands show corresponding results for our $m_{\mathrm{WDM}}=3~\mathrm{keV}$ model. For both WDM models, dark (light) bands show $68\%$ ($95\%$) confidence intervals from our fit, calculated by sampling from the posterior.}
    \label{fig:shmf_matched}
\end{figure*}

The \cite{Schive150804621} FDM suppression fit---corresponding to $\alpha_{\mathrm{FDM}}=1$, $\beta_{\mathrm{FDM}}=1.1$, $\gamma_{\mathrm{FDM}}=2.2$ with $M_0\approx M_{\mathrm{hm}}/3$ used in place of $M_{\mathrm{hm}}$---is often adopted in the literature. For all FDM models we simulate, this fit is much less suppressed at low $M_{\mathrm{sub,peak}}$ than ours, and its overall shape is shallower. \cite{Schive150804621} obtained this fit by measuring isolated halo mass functions in cosmological $N$-body DM-only simulations at $z=4$. Their simulations were initialized using the \cite{Hu008506} transfer function fit, which is steeper than our \textsc{axionCAMB} input at fixed $m_{\mathrm{FDM}}$; this may partly explain this difference with our results. In future work, it will be interesting to compare our results to FDM SHMF suppression predictions from recent simulations (e.g., \citealt{May220914886}) and semianalytic models (e.g., \citealt{Du189706,Kulkarni201102116}).

\subsection{Interacting Dark Matter}
\label{sec:idm_shmf}

\begin{figure*}[t!]
\centering
\includegraphics[width=\textwidth]{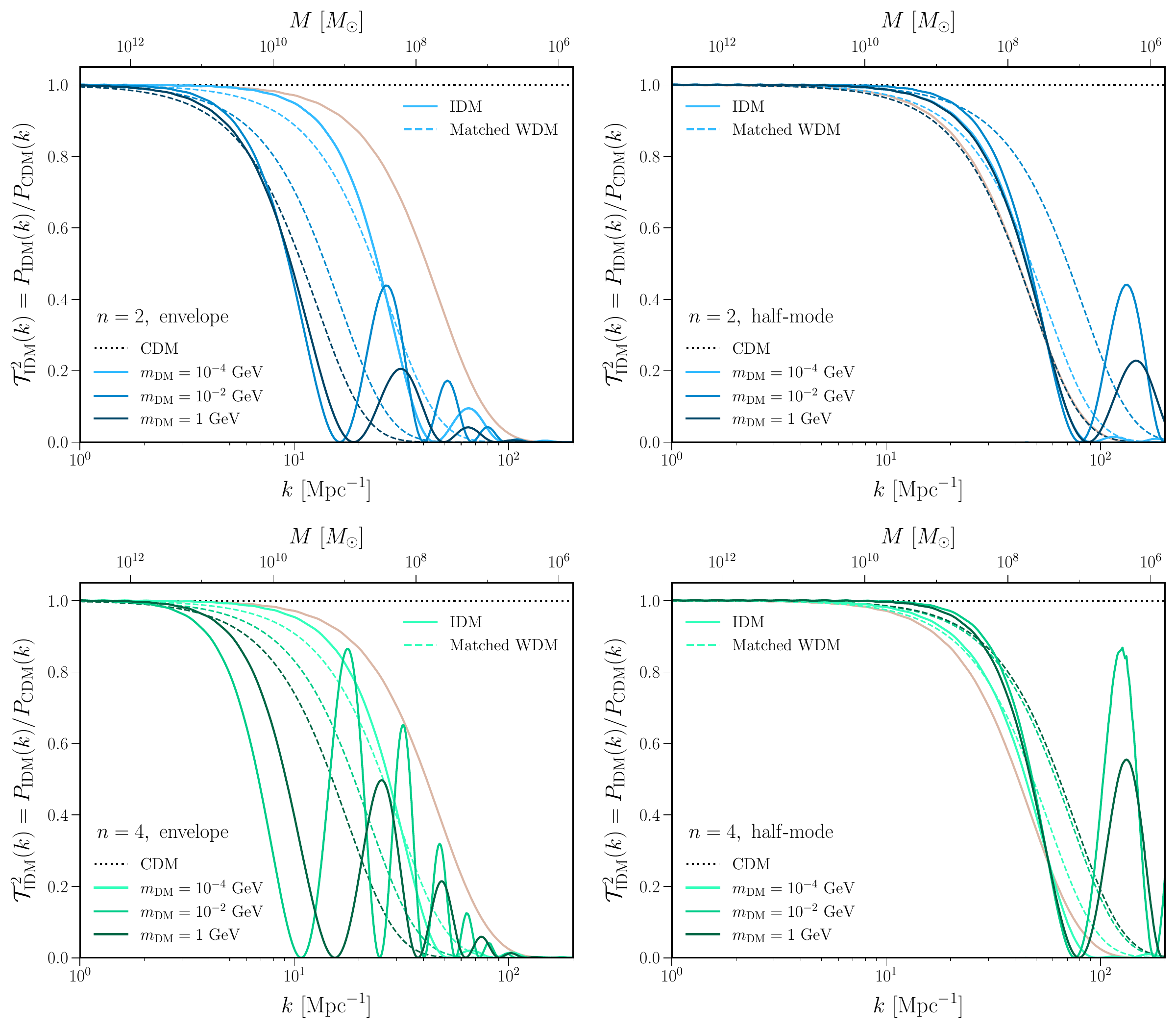}
    \caption{Ratio of the linear matter power spectrum for each IDM model we simulate (solid lines) relative to CDM (dotted black lines), compared to WDM models that produce the same total number of subhalos above the resolution limit (``Matched WDM''; dashed lines). The half-mode IDM models in the left column have the same half-mode scale as the limiting WDM scenario with $m_{\mathrm{WDM}}=6.5~\mathrm{keV}$, which is shown as a faint orange line in each panel. The IDM models in the left column are strictly more suppressed than the limiting WDM model.}
    \label{fig:transfers_matched}
\end{figure*}

Figure~\ref{fig:shmf_matched} shows the differential SHMF, as a function of $M_{\mathrm{sub,peak}}$ and subject to our fiducial-resolution cut of $M_{\mathrm{sub}}>1.2\times 10^8~M_{\mathrm{\odot}}$, for each IDM model we simulate. Several results are immediately apparent. First, the envelope IDM models produce SHMFs that are systematically more suppressed than the reference $m_{\mathrm{WDM}}=6.5~\mathrm{keV}$ model. This is not surprising, since the corresponding IDM transfer functions were chosen to be strictly more suppressed than this WDM model. Second, many of the envelope IDM models produce total subhalo abundances comparable to the most suppressed WDM model we simulate (i.e., $m_{\mathrm{WDM}}=3~\mathrm{keV}$). These results justify the procedure in \cite{Maamari201002936} to place bounds on the envelope models using the MW satellite population, as they would produce many fewer satellite galaxies than the limiting $m_{\mathrm{WDM}}=6.5~\mathrm{keV}$ WDM model.

The SHMFs in IDM models with the same half-mode scale as $m_{\mathrm{WDM}}=6.5~\mathrm{keV}$ are statistically consistent with the SHMFs that best fit the WDM simulations, reported in Section~\ref{sec:wdm_shmf}. As a result, we expect that upper bounds on the IDM cross section from the MW satellite population should be closer to the half-mode than the envelope cross section; this is further quantified below (see Section~\ref{sec:idm_bound}). IDM models yield slightly higher subhalo abundances than the $m_{\mathrm{WDM}}=6.5~\mathrm{keV}$ model, which indicates that these scenarios are not ruled out at $95\%$ confidence by the \cite{Nadler200800022} analysis, consistent with the reasoning in \cite{Maamari201002936}.

To facilitate comparisons between IDM and WDM SHMFs, we find the value of $m_{\mathrm{WDM}}$ that yields a \textit{total} number of subhalos with $M_{\mathrm{sub}}>1.2\times 10^8~M_{\mathrm{\odot}}$ that most closely matches the value we measure in IDM, after averaging over the three host halos in each scenario. Since our simulation suite only has a limited number of WDM mass benchmarks, to predict the total subhalo abundance as a function of $m_{\mathrm{WDM}}$, we apply the WDM SHMF suppression model from Equation~\ref{eq:f_beyond-CDM}, evaluated at the best-fit parameter values from Equation~\ref{eq:wdm_best_fit}, and modify the total count of subhalos measured in the CDM simulations accordingly. We match IDM to WDM based on total subhalo abundance rather than the full SHMF as a simple first-order comparison relevant for limits based on total MW satellite abundances. The results of this procedure are illustrated in Figure~\ref{fig:transfers_matched} and summarized in Table~\ref{tab:summary}.

Several interesting comparisons between IDM models emerge from Figure~\ref{fig:transfers_matched}. In the half-mode panels, the key takeaway is that all IDM transfer functions yield statistically indistinguishable subhalo abundances, as shown in Figure~\ref{fig:shmf_matched}. For example, compare the transfer functions for $m_{\mathrm{IDM}}=10^{-2}~\mathrm{GeV}$ and $1~\mathrm{GeV}$ half-mode models with $n=4$, which mainly differ in the height of the first DAO peak. The similarity of the resulting subhalo abundances implies that our measurement is not sensitive to the amplitude of the DAO peak for $k_{\mathrm{peak}}\gtrsim 100~\mathrm{Mpc}^{-1}$. This wavenumber corresponds to a mass of $\approx 5\times 10^6~M_{\mathrm{\odot}}$, well below our resolution limit.

DAOs play a more significant role for our envelope models because they appear on the scales of resolved halos. For $n=4$, the $m_{\mathrm{IDM}}=10^{-2}~\mathrm{GeV}$ case is matched to a less suppressed WDM transfer function than the $1~\mathrm{GeV}$ case, even though its initial cutoff occurs at smaller wavenumbers (compare the medium-green vs.\ dark-green lines in the bottom left panel of Figure~\ref{fig:transfers_matched}). Thus, SHMF suppression is reduced owing to the large DAOs in the $m_{\mathrm{IDM}}=10^{-2}~\mathrm{GeV}$ envelope model. A similar result holds for $n=2$ models (compare the medium-blue vs.\ dark-blue lines in the top left panel of Figure~\ref{fig:transfers_matched}). Note that the impact of DAOs varies over the range of IDM masses we study. For example, the $m_{\mathrm{IDM}}=10^{-4}~\mathrm{GeV}$ models behave fairly similarly to WDM because of their small DAOs, although minor differences persist even in these cases because of the steeper initial cutoffs in these models.

Several previous studies have simulated IDM models with DAOs, although (to our knowledge) no zoom-in simulations have been performed with ICs for the DM--baryon scattering models we consider. In particular, \cite{Schewtschenko151206774} ran simulations with ICs appropriate for DM--photon scattering models with typical DAO peak heights of $\mathcal{O}(1\%)$ relative to CDM (see also \citealt{Boehm14047012,Schewtschenko14124905,Akita230501913}). These authors find that WDM and DM--photon scattering models with matched $k_{\mathrm{hm}}$ yield similar subhalo populations, consistent with our findings for IDM models with small DAOs (i.e., the $m_{\mathrm{IDM}}=10^{-4}~\mathrm{GeV}$ cases for $n=2$ and $n=4$). These results lend confidence to studies that match such models to WDM to derive constraints (e.g., \citealt{Crumrine240619458}).

Other previous studies have simulated models with larger DAOs, comparable to our IDM ICs. For example, \cite{Vogelsberger151205349} ran zoom-in simulations of ETHOS models that include both $P(k)$ suppression and late-time self-interactions. Because these models feature nongravitational interactions, we defer a detailed comparison to Paper III, in which we consider similar scenarios. Meanwhile, \cite{Bohr200601842,Bohr210108790} ran cosmological simulations of ETHOS models at $z\gtrsim 5$, parameterized by $k_{\mathrm{peak}}$ and $h_{\mathrm{peak}}$, including cases with $\mathcal{O}(1)$ DAO peaks in the transfer function. That study finds that, at high redshifts, only models with $k_{\mathrm{peak}}\lesssim 70~\mathrm{Mpc}^{-1}$ and $h_{\mathrm{peak}}\gtrsim 0.2$ affect the abundance of $\approx 10^8~M_{\mathrm{\odot}}$ differently than WDM. This picture is qualitatively consistent with our MW subhalo population results at $z=0$, even though we do not directly observe oscillatory features in our mass functions. Thus, models with small DAOs ($h_{\mathrm{peak}}\lesssim 0.2$) are unlikely to be distinguished from WDM using their effects on (sub)halo abundances alone, consistent with the results of \cite{Schaeffer210112229}. It will be interesting to test whether (sub)halo density profiles can be used to differentiate these models from each other and from WDM, which we leave for future work.

\subsection{Summary of Key Results}
\label{sec:shmf_summary}

Finally, we summarize key results from our SHMF fits:
\begin{enumerate}
    \item We derive WDM SHMF suppression consistent with previous studies (Figure~\ref{fig:wdm_shmf_pred}). The best-fit parameters are given by Equation~\ref{eq:wdm_best_fit}.
    \item We derive FDM SHMF suppression that is significantly steeper than WDM owing to the sharper cutoff in $P(k)$ (Figure~\ref{fig:fdm_shmf_pred}). This suppression is more severe than commonly adopted models for the FDM SHMF. The best-fit parameters are given by Equation~\ref{eq:fdm_best_fit}.
    \item IDM SHMFs are consistent with the $m_{\mathrm{WDM}}=6.5~\mathrm{keV}$ case in the ``half-mode'' scenario and are severely suppressed in the ``envelope'' scenario. DAOs reduce the amplitude of the suppression relative to models with matched half-mode scales.
\end{enumerate}

\section{Bounds on beyond-CDM Scenarios}
\label{sec:limits}

We now apply our new SHMF models from Section~\ref{sec:shmf} to derive updated limits on WDM and FDM and to estimate updated limits on the IDM cross section, using the MW satellite galaxy population measured by DES and PS1 in \cite{Drlica-Wagner191203302}.

\subsection{Procedure}
\label{sec:procedure}

We use the inference framework from \cite{Nadler190410000,Nadler191203303, Nadler200800022}, which combines DM-only zoom-in simulations \citep{Mao150302637}, an empirical galaxy--halo connection model \citep{Nadler171204467,Nadler180905542,Nadler191203303}, and DES and PS1 MW satellite population observations and selection functions \citep{Drlica-Wagner191203302} to place constraints on beyond-CDM models. This framework allows us to conservatively marginalize over a wide range of galaxy--halo connection scenarios in order to obtain robust constraints. Following \cite{Nadler200800022}, we do not alter the galaxy--halo connection parameterization in our beyond-CDM scenarios. Thus, we assume that satellite abundances are affected in our beyond-CDM scenarios but that the observable properties of existing satellites are not. This assumption is consistent with the beyond-CDM hydrodynamic simulation results from \cite{Despali250112439}, who studied galaxies with larger stellar masses than we consider. We leave a study of the coupling between our galaxy--halo connection parameterization and beyond-CDM physics to future work (see \citealt{Nadler240110318} for further discussion).

In detail, we generate satellite population realizations using the eight galaxy--halo connection parameters from \cite{Nadler200800022} plus an additional parameter controlling the SHMF suppression. For WDM and FDM, this parameter is $M_{\mathrm{hm}}$, following our SHMF suppression model (Equation~\ref{eq:f_beyond-CDM}); for IDM, we will map to constraints on effective WDM models by matching only the total subhalo abundances. The number of predicted satellites in luminosity and surface brightness bin $i$ in a given realization is
\begin{equation}
    n_{\mathrm{pred,}i} = \sum_{s_{i}} p_{\mathrm{detect,}s_{i}}\times (1-p_{\mathrm{disrupt,}s_{i}})\times f_{\mathrm{gal,}s_{i}}\times f_{\mathrm{beyond-CDM},s_{i}},
\end{equation}
where $s_{i}$ denotes mock satellites in bin $i$, $p_{\mathrm{detect}}$ is the detection probability determined by the DES and PS1 selection functions \citep{Drlica-Wagner191203302}, $p_{\mathrm{disrupt}}$ is the disk disruption probability from \cite{Nadler171204467}, modified as in \cite{Nadler191203303}, $f_{\mathrm{gal}}$ is the galaxy occupation fraction modeled following \citep{Nadler191203303}, and $f_{\mathrm{beyond-CDM}}$ is the SHMF suppression (Equation~\ref{eq:f_beyond-CDM}), which we respectively evaluate at the best-fit parameters for WDM and FDM (Equations~\ref{eq:wdm_best_fit} and \ref{eq:fdm_best_fit}, respectively).\footnote{Note that $p_{\mathrm{disrupt}}$ captures enhanced tidal stripping due to the Galactic disk, which is distinct from the DM-only stripping effect that probabilistically relates $M_{\mathrm{peak}}$ and $M_{\mathrm{vir}}$ in our SHMF model.} Thus, we assume that the universal $f_{\mathrm{beyond-CDM}}$ measured across our three hosts applies to the subhalo populations used in \cite{Nadler200800022}. In our galaxy--halo connection model, the only subhalo property that $f_{\mathrm{gal}}$ and $f_{\mathrm{beyond-CDM}}$ depend on is $M_{\mathrm{sub,peak}}$; however, these terms suppress satellite abundances with a different dependence on $M_{\mathrm{sub,peak}}$ \citep{Nadler240110318}. Meanwhile, $p_{\mathrm{detect}}$ depends on each subhalo's three-dimensional position, peak $V_{\mathrm{max}}$, and virial radius at accretion, while $p_{\mathrm{disrupt}}$ depends on $V_{\mathrm{max}}$ and mass at accretion, infall scale factor, and distance and scale factor at first pericenter.

The probability of observing $n_{\mathrm{obs,i}}$ satellites in bin $i$ is then \citep{Nadler180905542,Nadler191203303,Nadler240110318}
\begin{align}
&P(n_{\mathrm{obs},i}|\boldsymbol{\theta}_{\mathrm{g-h}},M_{\mathrm{hm}})& \nonumber \\ &= \Big(\frac{N_{\mathrm{real.}}+1}{N_{\mathrm{real.}}}\Big)^{-({N}_{i} + 1)}\times (N_{\mathrm{real.}}+1)^{-n_{\mathrm{obs},i}}\frac{\Gamma({N}_{i} + n_{\mathrm{obs},i}+1)}{\Gamma(n_{\mathrm{obs},i}+1)\Gamma({N}_{i}+1)},&\label{eq:poisson_like}
\end{align}
where $\boldsymbol{\theta}_{\mathrm{g-h}}$ represents the eight galaxy--halo connection parameters from \cite{Nadler200800022}, ${N_\mathrm{real.}}$ is the number of model realizations at fixed galaxy--halo connection and beyond-CDM parameters, and ${N}_{i}\equiv \sum_{j=1}^{{N_\mathrm{real.}}}n_{\mathrm{pred},ij}$. Note that $N_{\mathrm{real.}}$ includes draws over different zoom-ins, different observer locations within each zoom-in (with the on-sky LMC position held fixed), and multiple realizations of the stochastic galaxy--halo connection model for each of these choices; we use $N_{\mathrm{real.}}=10$, following \cite{Nadler240110318}.

Finally, we calculate the likelihood of observing the DES and PS1 MW satellite population given a galaxy--halo connection and beyond-CDM model \citep{Nadler191203303,Nadler200800022},
\begin{equation}
    P(\mathbf{n}_{\mathrm{obs}}\lvert \boldsymbol{\theta}_{\mathrm{g-h}},M_{\mathrm{hm}}) = \prod_{\mathrm{bins\ } i} P(n_{\mathrm{obs},i}|\boldsymbol{\theta}_{\mathrm{g-h}},M_{\mathrm{hm}}),
\end{equation}
where $\mathbf{n}_{\mathrm{obs}}$ is a vector of observed satellite counts in luminosity and surface brightness bins over all realizations. We insert this likelihood into Bayes's theorem to calculate the posterior
\begin{equation}
    P(\boldsymbol{\theta}_{\mathrm{g-h}},M_{\mathrm{hm}}\lvert \mathbf{n}_{\mathrm{obs}}) = \frac{P(\mathbf{n}_{\mathrm{obs}}\lvert \boldsymbol{\theta}_{\mathrm{g-h}},M_{\mathrm{hm}})P(\boldsymbol{\theta}_{\mathrm{g-h}},M_{\mathrm{hm}})}{P(\mathbf{n}_{\mathrm{obs}})},\label{eq:bayes}
\end{equation}
where $P(\boldsymbol{\theta}_{\mathrm{g-h}},M_{\mathrm{hm}})$ is the galaxy--halo connection and beyond-CDM prior. We use the same galaxy--halo connection priors for all parameters as \cite{Nadler200800022}, and we sample $\log (M_{\mathrm{hm}}/M_{\mathrm{\odot}})$ uniformly over the interval $[7,10]$ for both WDM and FDM; this choice is conservative, since the marginalized posterior just begins to plateau at the lower limit of this prior. We sample the posterior by running \textsc{emcee} \citep{emcee} for $\sim 10^6$ steps with $36$ walkers, discarding $\sim 10^5$ burn-in steps, which yields well-converged posteriors with hundreds of independent samples. Marginalizing over $\boldsymbol{\theta}_{\mathrm{g-h}}$ yields $P(M_{\mathrm{hm}})$, which we use to derive our constraints. Crucially, this procedure marginalizes over a wide range of galaxy--halo connection parameter space, which is the dominant theoretical uncertainty in the inference; see \cite{Nadler240110318} for further discussion.

Note that \cite{Nadler200800022} used two zoom-in simulations from the original Symphony Milky Way suite \citep{Mao150302637}. Here we apply our new beyond-CDM SHMF models to this original pipeline to isolate how the constraints are affected by our new SHMF models. Thus, we assume that other subhalo population characteristics such as the radial distribution are unchanged in the beyond-CDM scenarios we study.

\subsection{Warm Dark Matter}
\label{sec:wdm_bound}

We present the full posterior from our WDM inference in Appendix~\ref{sec:full_posteriors}. Following \cite{Nadler200800022}, we multiply all values of $M_{\mathrm{hm}}$ in the posterior by the ratio of the $2\sigma$ upper limit on the MW mass from \cite{Callingham180810456} to the average mass of the two host halos used in the inference (i.e., by $1.8\times 10^{12}~M_{\mathrm{\odot}}/1.4\times 10^{12}~M_{\mathrm{\odot}}\approx 1.29$). From the marginalized $M_{\mathrm{hm}}$ posterior, we find $M_{\mathrm{hm}}<3.9\times 10^7~M_{\mathrm{\odot}}$ at $95\%$ confidence, corresponding to
\begin{equation}
    m_{\mathrm{WDM}}>5.9~\mathrm{keV}\ \mathrm{(95\%\ confidence).}
\end{equation}

This constraint is $0.6~\mathrm{keV}$ (or $\approx 10\%$) weaker than the $6.5~\mathrm{keV}$ limit derived in \cite{Nadler200800022} mainly because of our updated WDM transfer functions and $M_{\mathrm{hm}}(m_{\mathrm{WDM}})$ relation (see Section~\ref{sec:wdm_model}). Our new WDM SHMF model also slightly weakens the limit, since our best-fit $f_{\mathrm{WDM}}$ is marginally less suppressed than the \cite{Lovell13081399} model used in \cite{Nadler200800022}. In addition to using the best-fit WDM SHMF suppression parameters from Equation~\ref{eq:wdm_best_fit}, we tested marginalizing over $\alpha$, $\beta$, and $\gamma$ using our WDM posterior (Figure~\ref{fig:wdm_shmf_posterior}); the uncertainty contributed by these parameters is much smaller than from our galaxy--halo connection, and our $95\%$ confidence constraint is unchanged in this version of the analysis.

We emphasize that our $m_{\mathrm{WDM}}>5.9~\mathrm{keV}$ limit should be interpreted using the latest WDM transfer functions (see Section~\ref{sec:wdm_model} and \citealt{Vogel221010753}) and our new SHMF calibration (see Section~\ref{sec:wdm_shmf}), rather than using previous fits from \cite{Viel0501562} and \cite{Lovell13081399}, respectively. The updated transfer function of our ruled-out $5.9~\mathrm{keV}$ WDM model is nearly identical to the \cite{Viel0501562} transfer function fit for the ruled-out $6.5~\mathrm{keV}$ model from \cite{Nadler200800022}. Thus, bounds on other models derived by mapping to the ruled-out WDM transfer function are negligibly affected by our updated WDM analysis.

We also compute updated $n=0$ IDM bounds following the IDM--WDM half-mode matching procedure from \cite{Nadler190410000,Nadler200800022}, which is very accurate for these models. We derive $\sigma_0 < 1.3\times 10^{-29}$, $4.5\times 10^{-29}$, and $2.5\times 10^{-28}~\mathrm{cm}^2$ for $m_{\mathrm{IDM}}=10^{-4}$, $10^{-2}$, and $1~\mathrm{GeV}$, respectively. These constraints are comparable to the MW satellite limits from \cite{Nadler200800022} and are weaker than the Ly$\alpha$ forest bounds from \cite{Rogers211110386} by a factor of $\approx 2$.

\begin{figure*}[t!]
\includegraphics[width=0.5\textwidth]{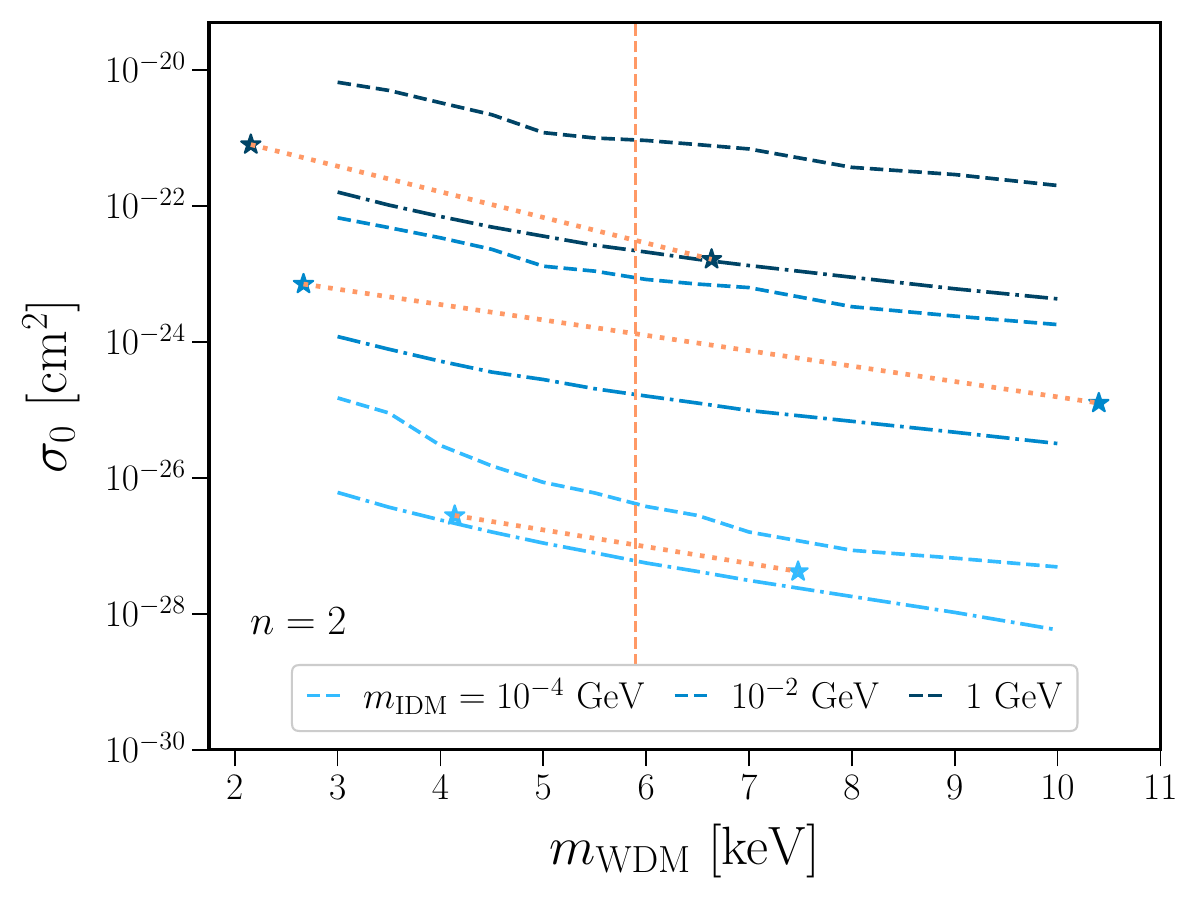}
\includegraphics[width=0.5\textwidth]{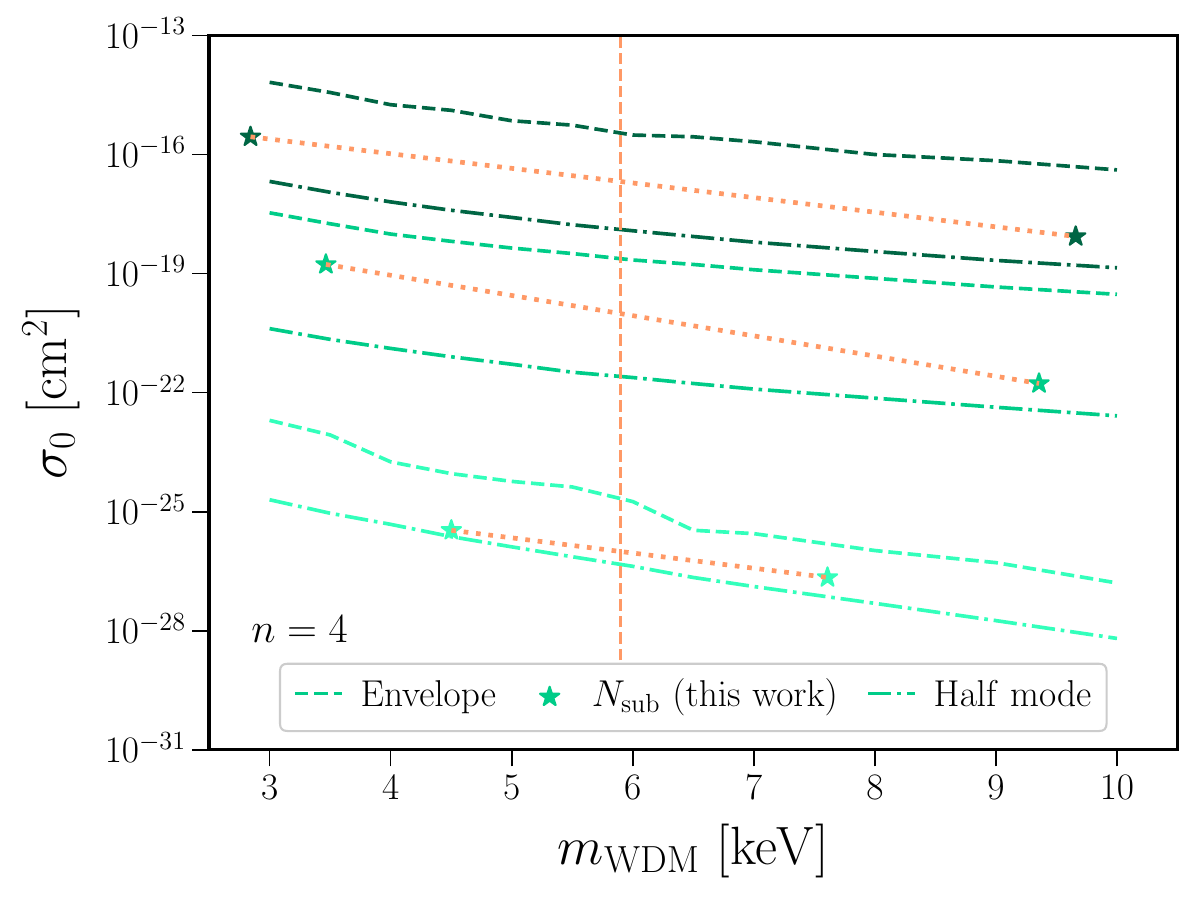}
    \caption{Approximate mapping between the $n=2$ (left) and $n=4$ (right) IDM cross section and $m_{\mathrm{WDM}}$, which we use to estimate updated IDM constraints. In both panels, relations are shown for $m_{\mathrm{IDM}}=10^{-4}$, $10^{-2}$, and $1~\mathrm{GeV}$, from lightest to darkest. For each mass, the upper dashed line shows the cross section that yields a $P(k)$ that is strictly more suppressed than WDM models, as a function of $m_{\mathrm{WDM}}$, and the lower dotted--dashed line shows cross sections that match the half-mode scales of WDM models. Stars show the effective WDM masses derived by matching to the total abundance of subhalos above the resolution limit in our envelope and half-mode IDM simulations (see Section~\ref{sec:idm_shmf}). Vertical orange dashed lines show our new $m_{\mathrm{WDM}}=5.9~\mathrm{keV}$ limit (Section~\ref{sec:wdm_bound}); our estimated IDM constraints are derived by linearly interpolating between the star symbols in $\log(\sigma_0/\mathrm{cm}^2)$ vs.\ $m_{\mathrm{WDM}}$, and finding the intersection between the resulting dotted orange lines and the vertical WDM bound.}
    \label{fig:idm_constraints}
\end{figure*}

\subsection{Fuzzy Dark Matter}
\label{sec:fdm_bound}

We present the full posterior from our FDM inference in Appendix~\ref{sec:full_posteriors}, where all values of $M_{\mathrm{hm}}$ are again scaled by the ratio of the observed versus simulated MW host mass. The marginalized posterior yields $M_{\mathrm{hm}}<4.3\times 10^7~M_{\mathrm{\odot}}$ at $95\%$ confidence, corresponding to
\begin{equation}
    m_{\mathrm{FDM}}>1.4\times 10^{-20}~\mathrm{eV}\ \mathrm{(95\%\ confidence).}
\end{equation}

This constraint is $\approx 5$ times stronger than the $m_{\mathrm{FDM}}>2.9\times 10^{-21}~\mathrm{eV}$ limit derived in \cite{Nadler200800022} because our best-fit FDM SHMF suppression is more suppressed than the \cite{Schive150804621} model used in \cite{Nadler200800022}. We note that our $f_{\mathrm{FDM}}$ model is more similar to (but still more suppressed than) the \cite{Du189706} SHMF fit; \cite{Nadler200800022} derived $m_{\mathrm{FDM}}>9.1\times 10^{-21}~\mathrm{eV}$ using this model, which is closer to (but still a factor of $\approx 1.5$ weaker than) our limit. Note that we directly use \textsc{axionCAMB} transfer functions rather than the \cite{Hu008506} fitting function as in \cite{Nadler200800022}; however, these yield similar $M_{\mathrm{hm}}(m_{\mathrm{FDM}})$ relations, so this difference does not affect our FDM limit. As for WDM, marginalizing over the FDM SHMF fit posterior (Figure~\ref{fig:fdm_shmf_posterior}) does not affect our constraint.

Interestingly, the half-mode scales of our ruled-out WDM and FDM models are similar. In the right panel of Figure~\ref{fig:fdm_shmf_pred}, our simulation with $m_{\mathrm{FDM,22}}=151$ (which has an FDM mass within $10\%$ of our $m_{\mathrm{FDM,22}}$ constraint) reaches an SHMF suppression at $M_{\mathrm{sub,peak}}=10^8~M_{\mathrm{\odot}}$ that is very similar to our ruled-out $m_{\mathrm{WDM}}=5.9~\mathrm{keV}$ model. This follows because the MW satellite constraint is largely driven by suppression near the minimum halo mass of observed satellites, i.e., at $M_{\mathrm{sub,peak}}\approx 10^8~M_{\mathrm{\odot}}$ \citep{Nadler200800022}. 

Our new FDM limit is $\approx 40\%$ weaker than the $2\times 10^{-20}~\mathrm{eV}$ Ly$\alpha$ forest bound from \cite{Rogers200712705}; this is reasonable given the similar WDM sensitivity of these probes. Meanwhile, our bound is a factor of $\approx 20$ weaker than the $3\times 10^{-19}~\mathrm{eV}$ bound from \cite{Dalal220305750} based on stellar heating in the ultrafaint dwarf galaxies Segue 1 and Segue 2, and a factor of $\approx 6$ stronger than the $2.2\times 10^{-21}~\mathrm{eV}$ bound from \cite{Zimmerman240520374} based on the stellar kinematics of Leo II. It will be interesting to explore the complementarity between these limits in future work, since the suppression of the linear matter power spectrum affects the formation histories and abundances of the same ultrafaint dwarf galaxies used to derive these bounds.

\subsection{Interacting Dark Matter}
\label{sec:idm_bound}

As discussed in Section~\ref{sec:idm_shmf}, we have not derived a general model for the IDM SHMF suppression owing to the varying impact of DAOs over the range of IDM masses and cross sections we simulate. Nonetheless, we mapped our half-mode and envelope models to effective WDM models by matching total subhalo abundances between the models. Here we use this mapping to estimate upper bounds on the IDM scattering cross section from the MW satellite population. We assume that the total abundance of subhalos above our resolution limit monotonically decreases with increasing cross section, consistent with our simulation results. We will show that the resulting estimated IDM limits significantly improve on the conservative envelope bounds from \cite{Maamari201002936}. Thus, our results strongly motivate further simulation and modeling work in these scenarios.

Figure~\ref{fig:idm_constraints} illustrates how we estimate IDM bounds for $n=2$ (left panel), $n=4$ (right panel), and for each $m_{\mathrm{IDM}}$. First, for reference, we calculate IDM cross sections with either ($i$) the same half-mode scale as or ($ii$) transfer functions that are strictly more suppressed than WDM models over a grid of $m_{\mathrm{WDM}}$ values. When evaluated at our new $m_{\mathrm{WDM}}=5.9~\mathrm{keV}$ bound, the envelope cross sections (dashed lines) are conservatively excluded because their $P(k)$ are strictly more suppressed than the ruled-out WDM model. Meanwhile, models along the lower end of the band (dotted--dashed lines) have $P(k)$ that match the half-mode scale of our newly constrained WDM transfer function. In Section~\ref{sec:idm_shmf}, we showed that IDM models produce subhalo abundances that are statistically consistent with WDM models with the same half-mode scale, so we cannot exclude them at $95\%$ confidence.

Instead, we linearly interpolate $\log(\sigma_0/\mathrm{cm}^2)$ as a function of $m_{\mathrm{WDM}}$ using the effective WDM masses derived in Section~\ref{sec:idm_shmf} by matching total subhalo abundances to our envelope and half-mode IDM simulation results. This yields a third relation (orange dotted lines) that, by construction, is shifted toward lower $\sigma_0$ than the envelope cross sections, since subhalo abundances in IDM envelope simulations are suppressed compared to the $m_{\mathrm{WDM}}=6.5~\mathrm{keV}$ case. The relation is also shifted toward higher $\sigma_0$ than the half-mode cross sections, since subhalo abundances in half-mode-matched IDM simulations are enhanced compared to the $m_{\mathrm{WDM}}=6.5~\mathrm{keV}$ case. We evaluate this subhalo abundance-matched relation at our new constraint of $m_{\mathrm{WDM}}=5.9~\mathrm{keV}$ to estimate updated IDM limits. For $n=2$, this procedure yields $\sigma_0 < 1.0\times 10^{-27}$, $1.3\times 10^{-24}$, and $3.1\times 10^{-23}~\mathrm{cm}^2$ for $m_{\mathrm{IDM}}=10^{-4}$, $10^{-2}$, and $1~\mathrm{GeV}$, respectively. For $n=4$, we find $\sigma_0 < 9.9\times 10^{-27}$, $9.8\times 10^{-21}$, and $2.1\times 10^{-17}~\mathrm{cm}^2$ for $m_{\mathrm{IDM}}=10^{-4}$, $10^{-2}$, and $1~\mathrm{GeV}$, respectively. 

These estimated limits improve on the envelope bounds from \cite{Maamari201002936} by one order of magnitude, on average, with a maximum improvement for $m_{\mathrm{IDM}}=1~\mathrm{GeV}$ and $n=2$ by a factor of $\approx 20$. To derive more rigorous limits, it is necessary to model the SHMF across the entire IDM parameter space and marginalize over the theoretical uncertainties in this model, including the varying impact of DAOs. This may be achieved by future work that expands and/or emulates our simulation suite.

\section{Discussion}
\label{sec:caveats}

We now discuss our results, focusing on our coverage of beyond-CDM ICs (Section~\ref{sec:coverage}), caveats in our treatment of beyond-CDM physics using $N$-body simulations (Section~\ref{sec:technique}), and areas for future work using the COZMIC simulation pipeline (Section~\ref{sec:futurework}).

\subsection{Coverage of Beyond-CDM Initial Conditions}
\label{sec:coverage}

We have considered three widely studied beyond-CDM scenarios that produce a cutoff in the linear matter power spectrum; specifically, the WDM, FDM, and IDM ICs we simulate span a variety of cutoff shapes, including DAOs (see Figure~\ref{fig:transfers}). Nonetheless, studying an even broader range of ICs is important to facilitate robust $P(k)$ measurements. Here we highlight directions for future work in this regard.

First, various beyond-CDM scenarios imprint $P(k)$ suppression that qualitatively differs from the sharp cutoffs we simulate. For example, $\mathcal{T}^2(k)$ plateaus in scenarios where a subcomponent of DM is a non-CDM species; we simulate such fractional beyond-CDM models in \citetalias{An241103431}. In future work, it will be interesting to directly simulate fractional WDM, IDM, and FDM models, which are theoretically motivated (e.g., \citealt{Arvanitaki09054720,Marsh11024851,Farrar220101334,Moore240303972}) and can potentially alleviate cosmological tensions on larger scales (e.g., \citealt{He230108260,Rogers230108361,Rogers231116377,Rubira220903974}). Meanwhile, models like millicharged DM can suppress power over a range of scales, with a distinctive shape compared to our IDM scenarios (e.g., \citealt{Boddy180800001,Driskell220904499}). The scenarios listed above are not well captured by generalized transfer function parameterizations (e.g., \citealt{Murgia170407838}), so it is important to include them in future beyond-CDM simulation efforts.

Second, many beyond-CDM scenarios produce $P(k)$ \emph{enhancement}, rather than suppression. For example, enhancement can arise from large-misalignment angle axion production (e.g., \citealt{Arvanitaki190911665}), an early period of matter domination (e.g., \citealt{Erickcek11060536}), and inflationary fluctuations that produce vector DM (e.g., \citealt{Graham150402102}), among many other possibilities (see, e.g., \citealt{Amin221109775}). Simulating such models can be challenging, particularly if there is appreciable power on scales that are not resolved. Thus, careful convergence testing and reassessment of issues like artificial disruption will be needed to robustly model subhalo populations in such scenarios.

Finally, we reiterate that we have only considered beyond-CDM physics insofar as it modifies the linear matter power spectrum. Small-scale structure can also be affected by a wide variety of late-time DM physics (see, e.g., \citealt{Bechtol220307354} for an overview). We consider one such scenario in \citetalias{Nadler241213065}, which self-consistently combines $P(k)$ suppression with strong, velocity-dependent SIDM.

\subsection{Treatment of Beyond-CDM Physics}
\label{sec:technique}

We have used standard codes for generating cosmological zoom-in ICs and running $N$-body simulations. While our simulations robustly map linear matter power spectra to DM-only subhalo populations, it is important to discuss how additional physics in each beyond-CDM scenario we consider might affect our simulations, beyond what is captured by $P(k)$ and our $N$-body simulation technique alone.

Beginning with WDM, we have shown that the $N$-body simulations yield well-resolved subhalo populations that are negligibly affected by artificial fragmentation. Nonetheless, several studies have simulated WDM (sub)halo populations using alternative simulation codes that better resolve halos near the free streaming scale and thus avoid this issue more directly. For example, \cite{Angulo13042406} simulated an $m_{\mathrm{WDM}}=0.25~\mathrm{keV}$ model using the phase-space sheet method, which mitigates artificial fragmentation and allows their analysis to probe halos well below $M_{\mathrm{hm}}$. That work demonstrates that halo finding is challenging and ambiguous in this regime. When restricting analyses to reliably identified halos, \cite{Angulo13042406} find a halo mass function suppression that is broadly consistent with extended Press--Schechter (ePS) predictions from \cite{Schneider11120330} and \cite{Benson12093018}, which are also consistent with our results. Meanwhile, \cite{Stucker210909760} use a hybrid phase-space sheet plus $N$-body solver, which inherits the benefits of the phase-space sheet method (before shell crossing) and of the traditional $N$-body solvers (after shell crossing). In general, while phase-space WDM simulations more accurately resolve halos that form from highly suppressed scales, our $N$-body approach is appropriate for the subhalo masses and WDM models we consider. Extending our framework to lower-mass subhalos will require reassessing the impact of potential spurious subhalos and subhalo finder issues.

Another consideration in WDM is the thermal velocity distribution of particles in the ICs. We use the standard version of \textsc{MUSIC} to generate ICs and thus do not include this effect. Thermal velocities are expected to be fairly small when our simulations are initialized. For example, following \cite{Bode0010389}, we obtain an rms thermal velocity of $\mathcal{O}(1)~\mathrm{km\ s}^{-1}$ at $z=99$ in the $m_{\mathrm{WDM}}=3~\mathrm{keV}$ model, much smaller than the virial velocities of the subhalos we resolve. Nonetheless, \cite{Benson12093018} show that ePS predictions for the WDM halo mass function are systematically more suppressed below $M_{\mathrm{hm}}$ when thermal velocities are included in the semianalytic model. Since we mainly analyze halos with $M_{\mathrm{sub,peak}}\gtrsim M_{\mathrm{hm}}$ owing to our resolution threshold and choice of WDM models, we do not expect thermal velocities to significantly affect our $f_{\mathrm{WDM}}$ measurements. However, higher-resolution simulations that resolve lower-mass (sub)halos may require a revision to this assumption.

Next, for FDM, we have performed $N$-body simulations that capture $P(k)$ suppression, but do not model the wave dynamics of ultralight particles. Various cosmological simulation codes have been developed for the purpose of capturing these additional effects, including SPH solvers of the Madelung equations (e.g., \citealt{Nori180108144,Elgamal231103591,Nori221008022}) and pseudospectral solvers of the Schr{\"o}dinger--Poisson equation (e.g., \citealt{Schive14066586,Schwabe200708256,May210101828,May220914886,Schwabe211009145}). These studies have not found significant differences between the (sub)halo mass function with and without the implementation of FDM dynamics. For example, \cite{May220914886} show that halo mass function suppression primarily depends on FDM ICs, rather than the wave dynamics: their results are in agreement with the halo mass function fit from \cite{Schive150804621}, which was derived using FDM ICs but $N$-body dynamics. However, semianalytic models like \cite{Du189706} find differences between the stripping of subhalo profiles, depending on the presence or absence of solitonic cores. For the relatively large values of $m_{\mathrm{FDM,22}}$ we consider, the impact of FDM wave dynamics on subhalos abundances above our resolution limit is expected to be small. However, this assumption warrants a dedicated check in future work.

Lastly, for IDM, we expect our $N$-body simulations to accurately capture subhalo abundances because the DM--baryon scattering rate drops rapidly with decreasing redshift in the models we consider. Indeed, kinetic and thermal decoupling between the DM and baryon fluids typically occurs at $z\gtrsim 10^5$, well before we initialize our simulations (e.g., \citealt{Nadler190410000,Maamari201002936}). Nonetheless, DM--baryon scattering may be enhanced in high-density environments at late times, leading to a variety of novel signatures in galaxy and (sub)halo density profiles (e.g., \citealt{Acevedo230908661}; see also \citealt{Heston240208718} for an example in the context of DM--neutrino scattering). Thus, it will be interesting to explore the potential impact of IDM scattering on subhalo populations as these modeling efforts develop.

\subsection{Future Work}
\label{sec:futurework}

We anticipate the following key areas for future work related to COZMIC. First, we have only simulated three independent MW systems: two MW analogs from the Milky Way-est suite \citep{Buch240408043}, and one MW-mass host from the Symphony Milky Way suite \citep{Nadler220902675}. For WDM and FDM, our derived SHMF suppression parameters are nearly identical across these three hosts (see Appendix~\ref{sec:shmf_details}). It will be important to understand this universality in future work. In addition, expanding COZMIC to include a larger sample of MW analogs would enable more robust forward modeling of the MW subhalo and satellite population. For example, we currently marginalize over the MW host halo mass using a conservative analytic approach to derive constraints. However, including more MW realizations in the inference would achieve this marginalization directly. Simulating a larger host sample would also allow us to quantify the potential environmental dependence of SHMF suppression, which we plan to explore in future work.

Second, our analysis is limited to subhalos at $z=0$. The zoom-in region surrounding each host extends several $\mathrm{Mpc}$ from the MW's center, and a large number of (sub)halos are resolved with high-resolution particles in this region (see Appendix~\ref{sec:simulation_table}). To fully exploit our simulations, contamination and convergence studies of (sub)halos throughout the zoom-in regions should be conducted. Meanwhile, many halos are well resolved in our simulations as early as $z\approx 10$. Thus, our simulations can be used to quantify the redshift dependence of (sub)halo mass function suppression in beyond-CDM scenarios, which we leave to a future study.

Third, we have used the standard halo finder and merger tree codes \textsc{Rockstar} and \textsc{consistent-trees} to post-process our results (see Section~\ref{sec:post-process}). We expect that implementing particle-tracking codes like \textsc{Symfind} \citep{Mansfield230810926} will yield even more robust subhalo catalogs, particularly for highly stripped objects, and we plan to apply such tools to COZMIC in future work. More generally, it will be important to revisit halo finder and merger tree code comparisons (e.g., \citealt{Onions12033695,Srisawat13073577}) in beyond-CDM scenarios, since most post-processing codes are calibrated to CDM simulations. For example, the (artificial) disruption of subhalos may differ for objects that form from suppressed linear matter power spectra owing to their delayed growth and reduced concentrations. This is an important issue for future work, and particularly for analyses of subhalo density profiles in our beyond-CDM simulations.

Fourth, we have quantified SHMF suppression using DM-only simulations. Baryons are expected to reduce the SHMF in MW-mass hosts relative to CDM via enhanced tidal stripping by the galactic disk (e.g., \citealt{Garrison-Kimmel170103792,Kelley181112413,Webb200606695,Wang240801487}). This effect may be enhanced owing to the reduced concentrations of subhalos in our beyond-CDM scenarios \citep{Du240309597}. Subhalos and halos with masses $\gtrsim 10^{10}~M_{\mathrm{\odot}}$, which host galaxies with stellar masses $\gtrsim 10^{6}~M_{\mathrm{\odot}}$ \citep{Nadler191203303}, can also develop cores owing to repeated episodes of supernova feedback (e.g., \citealt{Pontzen11060499,Dutton201111351}). Quantifying the correlated impact of baryons and beyond-CDM physics on subhalo populations is therefore an exciting area for future work. This could be achieved by embedding disk potentials in our beyond-CDM simulations, or by performing full hydrodynamic resimulations of our hosts (e.g., similar to \citealt{Forouhar220807376}).

Finally, by combining the simulations presented here with simulations of additional models in upcoming COZMIC papers, we aim to develop a general (sub)halo mass function model applicable to a broad range of beyond-CDM ICs. For the beyond-CDM scenarios in this paper, one key question is whether our SHMF fitting functions can be extended below the resolution limit, and/or to $M_{\mathrm{hm}}/M_{\mathrm{sub,peak}}$ ratios outside those covered by our simulations, or if they are only accurate in the range of beyond-CDM parameters and subhalo masses covered here. We plan to test this using semianalytic models and additional simulations. A general SHMF model will ultimately help enable model-independent $P(k)$ reconstruction using small-scale structure data.


\section{Conclusions}
\label{sec:conclusions}

As the first installment of the COZMIC simulation suite, we have presented $72$ cosmological DM-only zoom-in simulations of two MW-like hosts and one MW-mass host in three beyond-CDM scenarios that suppress the linear matter power spectrum: thermal-relic WDM, ultralight FDM, and DM--baryon scattering IDM (see Figure~\ref{fig:transfers}). To isolate the impact of ICs, we have assumed that these scenarios only impact $P(k)$ and subsequently evolve like CDM. Our work substantially expands the available sample of zoom-in simulations in beyond-CDM cosmologies and reveals that beyond-CDM SHMF suppression is fairly universal. Two companion studies will expand COZMIC: \citetalias{An241103431} presents beyond-CDM simulations with a plateau in the transfer function, characteristic of fractional beyond-CDM models, and \citetalias{Nadler241213065} presents simulations of SIDM models with self-consistently modeled $P(k)$ suppression.

Our main findings are as follows:
\begin{enumerate}
    \item The SHMF is significantly suppressed in many of our beyond-CDM simulations. The amount of suppression depends on both the wavenumber and shape of the $P(k)$ cutoff and scales differently with microphysical DM parameters in the WDM, FDM, and IDM scenarios we consider (Figure~\ref{fig:shmf}).
    \item We present a new model and likelihood framework to capture SHMF suppression as a function of the half-mode mass that characterizes the $P(k)$ cutoff; we fit this model to our WDM and FDM simulation data, with results summarized in Section~\ref{sec:shmf_summary}.
    \item These fits confirm that FDM SHMF suppression is significantly steeper than in WDM owing to the sharper $P(k)$ cutoff in this scenario (see Figure~\ref{fig:fdm_shmf_pred}).
    \item IDM SHMF suppression can be reduced due to DAOs in $P(k)$; the size of this effect varies over the range of IDM cross sections, masses, and velocity dependences we simulate (Figure~\ref{fig:transfers_matched}).
    \item We derive $m_{\mathrm{WDM}}>5.9~\mathrm{keV}$ at $95\%$ confidence from the MW satellite population using our new WDM SHMF suppression model; our result is $\approx 10\%$ weaker than a previous bound using the same framework owing to our updated $M_{\mathrm{hm}}(m_{\mathrm{WDM}})$ relation (Equation~\ref{eq:Mhm_mwdm}).
    \item We derive $m_{\mathrm{FDM}}>1.4\times 10^{-20}$ at $95\%$ confidence from the MW satellite population using our new FDM SHMF suppression model; this constraint is $\approx 5$ times stronger than a previous limit using the same framework because our FDM SHMFs are more suppressed than previous work.
    \item We estimate updated MW satellite population bounds on the DM--proton scattering cross section, with an $n=2$ ($n=4$) power-law velocity dependence, of $\sigma_0 < 1.0\times 10^{-27}$, $1.3\times 10^{-24}$, and $3.1\times 10^{-23}~\mathrm{cm}^2$ ($\sigma_0 < 9.9\times 10^{-27}$, $9.8\times 10^{-21}$, and $2.1\times 10^{-17}~\mathrm{cm}^2$) for IDM masses of $10^{-4}$, $10^{-2}$, and $1~\mathrm{GeV}$, respectively. These estimates are derived using an approximate mapping to total WDM subhalo abundances. The results motivate a future study leveraging a larger set of IDM simulations and the development of IDM SHMF models, which have the potential to improve IDM cross section bounds by up to a factor of $\sim 20$ with current data.
\end{enumerate}

Upcoming galaxy surveys will critically test our beyond-CDM predictions. For example, the Rubin Observatory Legacy Survey of Space and Time (\citealt{Ivezic08052366}) is expected to discover most of the remaining MW satellite population in the southern hemisphere (e.g., \citealt{Drlica-Wagner190201055,Tsiane240416203}). Combining these data with our beyond-CDM SHMF predictions and galaxy--halo connection framework will simultaneously advance our understanding of galaxy formation and DM physics; for example, see the forecasts in \cite{Nadler240110318}.

In parallel, upcoming data from the Nancy Grace Roman Space Telescope \citep{Spergel150303757,Gezari220212311} and other upcoming observational facilities \citep{Chakrabarti220306200} will provide complementary information about small-scale structure, both within the MW (e.g., by characterizing its stellar stream population) and beyond (e.g., by discovering new strong gravitational lens systems). We expect COZMIC to inform modeling efforts for all of these science targets. These same facilities will advance measurements of small-scale structure using strong gravitational lensing and stellar streams, which probe the beyond-CDM models we consider through their effects on both (sub)halo abundances and density profiles (see, e.g., \citealt{Vegetti230611781}; \citealt{Bonaca240519410} for reviews). It is therefore timely to expand our suite to higher-mass hosts relevant for strong lensing (e.g., using the Symphony Group suite from \citealt{Nadler220902675}) and to use it as a template for modeling the low-mass subhalos in the inner MW that perturb stellar streams.

Looking ahead, the development of small-scale structure emulators in beyond-CDM scenarios is well motivated, following recent efforts using hydrodynamic simulations of MW-mass systems (e.g., \citealt{Brown240311694,Rose240500766}). The advantage of our $N$-body approach is its relatively low computational cost, which allows us to simulate a wide range of beyond-CDM models. COZMIC thus facilitates predictions for the MW subhalo and satellite galaxy population across a wide range of possible linear matter power spectra. This effort will ultimately enable robust, model-independent reconstruction of the small-scale linear matter power spectrum using near-field cosmological observables.



\section*{Acknowledgements}

Halo catalogs, merger trees, and particle snapshots are distributed in Zenodo at doi: 10.5281/zenodo.14649137. Analysis code is available at \url{https://github.com/eonadler/COZMIC/}.

We are grateful to Deveshi Buch, Alex Drlica-Wagner, Yao-Yuan Mao, and Hai-Bo Yu for comments on the manuscript. We thank Kev Abazajian, Tom Abel, Sownak Bose, Francis-Yan Cyr-Racine, Cannon Vogel, and Risa Wechsler for helpful discussions. V.G.\ acknowledges the support from NASA through the Astrophysics Theory Program, award No.\ 21-ATP21-0135, the National Science Foundation (NSF) CAREER grant No. PHY2239205, and the Research Corporation for Science Advancement under the Cottrell Scholar Program. This research was supported in part by grant NSF PHY-2309135 to the Kavli Institute for Theoretical Physics (KITP).

The computations presented here were conducted through Carnegie's partnership in the Resnick High Performance Computing Center, a facility supported by Resnick Sustainability Institute at the California Institute of Technology. This
work used data from the Symphony and Milky Way-est suites of simulations, hosted at \url{https://web.stanford.edu/group/gfc/gfcsims/}, which were supported by the Kavli Institute for Particle Astrophysics and Cosmology at Stanford University, SLAC National Accelerator Laboratory, and the US Department of Energy under contract No.\ DE-AC02-76SF00515 to SLAC National Accelerator Laboratory.

\software{
\textsc{ChainConsumer} \citep{ChainConsumer},
{\sc consistent-trees} \citep{Behroozi11104370},
\textsc{emcee} \citep{emcee},
\textsc{h5py} (\http{www.h5py.org}),
\textsc{Healpy} (\http{healpy.readthedocs.io}),
\textsc{Helpers}, (\http{bitbucket.org/yymao/helpers/src/master/}),
\textsc{Pandas} \citep{pandas}, 
\textsc{Jupyter} (\http{jupyter.org}),
\textsc{Matplotlib} \citep{matplotlib},
\textsc{NumPy} \citep{numpy},
\textsc{pynbody} \citep{pynbody},
{\sc Rockstar} \citep{Behroozi11104372},
\textsc{Scikit-Learn} \citep{scikit-learn},
\textsc{SciPy} \citep{scipy},
\textsc{Seaborn} (\https{seaborn.pydata.org}).
}

\bibliographystyle{yahapj2}
\bibliography{references,software}


\appendix

\section{Transfer Function Calculations}
\label{sec:tk_details}

\subsection{Warm Dark Matter}

In our WDM \textsc{CLASS} runs, we consider a thermal relic particle with a Fermi--Dirac phase-space distribution characterized by the WDM-to-photon temperature ratio \citep{Colombi9505029}
\begin{equation}
    \frac{T_\mathrm{WDM}}{T_\gamma} = \Omega_\mathrm{WDM}h^2\left(\frac{4}{11}\right) ^{1/3}\left(\frac{m_\mathrm{WDM}}{94\ \mathrm{eV}}\right)^{-1},
\end{equation}
where $\Omega_\mathrm{WDM}$ is the present-day fractional WDM density, which corresponds to the entire DM density in our WDM simulations, i.e., $\Omega_{\mathrm{WDM}}=\Omega_{\rm m}$. We have compared the resulting transfer functions to the \cite{Decant211109321} and \cite{Vogel221010753} fitting functions for $\mathcal{T}^2_\mathrm{WDM}(k)$, finding that they both match our \textsc{CLASS} output very well.

Rather than using a fitting function for $\mathcal{T}^2_\mathrm{WDM}(k)$ to generate ICs, as done in many previous WDM simulations, we directly use \textsc{CLASS} transfer functions as input for \textsc{MUSIC} to ensure accurate ICs. The current version of \textsc{MUSIC} only features a \textsc{CAMB} transfer function plug-in. Thus, we convert \textsc{CLASS} density transfer functions into \textsc{CAMB} format by setting \texttt{format = camb}; we also convert \textsc{CLASS} Newtonian-gauge velocity transfer functions, $\theta$, to velocity transfer functions in \textsc{CAMB} format, $\mathcal{T}_{v}$, via
\begin{equation}
    \mathcal{T}_{v,i} = \theta_{i}/(\tilde{H}\tilde{k}^2).
\end{equation}
Here $i$ represents the particle species (DM or baryons), $\tilde{H}$ is the Hubble parameter in units of $\mathrm{Mpc}^{-1}$ at the initial redshift ($z=99$), and $\tilde{k}$ is the comoving wavenumber in units of $\mathrm{Mpc}^{-1}$.

\subsection{Fuzzy Dark Matter}

We make the following minor modifications to version v1.0 of \textsc{axionCAMB}.\footnote{Note that \textsc{axionCAMB} v2.0 includes several bug fixes. However, the resulting transfer functions only differ from v1.0 for $k\gg k_{\mathrm{hm}}$, so these differences do not affect our results.} By default, the velocity transfer function of FDM is not output by \textsc{axionCAMB}; we therefore modify the code to output this transfer function in the same format as for CDM. Furthermore, for $k\gg k_J$, the linear FDM density perturbation $\delta(k)$ oscillates with time instead of growing as in CDM, which can lead to negative values in the transfer functions. To avoid numerical errors, we therefore take the absolute values of these transfer functions before implementing them in \textsc{MUSIC}. This procedure does not affect our results, as the negative values only occur on extremely small scales. A patch with our changes to \textsc{axionCAMB} is available at \url{https://github.com/Xiaolong-Du/axionCAMB_patch}.

\subsection{Interacting Dark Matter}

To generate IDM transfer functions, we input $m_{\mathrm{IDM}}$, $\sigma_0$, and $n$ into our modified \textsc{CLASS} code. We then convert the transfer functions to \textsc{CAMB} format using the method described above for WDM. Note that the resulting half-mode and envelope IDM cross sections are slightly different than derived in \cite{Maamari201002936} because we match to an updated $m_{\mathrm{WDM}}=6.5~\mathrm{keV}$ transfer function.

\section{Additional Simulation Results}
\label{sec:simulation_table}

Here we present several relevant simulation results beyond the SHMFs. Table~\ref{tab:sims} summarizes the main properties of the host halo, subhalo population, and isolated halo population in each of our fiducial-resolution beyond-CDM simulations.

\begin{deluxetable*}{{ccccc}}[t!]
\centering
\tablecolumns{5}
\tablecaption{Properties of Our Standard Resolution Beyond-CDM Simulations}
\tablehead{\colhead{Scenario} & \colhead{Input Parameter(s)} & \colhead{Halo004} & \colhead{Halo113} & \colhead{Halo023}}
\startdata 
\hline \hline
\phantom{.} & 
-- &
$M_{\mathrm{host,12}},\ c_{\mathrm{host}},\ N_{\mathrm{sub}},\ N_{\mathrm{iso}}$ &
$M_{\mathrm{host,12}},\ c_{\mathrm{host}},\ N_{\mathrm{sub}},\ N_{\mathrm{iso}}$ &
$M_{\mathrm{host,12}},\ c_{\mathrm{host}},\ N_{\mathrm{sub}},\ N_{\mathrm{iso}}$ \\ 
\hline
CDM & 
-- & 
1.03, 10.8, 93, 630 & 
1.07, 10.6, 77, 882 &
1.14, 11.5, 69, 761
\\
\hline \hline
\phantom{.} & 
$m_{\mathrm{WDM}}~[\mathrm{keV}]$ &
$M_{\mathrm{host,12}},\ c_{\mathrm{host}},\ N_{\mathrm{sub}},\ N_{\mathrm{iso}}$ &
$M_{\mathrm{host,12}},\ c_{\mathrm{host}},\ N_{\mathrm{sub}},\ N_{\mathrm{iso}}$ &
$M_{\mathrm{host,12}},\ c_{\mathrm{host}},\ N_{\mathrm{sub}},\ N_{\mathrm{iso}}$ \\ 
\hline
\phantom{.} & 
3 & 
1.03, 10.6, 40, 224 & 
1.07, 10.6, 33, 334 &
1.17, 10.7, 38, 265
\\
\phantom{.} & 
4 & 
1.04, 10.4, 57, 346 & 
1.06, 10.4, 44, 488 & 
1.11, 12.5, 45, 420
\\
Thermal-relic & 
5 & 
1.04, 10.4, 66, 431 & 
1.06, 10.8, 58, 594 &
1.12, 12.4, 57, 519
\\
WDM & 
6 & 
1.03, 10.7, 79, 490 & 
1.06, 10.9, 66, 674 & 
1.11, 12.2, 64, 596
\\
\phantom{.} & 
6.5 & 
1.03, 10.4, 79, 518 & 
1.06, 10.9, 65, 709 & 
1.14, 11.7, 59, 629
\\
\phantom{.} & 
10 & 
1.03, 10.3, 90, 583 &
1.06, 11.0, 76, 814 &
1.14, 11.4, 61, 714
\\
\hline \hline
\phantom{.} & 
$m_{\mathrm{FDM}}~[10^{-22}~\mathrm{eV}]$ &
$M_{\mathrm{host,12}},\ c_{\mathrm{host}},\ N_{\mathrm{sub}},\ N_{\mathrm{iso}}$ &
$M_{\mathrm{host,12}},\ c_{\mathrm{host}},\ N_{\mathrm{sub}},\ N_{\mathrm{iso}}$ &
$M_{\mathrm{host,12}},\ c_{\mathrm{host}},\ N_{\mathrm{sub}},\ N_{\mathrm{iso}}$
\phantom{.} \\ 
\hline
\phantom{.} & 
25.9 & 
1.04, 12.4, 36, 213 & 
1.07, 10.4, 31, 308 &
1.16, 11.7, 37, 257
\\
\phantom{.} & 
69.4 & 
1.04, 11.1, 68, 419 & 
1.08, 9.8, 60, 560 & 
1.12, 12.2, 54, 495
\\
Ultralight & 
113 & 
1.04, 10.2, 82, 506 & 
1.08, 10.4, 72, 687 &
1.11, 12.5, 64, 617
\\
FDM & 
151 & 
1.04, 10.1, 89, 545 & 
1.07, 10.6, 74, 748 & 
1.11, 12.2, 69, 664
\\
\phantom{.} & 
185 & 
1.03, 10.1, 95, 576 & 
1.07, 10.7, 76, 792 & 
1.14, 11.6, 71, 698
\\
\phantom{.} & 
490 & 
1.03, 10.6, 97, 612 &
1.07, 10.7, 84, 852 &
1.13, 11.7, 71, 756
\\
\hline \hline
\phantom{.} & 
$m_{\mathrm{IDM}}~[\mathrm{GeV}]$, $\sigma_0~[\mathrm{cm}^2]$ &
$M_{\mathrm{host,12}},\ c_{\mathrm{host}},\ N_{\mathrm{sub}},\ N_{\mathrm{iso}}$ &
$M_{\mathrm{host,12}},\ c_{\mathrm{host}},\ N_{\mathrm{sub}},\ N_{\mathrm{iso}}$ &
$M_{\mathrm{host,12}},\ c_{\mathrm{host}},\ N_{\mathrm{sub}},\ N_{\mathrm{iso}}$
\phantom{.} \\ 
\hline
\phantom{.} & 
$10^{-4}$, $4.2\times 10^{-28}$ & 
1.03, 10.3, 84, 544 & 
1.07, 10.7, 74, 744 &
1.14, 11.7, 60, 659
\\
\phantom{.} & 
$10^{-4}$, $2.8\times 10^{-27}$ & 
1.03, 11.1, 60, 348 & 
1.07, 10.1, 52, 486 & 
1.16, 11.3, 44, 414
\\
DM--proton scattering & 
$10^{-2}$, $1.3\times 10^{-25}$ & 
1.03, 10.2, 86, 549 & 
1.07, 10.5, 77, 749 &
1.14, 11.7, 68, 678
\\
IDM ($n=2$) & 
$10^{-2}$, $7.1\times 10^{-24}$ & 
1.03, 10.8, 37, 238 & 
1.06, 11.1, 19, 345 & 
1.14, 11.9, 34, 269
\\
\phantom{.} & 
$1$, $1.6\times 10^{-23}$ & 
1.03, 9.9, 84, 538 & 
1.07, 10.8, 69, 737 & 
1.14, 11.7, 57, 658
\\
\phantom{.} & 
$1$, $8.0\times 10^{-22}$ & 
1.03, 9.8, 27, 168 &
1.06, 11.0, 12, 262 &
1.13, 11.2, 24, 216
\\
\hline \hline
\phantom{.} & 
$m_{\mathrm{IDM}}~[\mathrm{GeV}]$, $\sigma_0~[\mathrm{cm}^2]$ &
$M_{\mathrm{host,12}},\ c_{\mathrm{host}},\ N_{\mathrm{sub}},\ N_{\mathrm{iso}}$ &
$M_{\mathrm{host,12}},\ c_{\mathrm{host}},\ N_{\mathrm{sub}},\ N_{\mathrm{iso}}$ &
$M_{\mathrm{host,12}},\ c_{\mathrm{host}},\ N_{\mathrm{sub}},\ N_{\mathrm{iso}}$
\phantom{.} \\ 
\hline
\phantom{.} & 
$10^{-4}$, $2.2\times 10^{-27}$ & 
1.03, 10.3, 82, 536 & 
1.07, 10.9, 74, 735&
1.14, 11.5, 63, 642
\\
\phantom{.} & 
$10^{-4}$, $3.4\times 10^{-26}$ & 
1.04, 10.9, 63, 376 & 
1.08, 10.2, 59, 516 & 
1.14, 11.5, 46, 462
\\
DM--proton scattering & 
$10^{-2}$, $1.7\times 10^{-22}$ & 
1.03, 9.8, 85, 564 & 
1.08, 10.5, 80, 771 &
1.13, 11.6, 63, 695
\\
IDM ($n=4$) & 
$10^{-2}$, $1.7\times 10^{-19}$ & 
1.02, 12.7, 53, 373 & 
1.06, 9.8, 32, 490 & 
1.15, 11.1, 44, 418
\\
\phantom{.} & 
$1$, $8.6\times 10^{-19}$ & 
1.03, 10.1, 83, 557 & 
1.08, 10.6, 80, 758 & 
1.14, 11.5, 66, 680
\\
\phantom{.} & 
$1$, $2.8\times 10^{-16}$ & 
1.04, 12.2, 42, 238 &
1.07, 10.3, 22, 357 &
1.14, 11.7, 35, 308
\\
\hline \hline
\enddata
{\footnotesize \tablecomments{The first column lists the name of each DM scenario we consider, the second column lists input parameter(s) used to generate ICs for our simulations, and the third, fourth, and fifth columns respectively list simulation results for our MW-like hosts, Halo004 and Halo113, and our MW-mass host, Halo023. We list the host halo virial mass, $M_{\mathrm{host,12}}\equiv M_{\mathrm{host}}/10^{12}~M_{\mathrm{\odot}}$, and virial concentration, $c_{\mathrm{host}}\equiv R_{\mathrm{vir,host}}/R_{s\mathrm{,host}}$ (where $R_s$ is the NFW scale radius), along with the total number of subhalos within the MW host's virial radius ($N_{\mathrm{sub}}$) and isolated halos within $3~\mathrm{Mpc}$ of the MW host's center ($N_{\mathrm{iso}}$) above our fiducial-resolution threshold of $M_{\mathrm{sub}}>1.2\times 10^8~M_{\mathrm{\odot}}$. For each IDM mass, the first (second) row corresponds to our half-mode (envelope) models.}}
\label{tab:sims}
\end{deluxetable*}

\subsection{Host Halo Mass Accretion Histories}
\label{sec:basic_mah}

\begin{figure*}[t!]
\centering
    \includegraphics[width=\textwidth]{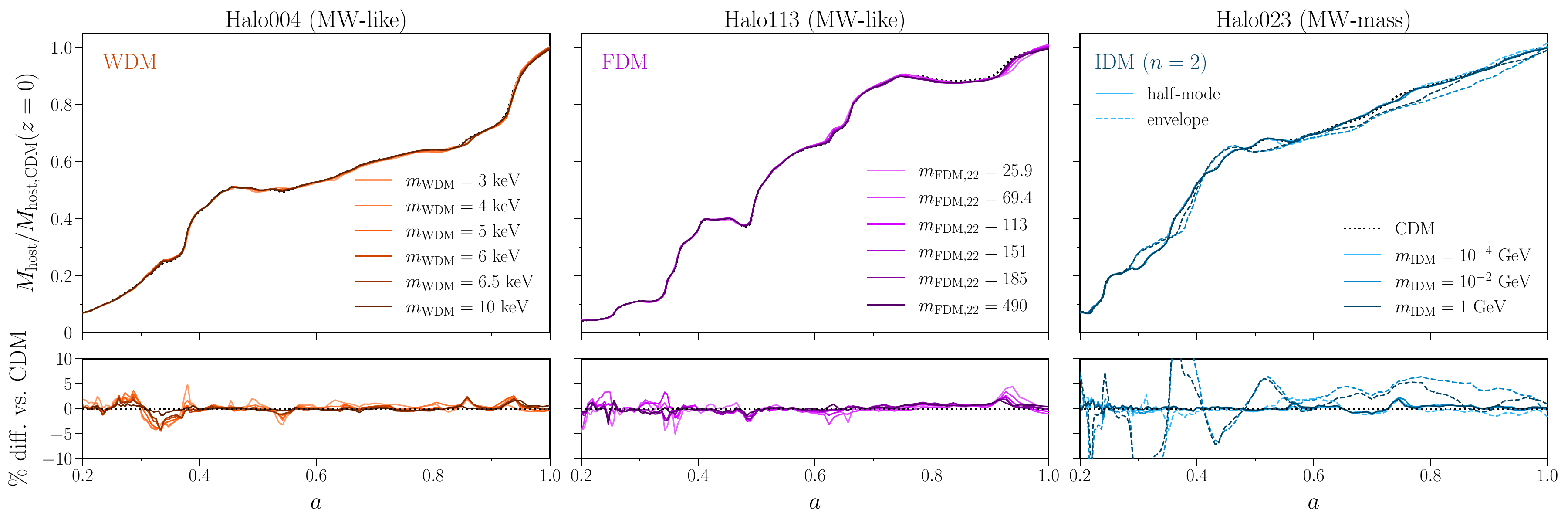}
    \caption{Host halo mass accretion histories for our MW-like hosts (left and middle panels) and MW-mass host (right panel) are shown as a function of the scale factor. In each panel, black dotted lines represent CDM, while the colored lines represent beyond-CDM models; each curve is normalized to the host mass in CDM at $z=0$. Bottom panels show the fractional differences. We note that host halo masses in beyond-CDM cosmologies match the CDM case at the percent level, at $z=0$. Note that we only show a subset of our IDM simulations here; the IDM results for $n=4$ are qualitatively similar to the $n=2$ case.}
    \label{fig:mah}
\end{figure*}

Figure~\ref{fig:mah} shows host halo mass accretion histories, normalized to each host's mass in CDM at $z=0$, for a subset of our beyond-CDM simulations. Mass accretion histories vary from host to host; in particular, the two MW-like hosts (Halo004 and Halo113) are selected to undergo an early merger with an analog of the GSE system and a recent merger with an LMC-like system. The GSE and LMC-analog mergers are reflected in these hosts' mass accretion histories at $a\approx 0.4$ and $a\approx 1$, respectively; for details, see \cite{Buch240408043}. Meanwhile, the MW-mass host (Halo023) experiences an early phase of rapid growth that ends at $a\approx 0.4$, typical of MW-mass halos' average formation histories (e.g., \citealt{Wechsler0108151,Lu0508624,Nadler220902675}).

Overall, hosts in beyond-CDM models undergo very similar mass growth relative to their CDM counterparts. Specifically, all hosts' $z=0$ masses match their CDM counterparts at the percent level. WDM, FDM, and half-mode IDM hosts' mass accretion histories further match CDM to within $\approx 5\%$ at all redshifts where they are well resolved, with the only differences noticeable at very early times. Meanwhile, in our envelope IDM models, mass accretion histories differ at the $\approx 10\%$ level from their CDM counterparts at early times for $m_{\mathrm{IDM}}=10^{-2}~\mathrm{GeV}$ and $m_{\mathrm{IDM}}=1~\mathrm{GeV}$. Nonetheless, these hosts' masses eventually converge to their CDM counterparts. This similarity is expected given that $P(k)$ is nearly identical to CDM on the scales corresponding to our MW host in all models we simulate.

We also measure virial concentrations of host halos $c_{\mathrm{vir}}\equiv R_{\mathrm{vir}}/R_s$, where $R_s$ is the Navarro--Frenk--White (NFW; \citealt{Navarro1997}) scale radius. We find that $z=0$ concentrations in most beyond-CDM simulations match those in CDM at the percent level. The most notable exceptions are (i) the most suppressed FDM model we consider (with $m_{\mathrm{FDM,22}}=25.9$), and (ii) the IDM envelope scenario, for which $c_{\mathrm{vir}}$ is increased by $\gtrsim 10\%$ relative to CDM. Certain FDM and WDM cases with similar half-mode scales (e.g., $m_{\mathrm{FDM,22}}=25.9$ and $m_{\mathrm{WDM}}=3~\mathrm{keV}$) yield different values of $c_{\mathrm{vir}}$, which could indicate that host concentration is sensitive to the shape of the transfer function; this would agree with previous findings using cosmological simulations (e.g., \citealt{Brown200512933}). We leave a study of halo and subhalo density profiles in our beyond-CDM simulations to future work.

\subsection{Subhalo Radial Distributions}
\label{sec:basic_rad}

\begin{figure*}[t!]
\centering
\includegraphics[width=\textwidth]{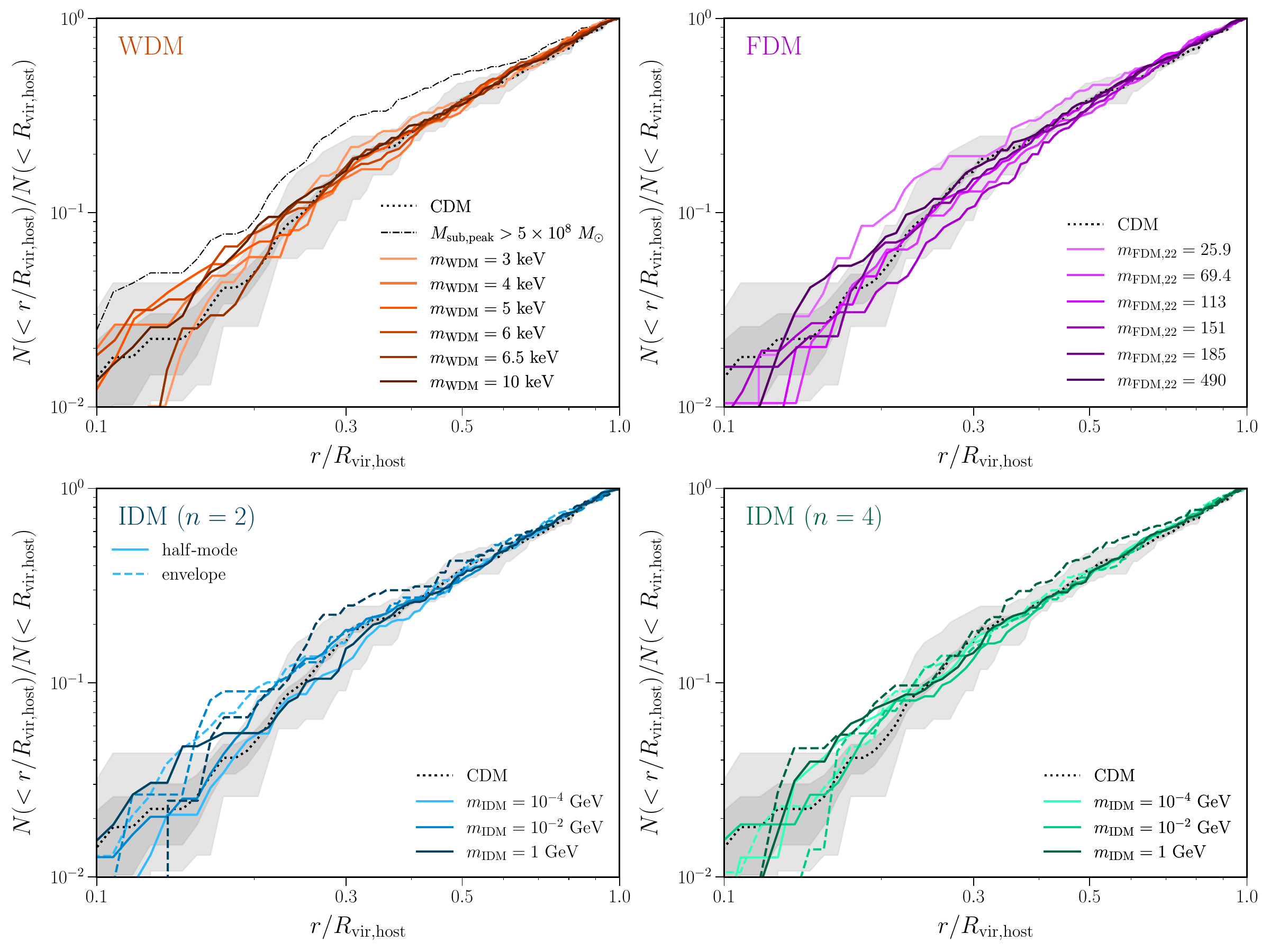}
    \caption{Mean subhalo radial distributions, averaged over our three MW hosts, in beyond-CDM cosmologies (solid colored lines) and in CDM (dotted black lines). Each radial distribution is normalized to the total number of subhalos within $R_{\mathrm{vir,host}}$. All results are restricted to subhalos with present-day virial masses $M_{\mathrm{sub}}>1.2\times 10^8~M_{\mathrm{\odot}}$ except for the dotted--dashed line in the top left panel, which shows the CDM radial distribution for subhalos with $M_{\mathrm{sub,peak}}>5\times 10^8~M_{\mathrm{\odot}}$. In the IDM panels, solid (dashed) lines show models with $P(k)$ matched to the half-mode scale of (strictly more suppressed than) the $m_{\mathrm{WDM}}=6.5~\mathrm{keV}$ model. Dark-gray bands show the $1\sigma$ Poisson uncertainty on the mean normalized radial distribution, and light-gray bands show the range of host-to-host variation.}
    \label{fig:radial}
\end{figure*}

Figure~\ref{fig:radial} shows subhalo radial distributions averaged over our three MW hosts. Before averaging, we normalize each radial distribution to the total number of subhalos within the corresponding host's virial radius, in order to highlight trends in the shape (since the total subhalo abundance differs in different beyond-CDM models). We apply the fiducial cut of $M_{\mathrm{sub}}>1.2\times 10^8~M_{\mathrm{\odot}}$ in all cases. The hosts all have virial radii of $R_{\mathrm{vir,host}}\approx 300~\mathrm{kpc}$, such that $r/R_{\mathrm{vir,host}}=0.1$, $0.3$, and $0.5$ roughly correspond to distances of $30$, $100$, and $150~\mathrm{kpc}$, respectively.

Normalized radial distributions in most of our beyond-CDM simulations converge to the CDM result for $r/R_{\mathrm{vir,host}}\gtrsim 0.3$. For $r/R_{\mathrm{vir,host}}\lesssim 0.3$, certain beyond-CDM radial distributions are more concentrated than in CDM, though the difference does not exceed the host-to-host scatter at any fixed radial distance. Since the abundances of low-mass subhalos are reduced in beyond-CDM cosmologies, and since the more massive subhalos tends to be more concentrated toward the center of the host (e.g., \citealt{Nagai0408273,Kravtsov09063295,Nadler220902675}), the mild enhancement of the inner radial distribution visible in some of the models could be a combination of these two effects. To illustrate this mass dependence, the top left panel of Figure~\ref{fig:radial} shows the radial distribution of subhalos with $M_{\mathrm{sub,peak}}>5\times 10^8~M_{\mathrm{\odot}}$ in CDM. This yields a more concentrated radial distribution, confirming previous results.

We note that radial distributions for individual resimulated hosts can be noisy owing to stochastic changes in subhalo orbits. In particular, particle trajectories in cosmological simulations are potentially chaotic (a form of the `butterfly effect'; \citealt{Genel180707084}); similar behavior has been noted for subhalo orbits in previous studies, even when the same host is resimulated in a fixed model at different resolution levels (e.g., \citealt{Frenk9906160,Springel08090898}). This kind of stochasticity may partly explain the nonmonotonic relation between radial distribution shape and the severity of beyond-CDM suppression shown in Figure~\ref{fig:radial}. Radial distributions can also be affected by orbital phase shifts of matched subhalo pairs that experience different gravitational potentials after falling into the MW host. However, density profiles of our MW hosts are not significantly altered relative to CDM, except in the most extreme beyond-CDM models (i.e., the envelope IDM scenarios).

To further unpack our radial distribution results, we match subhalos among our simulations, based on their pre-infall mass accretion histories and orbits, regardless of whether they survive to $z=0$. The most massive subhalos in our beyond-CDM simulations have orbits nearly identical to their CDM counterparts. While the orbital phases of lower-mass subhalos often shift, the resulting differences in present-day distance are not systematic and mainly introduce scatter in the comparison between CDM and beyond-CDM radial distributions. Thus, the radial distribution differences we identify are due to the suppression of low-mass subhalo abundances in beyond-CDM scenarios rather than differences in the orbits of matched subhalo pairs. Consistent with this result, \cite{Lovell210403322} find that WDM radial distributions are more concentrated in CDM at sufficiently low infall masses, indicating that recently accreted, low-mass CDM subhalos are largely responsible for this difference.

\subsection{Subhalo Formation Times}
\label{sec:formation_time}

Structure formation is delayed in models with suppressed linear matter power spectra. This effect has been quantified in cosmological WDM simulations (e.g., \citealt{Bose150701998}), but has not been studied extensively for subhalos in zoom-in simulations for all of the beyond-CDM scenarios we consider. To quantify this effect, we measure the distribution of half-mass formation scale factor, $a_{1/2}$---defined by $M_{\mathrm{sub}}(a_{1/2})\equiv0.5\times M_{\mathrm{sub}}(a=1)$---for subhalos with $M_{\mathrm{sub}}(z=0)>1.2\times 10^8~M_{\mathrm{\odot}}$ in all of our simulations and models.

\begin{figure*}[t!]
\centering
\includegraphics[width=\textwidth]{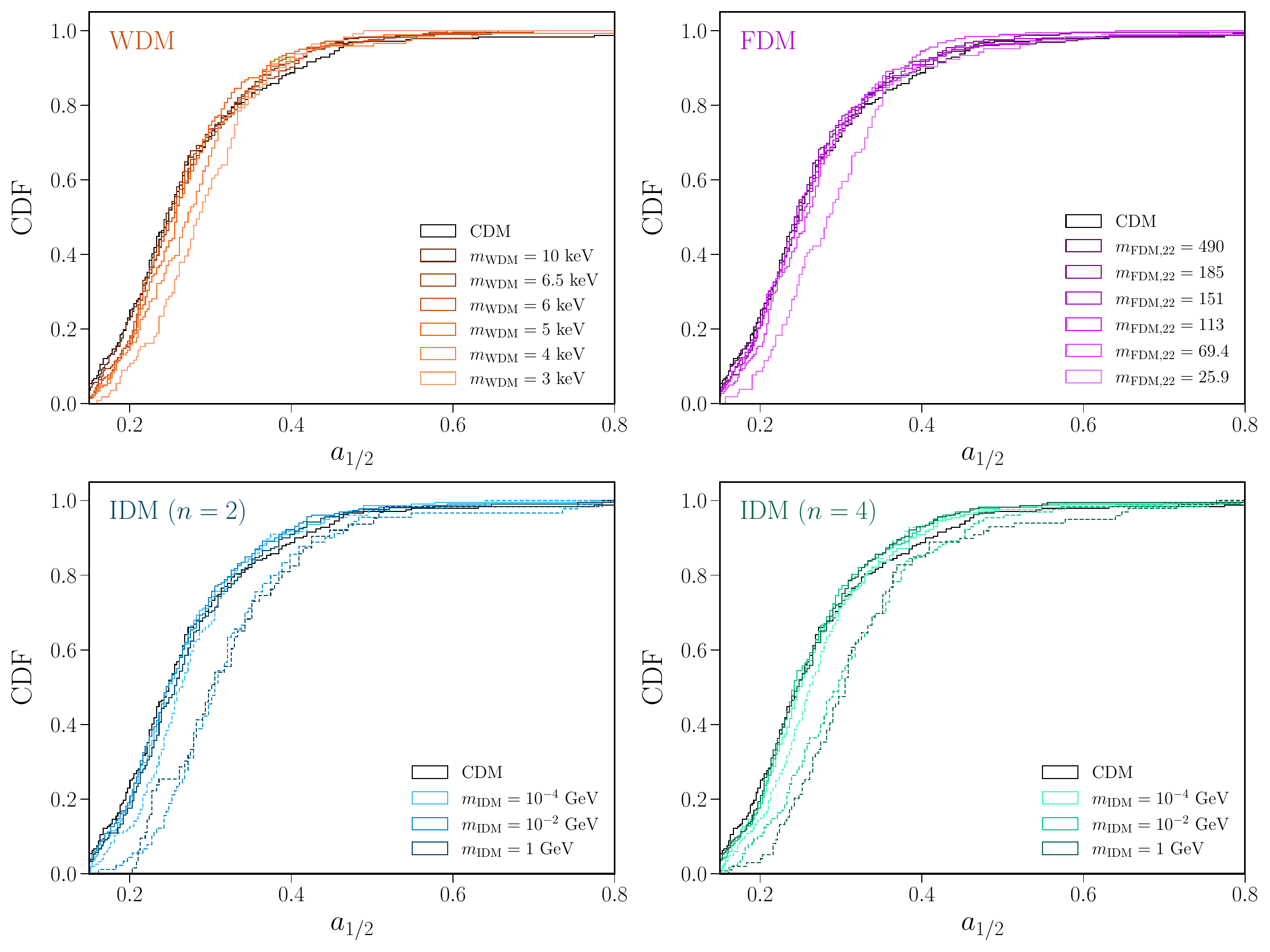}
    \caption{Normalized cumulative distributions of half-mass scale factor---defined by $M_{\mathrm{sub}}(a_{1/2})\equiv0.5\times M_{\mathrm{sub}}(a=1)$---for subhalos with $M_{\mathrm{sub}}>1.2\times 10^8~M_{\mathrm{\odot}}$ in four beyond-CDM cosmologies. In the IDM panels, solid (dashed) lines correspond to half-mode (envelope) models.}
    \label{fig:a12_all}
\end{figure*}

Figure~\ref{fig:a12_all} shows normalized cumulative distributions of $a_{1/2}$ in all of our simulations, stacked over subhalos in all three hosts. Subhalo half-mass formation times are systematically delayed in our beyond-CDM simulations, and this delay grows for models with linear matter power spectrum cutoffs on larger scales. For example, the median $a_{1/2}$ in our CDM ($m_{\mathrm{WDM}}=3~\mathrm{keV}$) simulation is $0.25$ ($0.29$), corresponding to an $\approx 0.5~\mathrm{Gyr}$ shift in half-mass formation time; we observe similar shifts in the $m_{\mathrm{FDM,22}}=25.9$ and envelope IDM models. Meanwhile, $a_{1/2}$ distributions for our WDM models with $m_{\mathrm{WDM}}\gtrsim 5~\mathrm{keV}$, FDM models with $m_{\mathrm{FDM,22}}\gtrsim 25.9$, and half-mode IDM models are very similar to that in CDM. Note that \cite{Bose150701998} found a similarly large shift in $a_{1/2}$ for (sub)halos in an $m_{\mathrm{WDM}}=3.3~\mathrm{keV}$ cosmological simulation.

\section{Convergence Tests}
\label{sec:convergence}

To test that our key results are converged, we run high-resolution (HR) resimulations in CDM and all beyond-CDM models for Halo004, for a total of 25 resimulations. These runs use five \textsc{MUSIC} refinement regions, i.e., an equivalent of 16,384 particles per side in the highest-resolution regions, and a particle mass of $m_{\mathrm{part,HR}}=5.0\times 10^4~M_{\mathrm{\odot}}$. We set the comoving Plummer-equivalent gravitational softening to $\epsilon_{\mathrm{HR}}=80~\mathrm{pc}~h^{-1}$, following \cite{Nadler220902675}.

The results of our convergence tests are in good agreement with Symphony and Milky Way-est CDM results \citep{Nadler220902675,Buch240408043}. Specifically, in both CDM and all beyond-CDM models we simulate, the present-day SHMF is converged at the $\approx 10\%$ level down to our fiducial $M_{\mathrm{sub}}>1.2\times 10^8~M_{\mathrm{\odot}}$ ($300$-particle) threshold. In the absence of an $M_{\mathrm{sub}}$ resolution cut, the $M_{\mathrm{sub,peak}}$ only reaches this level of convergence above an $M_{\mathrm{sub,peak}}\gtrsim 4\times 10^8~M_{\mathrm{\odot}}$ ($1000$-particle) threshold owing to subhalos that are stripped below the present-day mass resolution limit (see the discussion in \citealt{Nadler220902675}). However, our main results concern the suppression of the peak SHMF subject to an $M_{\mathrm{sub}}$ cut, and we find that this quantity is better converged both in CDM and across our beyond-CDM models.

\begin{figure}[t!]
\hspace{-5mm}
\includegraphics[width=0.49\textwidth]{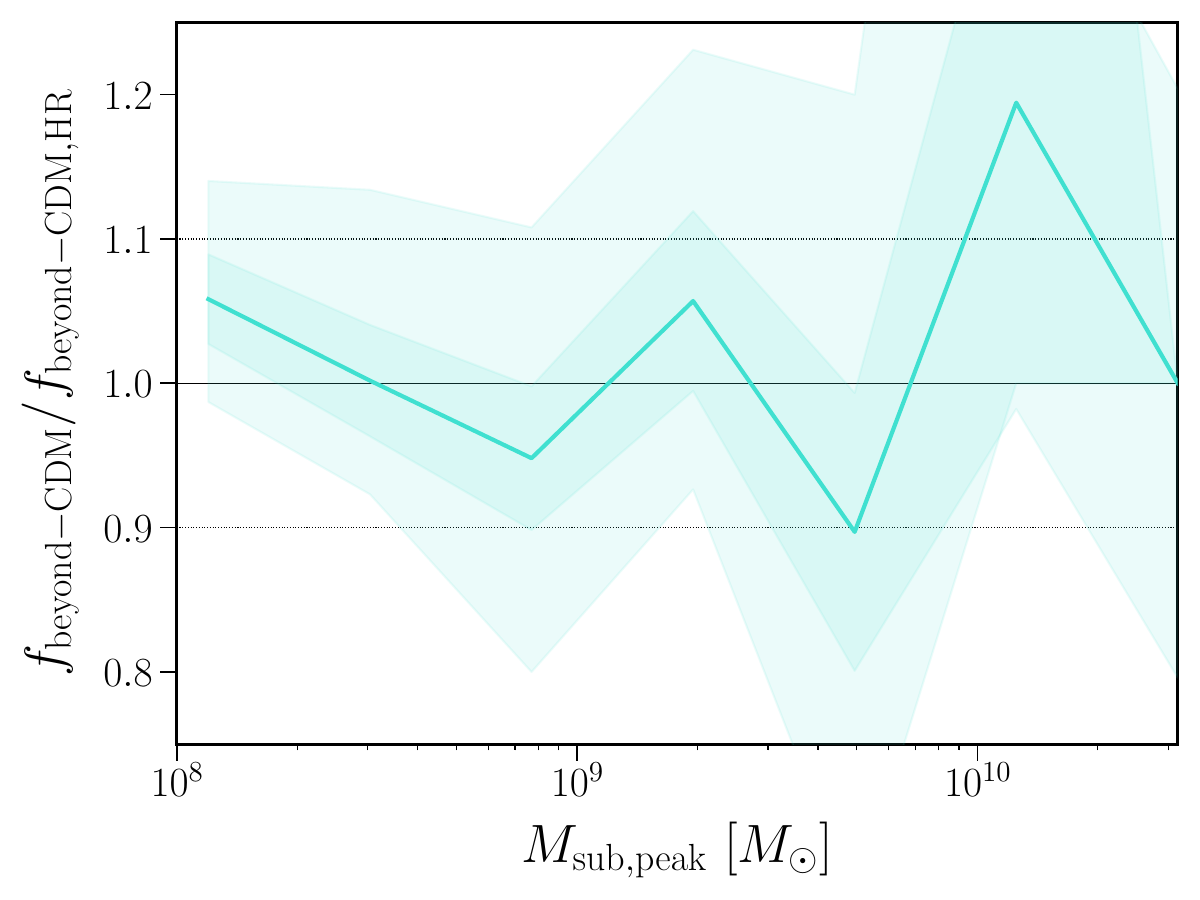}
    \caption{Ratio of the differential beyond-CDM peak SHMF relative to CDM, for subhalos with $M_{\mathrm{sub}}>1.2\times 10^8~M_{\mathrm{\odot}}$ in our fiducial-resolution simulations of Halo004, relative to the same quantity in our HR resimulations of the same host. The bold line shows the mean ratio stacked across all $24$ beyond-CDM models; dark (light) bands show $1\sigma$ Poisson uncertainty (model-to-model scatter).}
    \label{fig:shmf_convergence}
\end{figure}

Quantitatively, Figure~\ref{fig:shmf_convergence} shows the ratio of the peak SHMF suppression in our standard versus HR simulations; we do not find systematic differences in the behavior of this ratio across our beyond-CDM models, so we combine all beyond-CDM models in this figure. Our measurement of $f_{\mathrm{beyond-CDM}}$ is converged at the $\approx 10\%$ level for the lowest-mass subhalos we resolve in our fiducial-resolution simulations, given both the statistical uncertainty and model-to-model scatter. Thus, our SHMF suppression fits, which are driven by the lowest-mass subhalos we resolve, are well converged. We further verify this by refitting our WDM and FDM SHMF models to the HR simulations alone; the resulting posteriors are consistent with our fiducial results. Subhalo radial distributions are more variable between resolution levels, similar to the results in \cite{Nadler220902675}, which may reflect stochasticity in subhalo orbits and/or slower convergence of the radial distribution. Nonetheless, inner normalized radial distributions remain more concentrated than CDM in our beyond-CDM HR simulations, supporting our findings in Appendix~\ref{sec:basic_rad}.

The tests above are necessary (but not sufficient) to demonstrate convergence to a physical solution. For example, \cite{Mansfield230810926} use a particle-tracking method to show that \textsc{Rockstar} and {\sc consistent-trees} (the tools we use in this study) identify too few subhalos in Symphony CDM simulations, particularly for highly stripped systems in hosts' inner regions. We leave a particle-tracking analysis of COZMIC simulations to future work.

\section{Artificial Fragmentation}
\label{sec:fragmentation}

Previous analyses of beyond-CDM simulations have derived a resolution-dependent mass limit, $M_{\mathrm{lim}}$ below which a significant fraction of subhalos are formed spuriously, i.e., through artificial fragmentation. Specifically, they define the characteristic mass \citep{Wang0702575}
\begin{equation}
    M_{\mathrm{lim}} = 10.1\bar{\rho} d k_{\mathrm{peak}}^{-2},\label{eq:mlim}
\end{equation}
where $\bar{\rho}$ is the average matter density, $d$ is the mean interparticle spacing ($22~\mathrm{kpc}$ for our fiducial-resolution simulations; \citealt{Nadler220902675}), and $k_{\mathrm{peak}}$ is the wavenumber at which the dimensionless linear matter power spectrum, $\Delta^2(k)\equiv k^3 P(k)/2\pi$, is maximized. \cite{Lovell13081399} find that spurious subhalos significantly contribute to the total population at masses below $\kappa M_{\mathrm{lim}}$, where $\kappa \approx 0.5$.

For each beyond-CDM model, we obtain $k_{\mathrm{peak}}$ from our \textsc{CLASS} or \textsc{axionCAMB} output. Plugging these into Equation~\ref{eq:mlim} yields $M_{\mathrm{lim}}=[0.53,0.34,0.21,0.13,0.11,0.05]\times 10^8~M_{\mathrm{\odot}}$ for $m_{\mathrm{WDM}}=[3,4,5,6,6.5,10]~\mathrm{keV}$, $M_{\mathrm{lim}}=[0.31,0.13,0.09,0.07,0.05,0.02]\times 10^8~M_{\mathrm{\odot}}$ for $m_{\mathrm{FDM,22}}=[25.9,69.4,113,151,185,490]$, $M_{\mathrm{lim}}=[0.25,1.41,1.49]\times 10^8~M_{\mathrm{\odot}}$ ($[0.08,0.07,0.08]\times 10^8~M_{\mathrm{\odot}}$) for our $n=2$ envelope (half-mode) models with $m_{\mathrm{IDM}}=10^{-4}$, $10^{-2}$, and $1~\mathrm{GeV}$, and $M_{\mathrm{lim}}=[0.20,0.13,1.48]\times 10^8~M_{\mathrm{\odot}}$ ($[0.09,0.003,0.07]\times 10^8~M_{\mathrm{\odot}}$) for our $n=4$ envelope (half-mode) models with $m_{\mathrm{IDM}}=10^{-4}$, $10^{-2}$, and $1~\mathrm{GeV}$. Thus, as described in Section~\ref{sec:post-process}, $\kappa M_{\mathrm{lim}}<M_{\mathrm{res}} = 1.2\times 10^8~M_{\mathrm{\odot}}$ for all of our simulations; moreover, our only simulations with $M_{\mathrm{lim}}\gtrsim 10^8~M_{\mathrm{\odot}}$ are a subset of our envelope IDM runs.

\begin{figure}[t!]
\hspace{-5mm}
\includegraphics[width=0.49\textwidth]{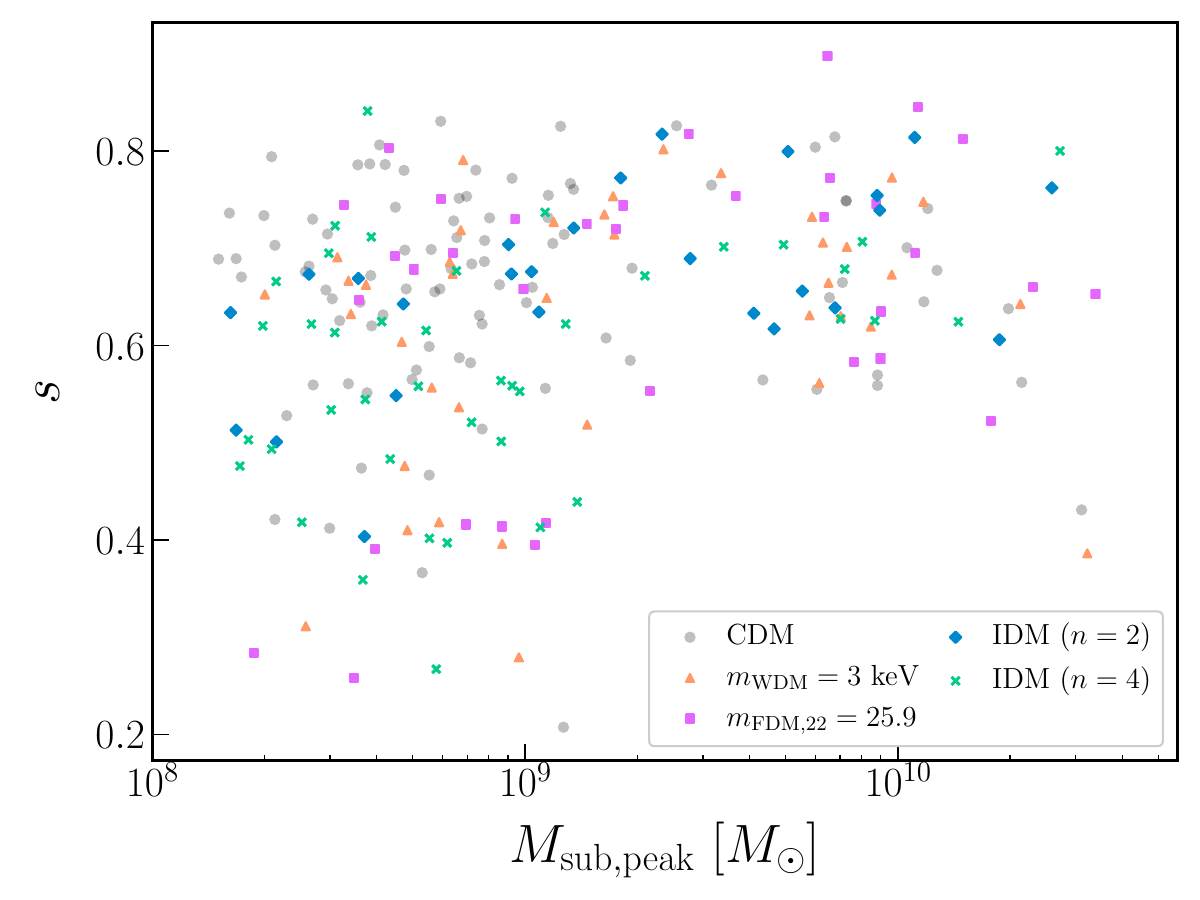}
    \caption{Protohalo sphericity versus peak virial mass for each subhalo with $M_{\mathrm{sub}}>1.2\times 10^8~M_{\mathrm{\odot}}$ in our CDM simulation of Halo004 (gray circles), compared to the same halo simulated with ICs for $m_{\mathrm{WDM}}=3~\mathrm{keV}$ (orange triangles), $m_{\mathrm{FDM,22}}=25.9$ (magenta squares), and $m_{\mathrm{IDM}}=1~\mathrm{GeV}$ envelope models with $n=2$ (blue diamonds) and $n=4$ (green crosses). Note that these models provide an upper limit on the spurious halo contribution in each beyond-CDM scenario we consider; thus, spurious halos contribute negligibly to our results.}
    \label{fig:s_mpeak}
\end{figure}

To more directly demonstrate that spurious halos do not impact our results, we study Halo004 simulations with the largest $M_{\mathrm{lim}}$ in each beyond-CDM scenario: $m_{\mathrm{WDM}}=3~\mathrm{keV}$, $m_{\mathrm{FDM,22}}=25.9$, and $n=2$ and $n=4$ envelope IDM models with $m_{\mathrm{IDM}}=1~\mathrm{GeV}$; these cases provide upper limits on the spurious halo contribution in each scenario. Following \cite{Lovell13081399}, we identify main-branch progenitors of present-day subhalos with $M_{\mathrm{sub}}>1.2\times 10^8~M_{\mathrm{\odot}}$ in each of these simulations. We select each progenitor at the redshift, $z_{1/2}$, when its mass first reaches $0.5M_{\mathrm{sub,peak}}$ before infall into any larger halo. We then find all high-resolution particles within each progenitor's virial radius at $z_{1/2}$ and identify a protohalo by tracing those particles back to the ICs. For each protohalo in the ICs, we calculate the inertia tensor
\begin{equation}
    I_{ij} = \sum_{\mathrm{protohalo\ particles}} m_{\mathrm{part}}\left(\delta_{ij}\lvert \mathbf{x}\rvert^2-x_ix_j\right),
\end{equation}
where $\delta_{ij}$ is the Kronecker delta and $x_i$ are position vector components. We calculate the eigenvalues $a \geq b \geq c$ of this tensor and define the sphericity
\begin{equation}
    s \equiv c/a.
\end{equation}
Note that all of our sphericity measurements are performed using at least $150$ particles, since we only consider subhalos with $M_{\mathrm{sub,peak}}\geq M_{\mathrm{sub}}(z=0)>M_{\mathrm{res}}=300m_{\mathrm{part}}$ and then identify their progenitors at $z_{1/2}$.

\begin{figure*}[t!]
\hspace{-5mm}
\includegraphics[trim={0 0.35cm 0 -1cm},width=0.5\textwidth]{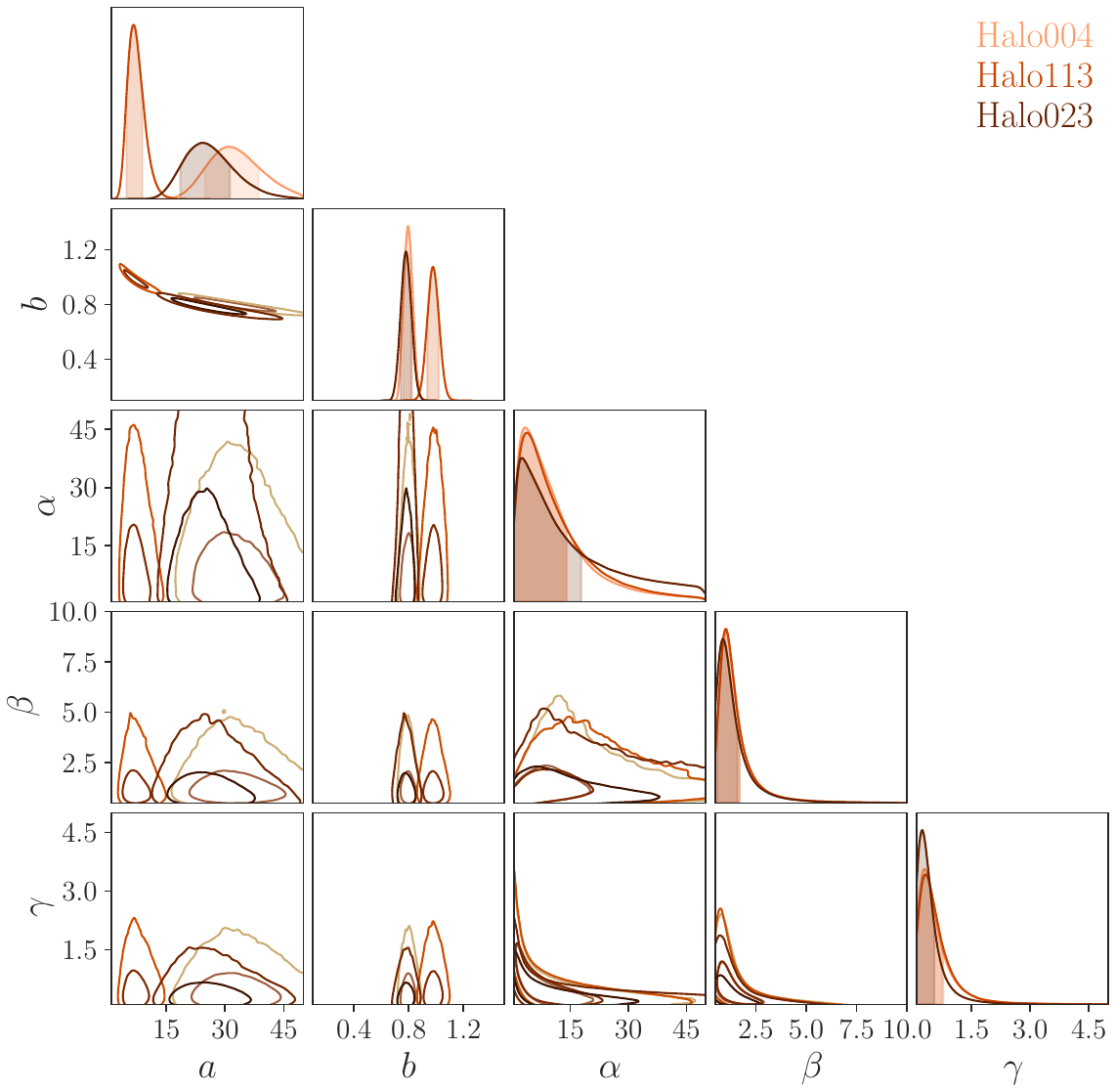}
\includegraphics[trim={0 0.35cm 0 -1cm},width=0.5\textwidth]{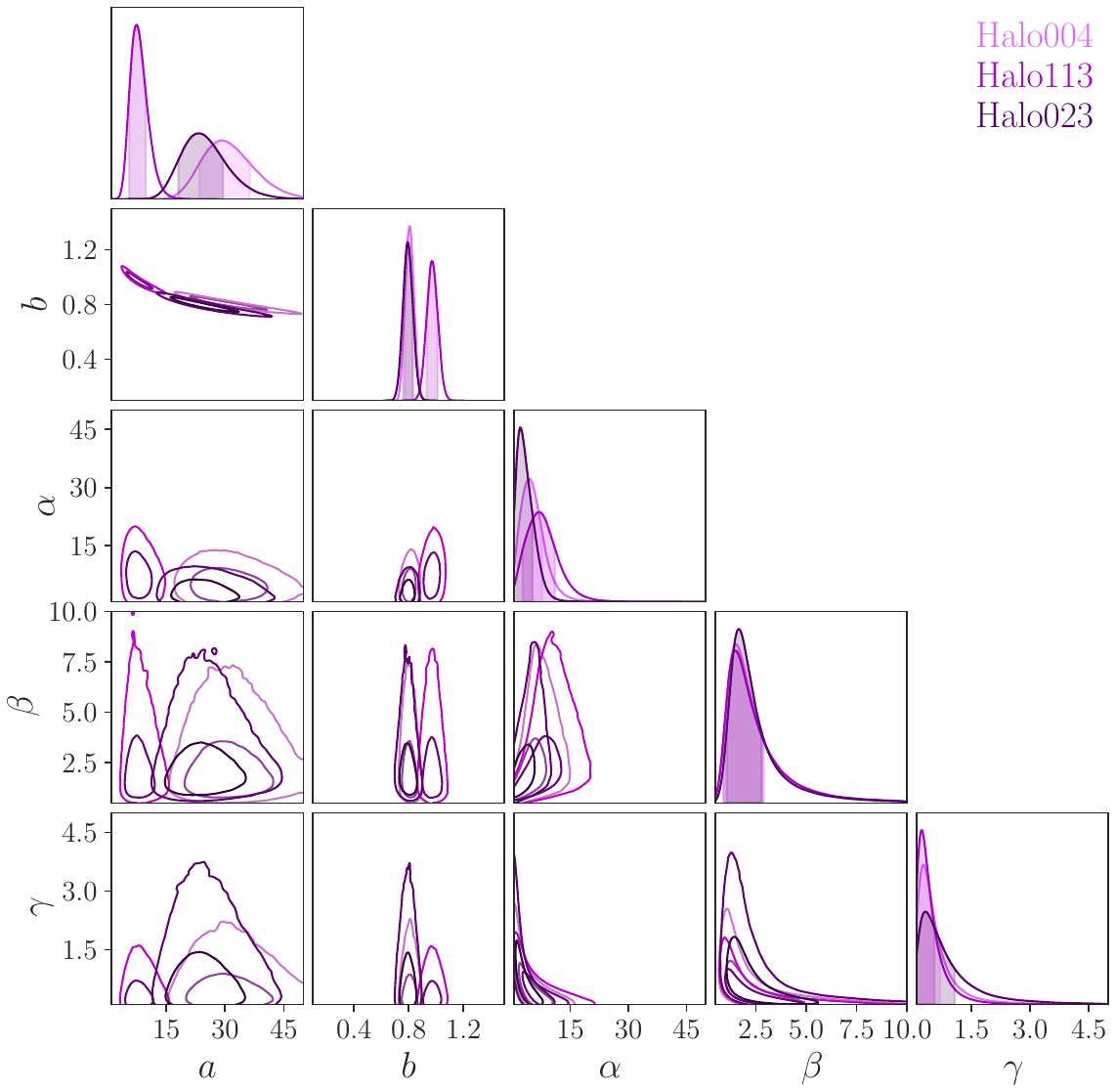}
    \caption{Marginalized posterior for our WDM (left) and FDM (right) SHMF model fit to Halo004, Halo113, and Halo023, from lightest to darkest colors. Dark (light) two-dimensional contours show $68\%$ ($95\%$) confidence intervals; top and side panels show marginal posteriors with shaded $68\%$ confidence intervals.}
    \label{fig:wdm_shmf_posterior_all}
\end{figure*}

\cite{Lovell13081399} find that $99\%$ of their CDM protohalos have sphericities $s>0.16$; the average sphericity of their CDM protohalo sample is $\langle s \rangle \approx 0.4$. These authors thus use $s>0.16$ as a threshold to remove spurious halos. Our CDM subhalos with $M_{\mathrm{sub}}>1.2\times 10^8~M_{\mathrm{\odot}}$ have similarly high sphericities in the ICs, with a minimum of $s_{\mathrm{min}}=0.21$ and an average of $\langle s \rangle \approx 0.66$ for the corresponding protohalos. Figure~\ref{fig:s_mpeak} shows that the sphericity distributions of resolved subhalos in our most extreme beyond-CDM simulations are very similar to CDM. In particular, we find $s_{\mathrm{min}}=[0.28,0.26,0.40,0.27]$ and $\langle s \rangle \approx [0.62,0.64,0.67,0.59]$ for protohalos in $m_{\mathrm{wdm}}=3~\mathrm{keV}$, $m_{\mathrm{FDM,22}}=25.9$, and $n=2$ and $n=4$ envelope IDM models with $m_{\mathrm{IDM}}=1~\mathrm{GeV}$, respectively.

Thus, even in our most extreme beyond-CDM simulations, the subhalos we analyze form from particle distributions with similar shapes to CDM. This implies that spurious halos contribute negligibly to our results. Furthermore, we show in Appendix~\ref{sec:formation_time} that the formation time distribution of subhalos in our beyond-CDM subhalos is a smooth function of parameters like $m_{\mathrm{WDM}}$, which can be violated if a significant fraction of subhalos form spuriously (e.g., \citealt{Bose150701998}).

\section{Host-to-host Variation in Subhalo Mass Function and Suppression Parameters}
\label{sec:shmf_details}

To derive our SHMF suppression results in Section~\ref{sec:shmf}, we simultaneously fit our SHMF model in Equation~\ref{eq:shmf_model} to all three hosts we simulate. This model has independent normalizations ($a_i$) and slopes ($b_i$) for the three hosts but shared suppression parameters ($\alpha$, $\beta$, and $\gamma$). Here we show that fitting our MW-like (Halo004 and Halo113) and MW-mass (Halo023) hosts individually yields consistent results for the WDM and FDM SHMF suppression parameters, and we compare the derived normalizations and slopes.

Specifically, the left panel of Figure~\ref{fig:wdm_shmf_posterior_all} shows the posterior for each host from our WDM SHMF fit. Critically, both WDM and FDM SHMF suppression parameters are very similar among all three hosts we study, which sample different environments, accretion histories, and subhalo populations. It will be interesting to assess this universality further using a wider range of environments and hosts (e.g., across Symphony suites; \citealt{Nadler220902675}) and for lower-mass (sub)halos that probe $P(k)$ on deeply suppressed scales.

Meanwhile, for Halo004, Halo113, and Halo023, we respectively derive a normalization (slope) of $a_1=31.1^{+7.6}_{-6.4}$ ($b_1=0.80^{+0.03}_{-0.03}$), $a_2=6.6^{+2.3}_{-1.8}$ ($b_2=0.98^{+0.04}_{-0.04}$), and $a_3=24.2^{+7.1}_{-5.6}$ ($b_3=0.78^{+0.04}_{-0.04}$). Thus, the inferred normalization and slope is consistent between Halo004 and Halo023, but the normalization (slope) is significantly lower (higher) in Halo113. We speculate that these differences between Halo004 and Halo113, which are both drawn from the Milky Way-est suite, are due to the larger LMC-associated subhalo contribution in Halo004 (which hosts 10 LMC-associated subhalos down to our resolution limit) versus Halo113 (which hosts zero; \citealt{Buch240408043}). Meanwhile, Halo023 does not contain an LMC analog and is thus expected to have a lower SHMF normalization than a ``typical'' Milky Way-est host like Halo004, as shown by \cite{Buch240408043}, consistent with our findings.\footnote{Among the three systems, Halo023 hosts the fewest subhalos with $M_{\mathrm{sub}}>1.2\times 10^8~M_{\mathrm{\odot}}$ in CDM despite having a higher host mass than Halo004 and Halo113, consistent with this picture; see Table~\ref{tab:sims}.} Similar conclusions hold when comparing the derived normalizations and slopes from our FDM SHMF fits of individual hosts, shown in the right panel of Figure~\ref{fig:wdm_shmf_posterior_all}.

For each host, Figure~\ref{fig:wdm_shmf_posterior_all} also demonstrates that the inferred CDM SHMF normalization and slope are consistent between our WDM and FDM fits, indicating that these fits capture the CDM limit of our simulation results consistently.

\section{Milky Way Satellite Inference Posteriors}
\label{sec:full_posteriors}

\begin{figure*}[t!]
\centering
    \includegraphics[scale=0.61]{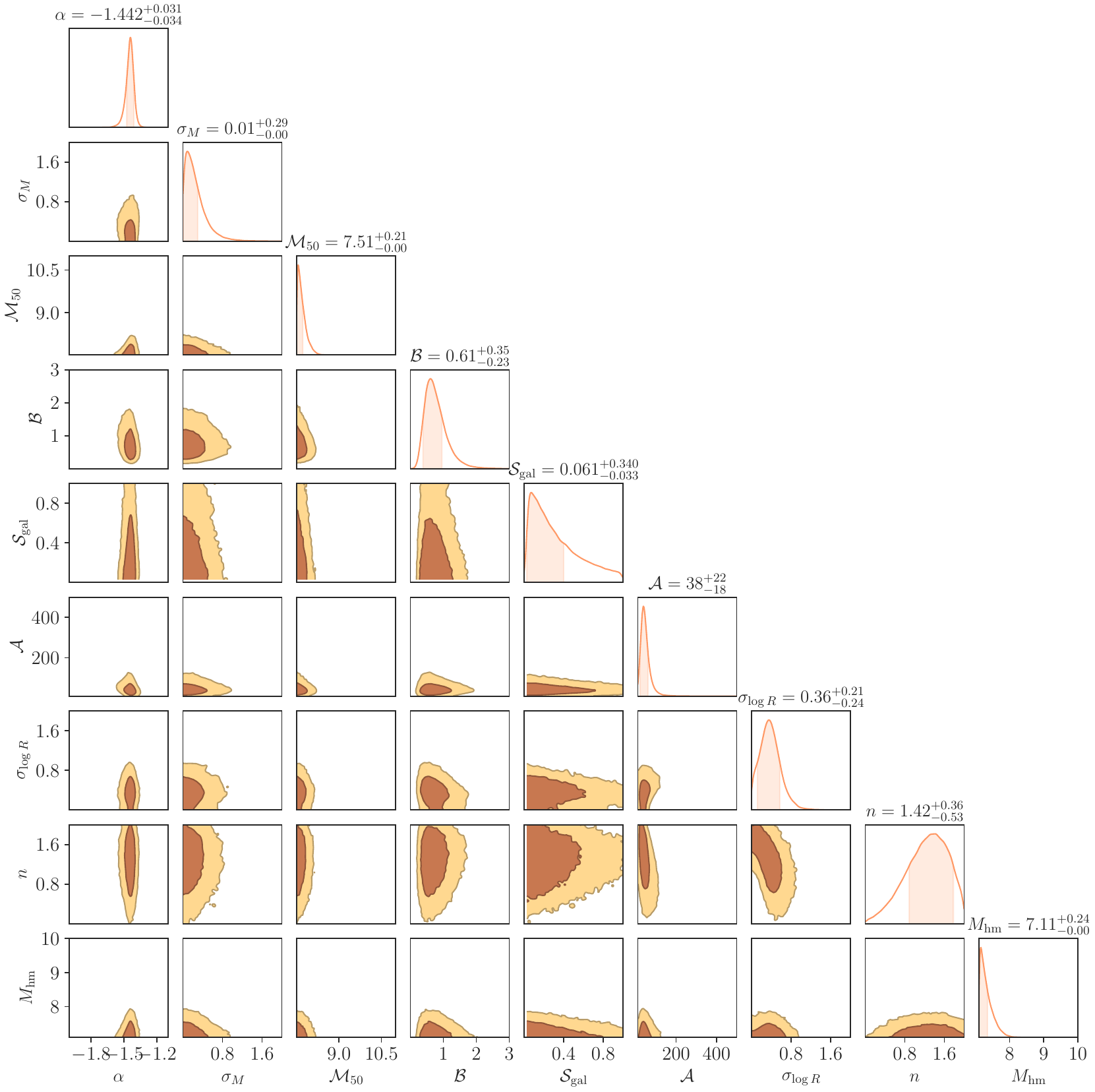}
    \caption{Posterior distribution for our WDM MW satellite population inference from Section~\ref{sec:limits}. Dark (light) contours represent $68\%$ ($95\%$) confidence intervals. We report the galaxy--halo luminosity scatter $\sigma_M$ and size scatter $\sigma_{\log R}$ in $\rm{dex}$, the $50\%$ galaxy occupation peak virial mass $\mathcal{M}_{50}$ as $\log(\mathcal{M}_{50}/M_{\rm{\odot}})$, the amplitude of the galaxy--halo size relation $\mathcal{A}$ in parsecs, and the half-mode mass $M_{\mathrm{hm}}$ as $\log(M_{\mathrm{hm}}/M_{\rm{\odot}})$; the faint-end luminosity function slope $\alpha$ (note that this is different from the scale $\alpha$ used in our WDM and FDM SHMF suppression fits), disk disruption efficiency $\mathcal{B}$, galaxy occupation fraction tilt $\mathcal{S}_{\mathrm{gal}}$, and galaxy--halo size relation slope $n$ are dimensionless.} \label{fig:wdm_full_posterior}
\end{figure*}

\begin{figure*}[t!]
\centering
    \includegraphics[scale=0.61]{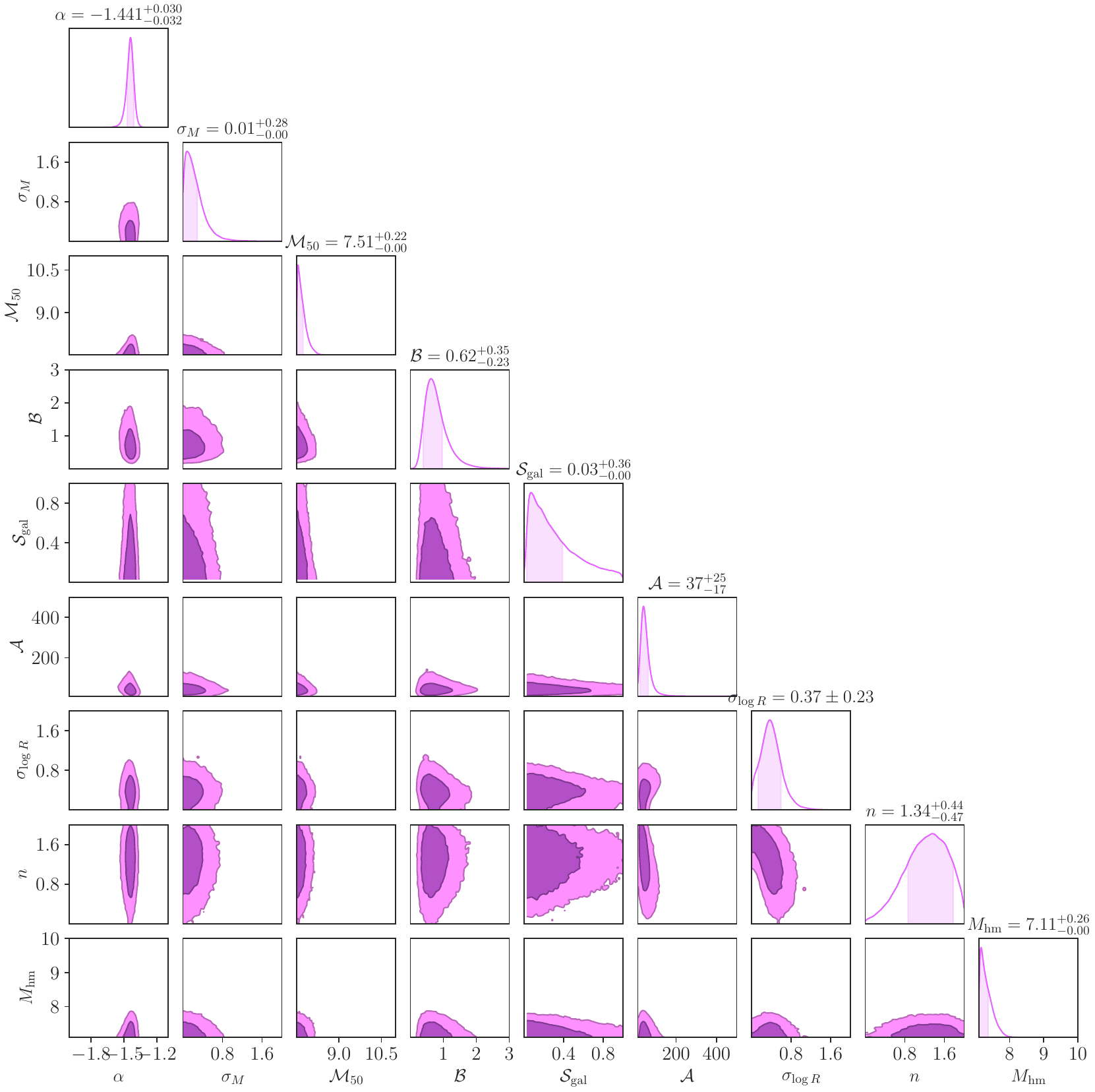}
    \caption{Same as Figure~\ref{fig:wdm_full_posterior}, but for our FDM MW satellite population inference.} \label{fig:fdm_full_posterior}
\end{figure*}

We present the posteriors from our WDM and FDM MW satellite fits in Figures~\ref{fig:wdm_full_posterior} and \ref{fig:fdm_full_posterior}, respectively. Degeneracies between galaxy--halo connection parameters and $M_{\mathrm{hm}}$ are mild, and the galaxy--halo connection constraints we derive are consistent with those reported in \cite{Nadler200800022}.

\end{document}